\pdfoutput=1
\documentclass[12pt,a4paper]{article}

\usepackage{ifthen} 
\newboolean{pdflatex}
\setboolean{pdflatex}{true} 

\newboolean{articletitles}
\setboolean{articletitles}{true} 

\newboolean{uprightparticles}
\setboolean{uprightparticles}{false} 

\def\paperauthors{LHCb collaboration} 
\def\paperasciititle{Measurement of the branching fractions and longitudinal polarisations of B(s)0 -> K*(892)0 K*(892)0 decays} 
\def\papertitle{Measurement of the branching fractions and longitudinal polarisations of $B^0_{(s)} \to K^{*0} \kern 0.18em \overline{\kern -0.18em K}{}^{*0}$ decays} 
\def\paperkeywords{{flavour physics}, {B physics}, {branching fraction}, {polarisation}} 
\def\papercopyright{\the\year\ CERN for the benefit of the LHCb collaboration} 
\def\paperlicence{CC BY 4.0 licence}
\def\paperlicenceurl{https://creativecommons.org/licenses/by/4.0/}

\newif\ifEnableSectionTOCLinks
\EnableSectionTOCLinksfalse 

\usepackage[top=1in, bottom=1.25in, left=1in, right=1in]{geometry}

\usepackage{microtype}
\usepackage{lineno}  
\usepackage{xspace} 
\usepackage{caption} 

\usepackage{graphicx}  
\usepackage{color}
\usepackage{colortbl}
\graphicspath{{./figs/}} 

\usepackage{amsmath} 
\usepackage{amssymb}
\usepackage{amsfonts}
\usepackage{upgreek} 

\newcommand*\patchAmsMathEnvironmentForLineno[1]{%
\expandafter\let\csname old#1\expandafter\endcsname\csname #1\endcsname
\expandafter\let\csname oldend#1\expandafter\endcsname\csname
end#1\endcsname
 \renewenvironment{#1}%
   {\linenomath\csname old#1\endcsname}%
   {\csname oldend#1\endcsname\endlinenomath}%
}
\newcommand*\patchBothAmsMathEnvironmentsForLineno[1]{%
  \patchAmsMathEnvironmentForLineno{#1}%
  \patchAmsMathEnvironmentForLineno{#1*}%
}
\AtBeginDocument{%
\patchBothAmsMathEnvironmentsForLineno{equation}%
\patchBothAmsMathEnvironmentsForLineno{align}%
\patchBothAmsMathEnvironmentsForLineno{flalign}%
\patchBothAmsMathEnvironmentsForLineno{alignat}%
\patchBothAmsMathEnvironmentsForLineno{gather}%
\patchBothAmsMathEnvironmentsForLineno{multline}%
\patchBothAmsMathEnvironmentsForLineno{eqnarray}%
}

\usepackage[pdftex,
            pdfauthor={\paperauthors},
            pdftitle={\paperasciititle},
            pdfkeywords={\paperkeywords}]{hyperref}
\usepackage{hyperxmp}
\hypersetup{
    pdfcopyright={Copyright (C) \papercopyright},
    pdflicenseurl={\paperlicenceurl}
}

\usepackage[colorinlistoftodos,textsize=scriptsize]{todonotes}

\usepackage[bottom,flushmargin,hang,multiple]{footmisc}

\usepackage[all]{hypcap} 

\usepackage{xspace} 
\usepackage{upgreek}

\def\lhcb   {\mbox{LHCb}\xspace}

\def\MagUp {\mbox{\em Mag\kern -0.05em Up}\xspace}

\ifthenelse{\boolean{uprightparticles}}%
{
 
 \def\Pgamma      {\ensuremath{\upgamma}\xspace}

 \def\Pmu         {\ensuremath{\upmu}\xspace}

 \def\Ppi         {\ensuremath{\uppi}\xspace}                 
                  
 \def\Prho        {\ensuremath{\uprho}\xspace}

 \def\Pphi        {\ensuremath{\upphi}\xspace}

 \def\Ppsi        {\ensuremath{\uppsi}\xspace}

 \def\PDelta      {\ensuremath{\Delta}\xspace}                 
 \def\PXi         {\ensuremath{\Xi}\xspace}                 
 \def\PLambda     {\ensuremath{\Lambda}\xspace}                 
 \def\PSigma      {\ensuremath{\Sigma}\xspace}                 
 \def\POmega      {\ensuremath{\Omega}\xspace}                 
 \def\PUpsilon    {\ensuremath{\Upsilon}\xspace}
 \let\oldPi\Pi
 \def\PPi         {\ensuremath{\oldPi}\xspace}

 \def\PB      {\ensuremath{\mathrm{B}}\xspace}                 
 \def\PD      {\ensuremath{\mathrm{D}}\xspace}                 
 \def\PJ      {\ensuremath{\mathrm{J}}\xspace}                 
 \def\PK      {\ensuremath{\mathrm{K}}\xspace}                 
 \def\Pb      {\ensuremath{\mathrm{b}}\xspace}                 
 \def\Pc      {\ensuremath{\mathrm{c}}\xspace}                 
 \def\Pd      {\ensuremath{\mathrm{d}}\xspace}                 

 \def\Pp      {\ensuremath{\mathrm{p}}\xspace}                 

 \def\Ps      {\ensuremath{\mathrm{s}}\xspace}

 \def\thebaroffset{0.0em}
}
{
 
 \def\Pgamma      {\ensuremath{\gamma}\xspace}

 \def\Pmu         {\ensuremath{\mu}\xspace}

 \def\Ppi         {\ensuremath{\pi}\xspace}                 
                  
 \def\Prho        {\ensuremath{\rho}\xspace}

 \def\Pphi        {\ensuremath{\phi}\xspace}

 \def\Ppsi        {\ensuremath{\psi}\xspace}                 
                  
 \mathchardef\PDelta="7101
 \mathchardef\PXi="7104
 \mathchardef\PLambda="7103
 \mathchardef\PSigma="7106
 \mathchardef\POmega="710A
 \mathchardef\PUpsilon="7107
 \mathchardef\PPi="7105
 \def\PB      {\ensuremath{B}\xspace}                 
 \def\PD      {\ensuremath{D}\xspace}                 
 \def\PJ      {\ensuremath{J}\xspace}                 
 \def\PK      {\ensuremath{K}\xspace}                 
 \def\Pb      {\ensuremath{b}\xspace}                 
 \def\Pc      {\ensuremath{c}\xspace}                 
 \def\Pd      {\ensuremath{d}\xspace}                 

 \def\Pp      {\ensuremath{p}\xspace}                 

 \def\Ps      {\ensuremath{s}\xspace}

 \def\thebaroffset{0.18em}
}
\newcommand{\offsetoverline}[2][\thebaroffset]{\kern #1\overline{\kern -#1 #2}}%

\makeatletter
\ifcase \@ptsize \relax
  \newcommand{\miniscule}{\@setfontsize\miniscule{4}{5}}
\or
  \newcommand{\miniscule}{\@setfontsize\miniscule{5}{6}}
\or
  \newcommand{\miniscule}{\@setfontsize\miniscule{5}{6}}
\fi
\makeatother

\DeclareRobustCommand{\optbar}[1]{\shortstack{{\miniscule (\rule[.5ex]{1.25em}{.18mm})}
  \\ [-.7ex] $#1$}}

\def\mup        {{\ensuremath{\Pmu^+}}\xspace}
\def\mun        {{\ensuremath{\Pmu^-}}\xspace} 

\def\mumu       {{\ensuremath{\Pmu^+\Pmu^-}}\xspace}

\def\g      {{\ensuremath{\Pgamma}}\xspace}

\def\dquark    {{\ensuremath{\Pd}}\xspace}
\def\dquarkbar {{\ensuremath{\overline \dquark}}\xspace}

\def\squark    {{\ensuremath{\Ps}}\xspace}
\def\squarkbar {{\ensuremath{\overline \squark}}\xspace}

\def\cquark    {{\ensuremath{\Pc}}\xspace}

\def\bquark    {{\ensuremath{\Pb}}\xspace}
\def\bquarkbar {{\ensuremath{\overline \bquark}}\xspace}

\def\pion   {{\ensuremath{\Ppi}}\xspace}
\def\piz    {{\ensuremath{\pion^0}}\xspace}
\def\pip    {{\ensuremath{\pion^+}}\xspace}
\def\pim    {{\ensuremath{\pion^-}}\xspace}
\def\pipm   {{\ensuremath{\pion^\pm}}\xspace}
\def\pimp   {{\ensuremath{\pion^\mp}}\xspace}

\def\rhomeson {{\ensuremath{\Prho}}\xspace}
\def\rhoz     {{\ensuremath{\rhomeson^0}}\xspace}

\def\kaon    {{\ensuremath{\PK}}\xspace}
\def\Kbar    {{\ensuremath{\offsetoverline{\PK}}}\xspace}
\def\Kb      {{\ensuremath{\Kbar}}\xspace}
\def\KorKbar {\kern \thebaroffset\optbar{\kern -\thebaroffset \PK}{}\xspace}
\def\Kz      {{\ensuremath{\kaon^0}}\xspace}

\def\Kp      {{\ensuremath{\kaon^+}}\xspace}
\def\Km      {{\ensuremath{\kaon^-}}\xspace}
\def\Kpm     {{\ensuremath{\kaon^\pm}}\xspace}
\def\Kmp     {{\ensuremath{\kaon^\mp}}\xspace}

\def\Kstarz  {{\ensuremath{\kaon^{*0}}}\xspace}
\def\Kstarzb {{\ensuremath{\Kbar{}^{*0}}}\xspace}
\def\Kstar   {{\ensuremath{\kaon^*}}\xspace}
\def\Kstarb  {{\ensuremath{\Kbar{}^*}}\xspace}

\newcommand{\phiz}{\ensuremath{\Pphi}\xspace}

\def\D       {{\ensuremath{\PD}}\xspace}

\def\DorDbar {\kern \thebaroffset\optbar{\kern -\thebaroffset \PD}\xspace}
\def\Dz      {{\ensuremath{\D^0}}\xspace}

\def\Dp      {{\ensuremath{\D^+}}\xspace}
\def\Dm      {{\ensuremath{\D^-}}\xspace}

\def\DpDm    {\ensuremath{\Dp {\kern -0.16em \Dm}}\xspace}

\def\Dstarp  {{\ensuremath{\D^{*+}}}\xspace}

\def\Dsp     {{\ensuremath{\D^+_\squark}}\xspace}
\def\Dsm     {{\ensuremath{\D^-_\squark}}\xspace}

\def\Dsmp    {{\ensuremath{\D^{\mp}_\squark}}\xspace}

\def\Dssm    {{\ensuremath{\D^{*-}_\squark}}\xspace}

\def\DmorDsm {{\ensuremath{\D{}_{(\squark)}^-}}\xspace}

\def\B       {{\ensuremath{\PB}}\xspace}
\def\Bbar    {{\ensuremath{\offsetoverline{\PB}}}\xspace}

\def\BorBbar {\kern \thebaroffset\optbar{\kern -\thebaroffset \PB}\xspace}
\def\Bz      {{\ensuremath{\B^0}}\xspace}

\def\Bd      {{\ensuremath{\B^0}}\xspace}

\def\BdorBdbar {\kern \thebaroffset\optbar{\kern -\thebaroffset \Bd}\xspace}

\def\Bs      {{\ensuremath{\B^0_\squark}}\xspace}
\def\Bsb     {{\ensuremath{\Bbar{}^0_\squark}}\xspace}
\def\BsorBsbar {\kern \thebaroffset\optbar{\kern -\thebaroffset \Bs}\xspace}

\def\Bds     {{\ensuremath{\B_{(\squark)}^0}}\xspace}
\def\Bdsb    {{\ensuremath{\Bbar{}_{(\squark)}^0}}\xspace}
\def\BdorBs  {\Bds}
\def\BdorBsbar  {\Bdsb}

\def\jpsi     {{\ensuremath{{\PJ\mskip -3mu/\mskip -2mu\Ppsi}}}\xspace}

\def\Y#1S{\ensuremath{\PUpsilon{(#1S)}}\xspace}

\def\proton      {{\ensuremath{\Pp}}\xspace}

\def\Lz          {{\ensuremath{\PLambda}}\xspace}

\def\LorLbar     {\kern \thebaroffset\optbar{\kern -\thebaroffset \PLambda}\xspace}

\def\Lb           {{\ensuremath{\Lz^0_\bquark}}\xspace}

\newcommand{\decay}[2]{\ensuremath{\mathinner{#1\!\to #2}}\xspace}

\def\to                 {\ensuremath{\rightarrow}\xspace}

\def\CP                {{\ensuremath{C\!P}}\xspace}

\newcommand{\dm}{{\ensuremath{\Delta m}}\xspace}

\newcommand{\DG}{{\ensuremath{\Delta\Gamma}}\xspace}
\newcommand{\DGs}{{\ensuremath{\Delta\Gamma_{\squark}}}\xspace}

\newcommand{\Gs}{{\ensuremath{\Gamma_{\squark}}}\xspace}

\def\AT#1     {\ensuremath{A_{\mathrm{T}}^{#1}}\xspace}           

\def\C#1      {\ensuremath{\mathcal{C}_{#1}}\xspace}                       
\def\Cp#1     {\ensuremath{\mathcal{C}_{#1}^{'}}\xspace}                    
\def\Ceff#1   {\ensuremath{\mathcal{C}_{#1}^{\mathrm{(eff)}}}\xspace}        
\def\Cpeff#1  {\ensuremath{\mathcal{C}_{#1}^{'\mathrm{(eff)}}}\xspace}       
\def\Ope#1    {\ensuremath{\mathcal{O}_{#1}}\xspace}                       
\def\Opep#1   {\ensuremath{\mathcal{O}_{#1}^{'}}\xspace}                    

\newcommand{\nospaceunit}[1]{\ensuremath{\text{#1}}}       
\newcommand{\aunit}[1]{\ensuremath{\text{\,#1}}}       

\newcommand{\tev}{\aunit{Te\kern -0.1em V}\xspace}
\newcommand{\gev}{\aunit{Ge\kern -0.1em V}\xspace}
\newcommand{\mev}{\aunit{Me\kern -0.1em V}\xspace}
\newcommand{\kev}{\aunit{ke\kern -0.1em V}\xspace}
\newcommand{\ev}{\aunit{e\kern -0.1em V}\xspace}
 
\newcommand{\mevc}{\ensuremath{\aunit{Me\kern -0.1em V\!/}c}\xspace}
\newcommand{\gevc}{\ensuremath{\aunit{Ge\kern -0.1em V\!/}c}\xspace}
\newcommand{\mevcc}{\ensuremath{\aunit{Me\kern -0.1em V\!/}c^2}\xspace}
\newcommand{\gevcc}{\ensuremath{\aunit{Ge\kern -0.1em V\!/}c^2}\xspace}

\def\mum  {\ensuremath{\,\upmu\nospaceunit{m}}\xspace}

\def\fm   {\aunit{fm}\xspace}

\def\fb   {\ensuremath{\aunit{fb}}\xspace}
\def\invfb   {\ensuremath{\fb^{-1}}\xspace}

\def\ps   {\ensuremath{\aunit{ps}}\xspace}

\newcommand{\stat}{\aunit{(stat)}\xspace}
\newcommand{\syst}{\aunit{(syst)}\xspace}

\newcommand{\chisq}{\ensuremath{\chi^2}\xspace}

\def\gsim{{~\raise.15em\hbox{$>$}\kern-.85em
          \lower.35em\hbox{$\sim$}~}\xspace}
\def\lsim{{~\raise.15em\hbox{$<$}\kern-.85em
          \lower.35em\hbox{$\sim$}~}\xspace}

\newcommand{\Real}{\ensuremath{\mathcal{R}e}\xspace}
\newcommand{\Imag}{\ensuremath{\mathcal{I}m}\xspace}

\def\sPlot{\mbox{\em sPlot}\xspace}
\def\sFit{\mbox{\em sFit}\xspace}

\def\pt         {\ensuremath{p_{\mathrm{T}}}\xspace}

\def\ptot       {\ensuremath{p}\xspace}

\def\evtgen     {\mbox{\textsc{EvtGen}}\xspace}

\def\geant      {\mbox{\textsc{Geant4}}\xspace}

\def\photos     {\mbox{\textsc{Photos}}\xspace}

\def\pythia     {\mbox{\textsc{Pythia}}\xspace}

\def\hepml      {\mbox{\textsc{HEP\_ML}}\xspace}

\def\tell1  {TELL1\xspace}
\def\ukl1   {UKL1\xspace}

\newcommand{\ie}{\mbox{\itshape i.e.}\xspace}

\newcommand{\lhcborcid}[1]{\href{https://orcid.org/#1}{\hspace*{0.1em}\raisebox{-0.45ex}{\includegraphics[width=1em]{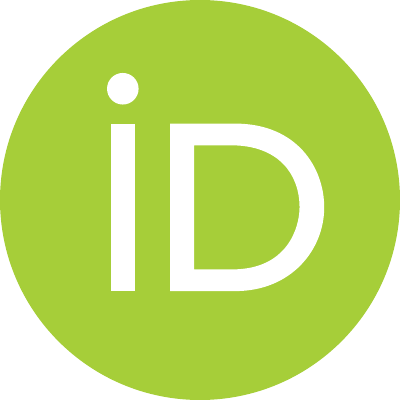}}}}

\hypersetup{
  colorlinks   = true, 
  urlcolor     = blue, 
  linkcolor    = blue, 
  citecolor    = red   
}

\ifEnableSectionTOCLinks
    \usepackage[explicit]{titlesec} 
    
    \let\oldcontentsline\contentsline
    \renewcommand

    \titleformat{\section}{\normalfont\Large\bf}{\hyperlink{tocsection.\thesection}{{\thesection} \parbox[t]{\dimexpr\textwidth-1pc}{#1}}}{1pc}{}

    \titleformat{\subsection}{\normalfont\bf}{\hyperlink{tocsubsection.\thesubsection}{{\thesubsection} \parbox[t]{\dimexpr\textwidth-1pc}{#1}}}{1pc}{}

    \titleformat{name=\section,numberless}[display]{}{}{0pt}{\normalfont\Huge\bfseries #1}
\fi

\usepackage{cite} 
\usepackage{LHCb/mciteplus}

\usepackage{longtable}
\usepackage{subcaption}
\usepackage{placeins}
\usepackage{multirow}

\newcommand{\mc}[3]{\multicolumn{#1}{#2}{#3}}

\usepackage{array}
\newcolumntype{L}[1]{>{\raggedright\let\newline\\\arraybackslash\hspace{0pt}}m{#1}}
\newcolumntype{C}[1]{>{\centering\let\newline\\\arraybackslash\hspace{0pt}}m{#1}}
\newcolumntype{R}[1]{>{\raggedleft\let\newline\\\arraybackslash\hspace{0pt}}m{#1}}

\definecolor{darkgreen}{rgb}{0.0, 0.5, 0.0}
\definecolor{lightgray}{rgb}{0.8,0.8,0.8}
\definecolor{darkgray}{rgb}{0.4,0.4,0.4}
\definecolor{lightblue}{rgb}{0.63,0.79,0.95}
\definecolor{darkblue}{rgb}{0.0, 0.0, 0.55}
\definecolor{yellow}{rgb}{1.0,0.75,0.}

\newcommand{\LKK}{\ensuremath{L_{\Kst\Kstb}}\xspace}

\newcommand{\Dgtwo}{\ensuremath{\frac{\DG}{2}}\xspace}

\newcommand{\coshdgt}{\cosh \Big( \Dgtwo \, t \Big)\xspace}
\newcommand{\sinhdgt}{\sinh \Big( \Dgtwo \, t \Big)\xspace}
\newcommand{\cosdmt}{\cos \big( \dm \, t \big)\xspace}
\newcommand{\sindmt}{\sin \big( \dm \, t \big)\xspace}

\newcommand{\Ab}{\ensuremath{\offsetoverline{A}}\xspace}

\newcommand{\fLBs}{\ensuremath{f_L^s}\xspace}
\newcommand{\fLBd}{\ensuremath{f_L^d}\xspace}

\newcommand{\Kst}{\Kstarz}
\newcommand{\Kstb}{\Kstarzb}

\newcommand{\BdorBsorBar}{\ensuremath{\kern \thebaroffset \optbar{ \kern - \thebaroffset \BdorBs}}\xspace}

\newcommand{\theKstarzTwo}{\ensuremath{\kaon^{*}_2(1430)^0}\xspace}

\newcommand{\BVV}{\decay{B}{VV}}
\newcommand{\BsKstKst}{\decay{\Bs}{\Kstarz\Kstarzb}}

\newcommand{\BdKstKst}{\decay{\Bd}{\Kstarz\Kstarzb}}

\newcommand{\BdsKstKst}{\decay{\BdorBs}{\Kstarz\Kstarzb}}

\newcommand{\BsKpiKpi}{\decay{\Bs}{(\Kp\pim)(\Km\pip)}}
\newcommand{\BdKpiKpi}{\decay{\Bd}{(\Kp\pim)(\Km\pip)}}
\newcommand{\BdsKpiKpi}{\decay{\BdorBs}{(\Kp\pim)(\Km\pip)}}

\newcommand{\BsDspi}{\decay{\Bs}{\Dsm\pip}}
\newcommand{\BsDsK}{\decay{\Bs}{\Dsm\Kp}}
\newcommand{\BdDspi}{\decay{\Bd}{\Dsp\pim}}
\newcommand{\BsDsstpi}{\decay{\Bs}{\Dssm[\to{\Dsm\g}]\pip}}

\newcommand{\BdDpi}{\decay{\Bd}{\Dm\pip}}

\newcommand{\BdDK}{\decay{\Bd}{\Dm\Kp}}

\newcommand{\DsKKpi}{\decay{\Dsm}{\Kp\Km\pim}}
\newcommand{\DKKpi}{\decay{\Dm}{\Kp\Km\pim}}
\newcommand{\BdorBsDpi}{\decay{\BdorBs}{\DmorDsm\pip}}

\newcommand{\DmorDsmKKpi}{\decay{\DmorDsm}{\Kp\Km\pim}}

\newcommand{\br}[1]{{\ensuremath{\mathcal{B}(#1)}}\xspace}
\newcommand{\BfBsKst}{\br{\decay{\Bs}{\Kstarz\Kstarzb}}}
\newcommand{\BfBdKst}{\br{\decay{\Bd}{\Kstarz\Kstarzb}}}

\newcommand{\BfBdorBsKst}{\br{\decay{\BdorBs}{\Kstarz\Kstarzb}}}

\newcommand{\Sb}{\ensuremath{\offsetoverline{S}}\xspace}
\newcommand{\Vb}{\ensuremath{\offsetoverline{V}}\xspace}
\newcommand{\VSb}{\ensuremath{V\!\Sb}\xspace}
\newcommand{\SVb}{\ensuremath{S\Vb}\xspace}
\newcommand{\VS}{\ensuremath{{V\!S}\xspace}}

\usepackage{floatflt}
\usepackage{wrapfig}
\usepackage{pgfgantt}
\usepackage{tikz}
\usetikzlibrary{patterns}
\usepackage{tikzscale}
\usepackage{tikz-feynman}

\usepackage{longtable} 

\begin{document}

\renewcommand{\thefootnote}{\fnsymbol{footnote}}
\setcounter{footnote}{1}

\begin{titlepage}
\pagenumbering{roman}

\vspace*{-1.5cm}
\centerline{\large EUROPEAN ORGANIZATION FOR NUCLEAR RESEARCH (CERN)}
\vspace*{1.5cm}
\noindent
\begin{tabular*}{\linewidth}{lc@{\extracolsep{\fill}}r@{\extracolsep{0pt}}}
\ifthenelse{\boolean{pdflatex}}
{\vspace*{-1.5cm}\mbox{\!\!\!\includegraphics[width=.14\textwidth]{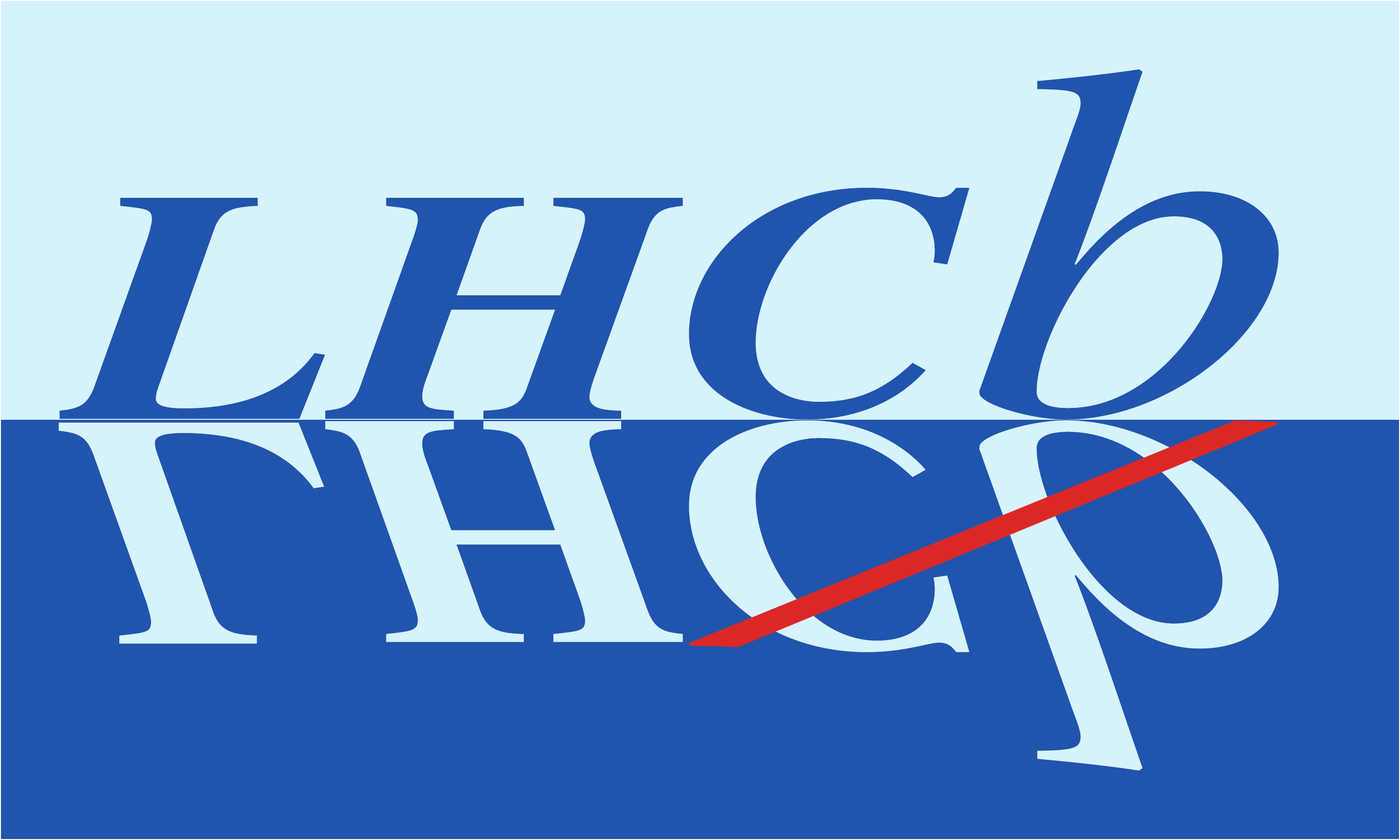}} & &}%
{\vspace*{-1.2cm}\mbox{\!\!\!\includegraphics[width=.12\textwidth]{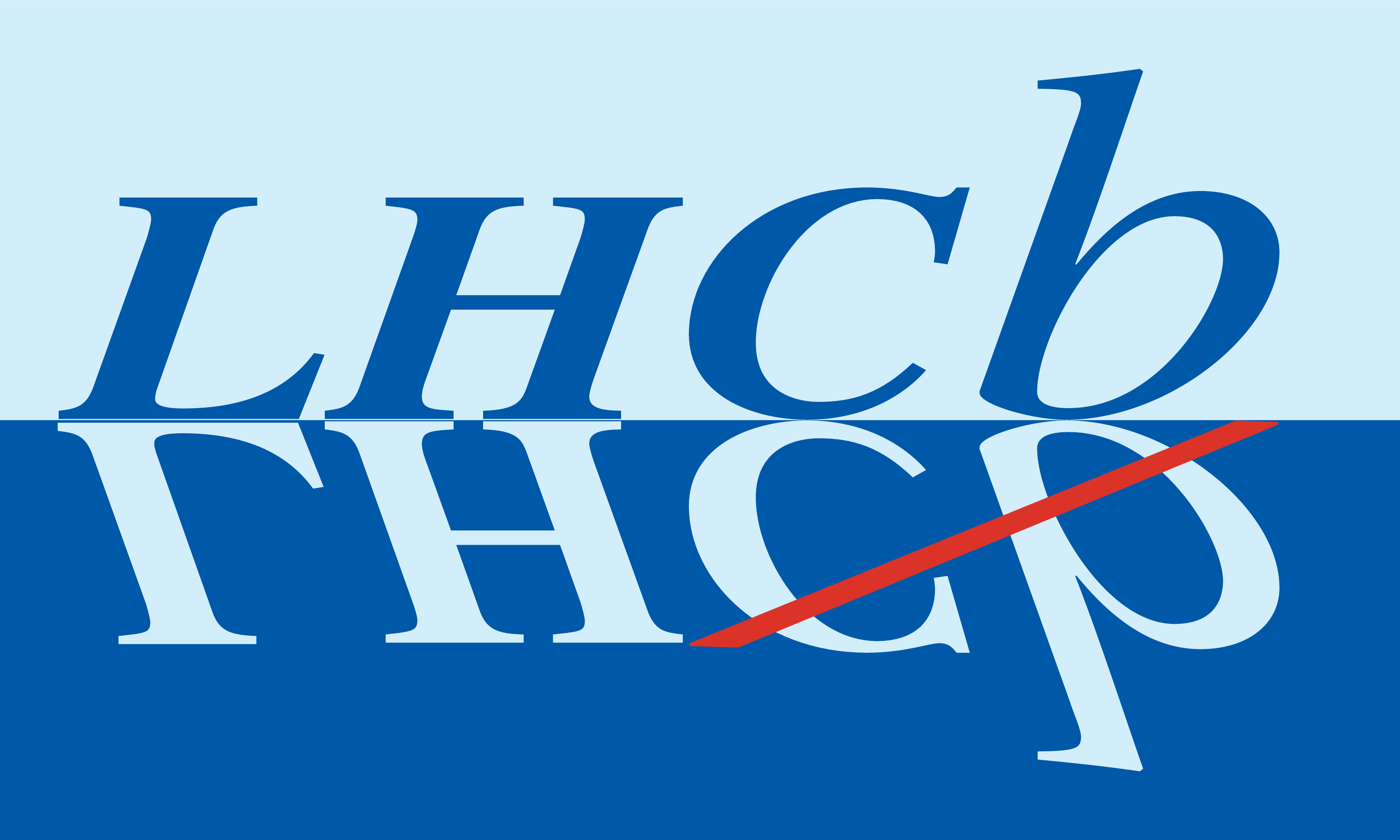}} & &}%
\\
 & & CERN-EP-2025-265 \\  
 & & LHCb-PAPER-2025-046 \\  
 & & May 11, 2026 \\
 \end{tabular*}

\vspace*{3.0cm}

{\normalfont\bfseries\boldmath\huge
\begin{center}
  \papertitle 
\end{center}
}

\vspace*{1.0cm}

\begin{center}
\paperauthors\footnote{Authors are listed at the end of this paper.}
\end{center}

\vspace{\fill}

\begin{abstract}
  \noindent

  A time- and flavour-integrated amplitude analysis of $B^0$ and $B^0_s$ decays to the $(K^+\pi^-)(K^-\pi^+)$ final state in the $K^*(892)^0 \kern 0.18em \overline{\kern -0.18em K}{}^{*}(892)^0$ region is presented, using $pp$ collision data recorded with the LHCb detector in 2011--2018, corresponding to an integrated luminosity of $9\,\text{fb}^{-1}$. The branching fractions of the $B^0$ and $B^0_s$ decays are measured relative to the $B^0 \to D^-\pi^+$ and $B^0_s \to D^-_s \pi^+$ modes, respectively. The corresponding longitudinal polarisation fractions are found to be $f_L^{d} = 0.600 \pm 0.022 \pm 0.017$ and $f_L^{s} = 0.159 \pm 0.010 \pm 0.007$, 
  where the uncertainties are statistical and systematic, respectively.
  The theory-motivated ratio of the squared $B^0_s$ to $B^0$ longitudinally polarised decay amplitudes is found to be \mbox{$L_{K^{*0} \kern 0.18em \overline{\kern -0.18em K}{}^{*0}} = 4.92 \pm 0.55 \pm 0.48 \pm 0.02 \pm 0.10$}, where the uncertainties are statistical, systematic, due to uncertainty of external mass and lifetime measurements, and due to knowledge of the fragmentation fraction ratio, respectively. 
  This confirms the previously reported tension between experimental determinations and theoretical predictions of longitudinal polarisation in $B \to VV$ decays at the level of 4.4 standard deviations.
\end{abstract}

\vspace*{1.0cm}

\begin{center}
  Published in Physical Review D {\bf 113}\ (2026) 092002
\end{center}

\vspace{\fill}

{\footnotesize 
\centerline{\copyright~\papercopyright. \href{\paperlicenceurl}{\paperlicence}.}}
\vspace*{2mm}

\end{titlepage}


\newpage
\setcounter{page}{2}
\mbox{~}


\renewcommand{\thefootnote}{\arabic{footnote}}
\setcounter{footnote}{0}


\cleardoublepage


\pagestyle{plain} 
\setcounter{page}{1}
\pagenumbering{arabic}


\section{Introduction}
\label{sec:introduction}

\begin{figure}[bp]
    \centering
    \includegraphics[width=.49\linewidth]{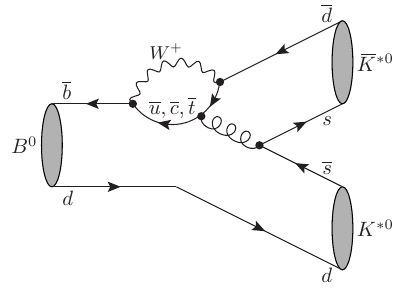}
    \centering
    \includegraphics[width=.49\linewidth]{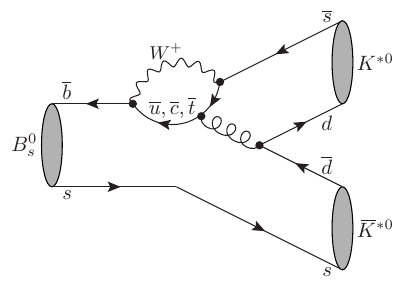}
    \caption{Hadronic penguin diagrams contributing to (a)~$B^0 \to K^{*0} \kern 0.18em \overline{\kern -0.18em K}{}^{*0}$ and (b)~$B_s^0 \to K^{*0} \kern 0.18em \overline{\kern -0.18em K}{}^{*0}$~decays in the SM.}
    \label{fig:feyns}
\end{figure}

In the Standard Model (SM) of particle physics, the \BdKstKst and \BsKstKst decays proceed predominantly via the \decay{\bquarkbar}{\dquarkbar\squark \squarkbar} and \decay{\bquarkbar}{\squarkbar\dquark \dquarkbar} gluonic loop (penguin) transitions, respectively, as illustrated in Fig.~\ref{fig:feyns}.\footnote{Throughout this paper, the inclusion of charge-conjugate processes is implied, and the symbol $\Kst$ specifically denotes the $\Kstar(892)^0$ state.}
Because of their loop-suppressed nature, these processes offer a sensitive probe for New Physics~(NP), as contributions from particles beyond the SM can lead to sizeable deviations from the SM predictions for certain observables~\cite{Fleischer:1999zi,Fleischer:2007wg,Descotes-Genon:2007iri,Descotes-Genon:2011rgs,Bhattacharya:2012hh,Bhattacharya:2013sga,Gronau:2000zy,Grossman:2024amc}.

Quantities of particular interest are those for which theoretical uncertainties from nonperturbative hadronic interactions can be well controlled.  
The \BdKstKst and \BsKstKst decays provide a set of such quantities and are thus considered in the literature as \textit{golden channels} for NP searches~\cite{Ciuchini:2007hx, Alguero:2020xca}.  
Given the identical final state of both decays, 
U-spin symmetry, which reflects the approximate invariance of the amplitudes under the interchange of $\squark$ and $\dquark$ quarks, may be applied. 
One particular theoretically motivated quantity is the amplitude-squared ratio of the longitudinal components contributing to \Bs and \Bd decays~\cite{Alguero:2020xca,Biswas:2023pyw,Biswas:2025drz}, defined as
\begin{equation}
\label{eq:LKK}
L_{\Kst\Kstb} \equiv \mathcal{G} \frac{\mathcal{B}(\BsKstKst)}{\mathcal{B}(\BdKstKst)} \frac{
\fLBs}{\fLBd},
\end{equation}
where $\mathcal{B}$ denotes the branching fraction of the indicated decay and $\mathcal{G} = 1.014 \pm 0.004$ is a factor accounting for the differences in the measured mass and lifetime between the \Bd and \Bs mesons~\cite{Alguero:2020xca}. 
Respectively, the longitudinal polarisation fractions, \fLBd and \fLBs, are the proportions of \BdKstKst and \BsKstKst decays in which the spins of the two produced vector mesons are aligned along their directions of motion. Hadronic uncertainties largely cancel in this ratio in theoretical calculations. From an experimental perspective, the observable $L_{\Kst\Kstb}$ is also subject to relatively small systematic uncertainties through cancellation effects, as the two decays share identical final-state particles.
The time-dependent \CP-violating parameters in the \Bs channel constitute another set of observables with considerable interest. An overall weak-phase difference is absent in the dominant top-quark loop diagram within the SM and thus a significant \CP asymmetry observed in these decays would provide strong evidence for NP. Potential pollution from subleading charm- and up-quark loop diagrams can be well controlled by relating to similar diagrams for the \Bd decay under U-spin symmetry~\cite{Ciuchini:2007hx}.

Previous measurements of these decays have been performed by the LHCb collaboration using $pp$ collision data, corresponding to an integrated luminosity of 3\invfb~\cite{LHCb-PAPER-2019-004,LHCb-PAPER-2017-048}.
The key finding was a significant difference between the measured longitudinal polarisation fractions: $\fLBd = 0.724 \pm 0.051 \pm 0.016$ and $\fLBs = 0.240 \pm 0.031 \pm 0.025$~\cite{LHCb-PAPER-2019-004}, where the uncertainties are statistical and systematic, respectively.
These can be used to determine an experimental value of $L_{\Kst\Kstb} = 4.43 \pm 0.92$~\cite{Alguero:2020xca}.
The latest theoretical prediction is $L_{\Kst\Kstb} = 26.08 ^{+5.70}_{-4.72}$, when both light-cone sum rules and lattice QCD predictions for the form factors are combined~\cite{Biswas:2025drz}.
Naive factorisation suggests that in a $B\to VV$ decay, where $V$ represents a vector state, the longitudinal component typically dominates due to helicity suppression of the transversely polarised amplitudes~\cite{Korner:1979ci}. 
The observation of small values of $f_L$ in \BsKstKst decays, as well as other related modes~\cite{Belle:2003lsm,Belle:2004uca,Belle:2005lvd,Belle:2008vwd,Belle:2012ayg,Belle:2013vat,Belle:2015yjj,Belle:2015xfb,LHCb-PAPER-2013-012,LHCb-PAPER-2014-005,LHCb-PAPER-2015-006,LHCb-PAPER-2018-042,LHCb-PAPER-2019-019,LHCb-PAPER-2025-026}, has led to several possible explanations for this so-called \textit{polarisation puzzle}. These include enhanced contributions from weak-annihilation amplitudes~\cite{Li:2004mp,Huang:2005if,Kagan:2004uw,Su:2010vt,Wang:2017hxe}, charm-loop diagrams~\cite{Bauer:2004tj,Wang:2017rmh}, higher-order corrections~\cite{Zou:2015iwa,Yan:2018fif}, final-state interactions~\cite{Colangelo:2004rd,Li:2004ti,Cheng:2004ru,Ladisa:2004bp,Yu:2024vlz}, and NP contributions~\cite{Yang:2004pm,Baek:2005jk,Huang:2005qb,Bao:2008hd,Kim:2007eea,Datta:2003mj}.

This paper presents a time- and flavour-integrated amplitude analysis of \BdKstKst and \BsKstKst decays, with $\Kst\to\Kp\pim$, using $pp$ collision data collected by the LHCb experiment at centre-of-mass energies of $\sqrt{s}=7$, $8$ and $13$\tev, during LHC Runs 1 and 2, which took data between 2011 and 2018, corresponding to an integrated luminosity of $9\invfb$. 
Both of the $\kaon\pion$ pairs are reconstructed in the invariant-mass region around the vector \Kst resonance. Decays with the identical final-state particles, namely $\Bd \to \Dm[\to \Kp\Km\pim]\pip$ and $\Bs\to \Dsm[\to\Kp\Km\pim]\pip$, are used as normalisation modes.
Results for the amplitude parameters are presented, with particular focus on the branching fractions, longitudinal polarisation fractions and the $L_{\Kst\Kstb}$ observable. 
These measurements are the most precise to date and supersede the previous LHCb results~\cite{LHCb-PAPER-2019-004,LHCb-PAPER-2017-048}. 
Compared to previous measurements, this analysis improves the spin description of the contributing amplitudes, models the nonresonant $(K\pi)$ contribution down to the production threshold, and uses a more robust method for computing the efficiency.

\section{Detector and simulation}
\label{sec:detector_and_sim}

The \lhcb detector~\cite{LHCb-DP-2008-001,LHCb-DP-2014-002} is a single-arm forward
spectrometer covering the \mbox{pseudorapidity} range $2<\eta <5$, designed for the study of particles containing \bquark or \cquark quarks. The detector used to collect the data analysed in this paper includes a high-precision tracking system consisting of a silicon-strip vertex detector surrounding the $pp$ interaction region, a large-area silicon-strip detector located upstream of a dipole magnet with a bending power of about $4{\mathrm{\,T\,m}}$, and three stations of silicon-strip detectors and straw drift tubes placed downstream of the magnet.
The tracking system provides a measurement of the momentum, \ptot, of charged particles with a relative uncertainty that varies from 0.5\% at low momentum to 1.0\% at 200\gevc.
The minimum distance of a track to a primary $pp$ collision vertex (PV), the impact parameter (IP), 
is measured with a resolution of $(15+29/\pt)\mum$,
where \pt is the component of the momentum transverse to the beam, in\,\gevc.
Different types of charged hadrons are distinguished using information
from two ring-imaging Cherenkov (RICH) detectors. 
Photons, electrons and hadrons are identified by a calorimeter system consisting of
scintillating-pad and preshower detectors, an electromagnetic
and a hadronic calorimeter. Muons are identified by a
system composed of alternating layers of iron and multiwire
proportional chambers.
The online event selection is performed by a trigger, 
which consists of a hardware stage, based on information from the calorimeter and muon
systems, followed by a software stage, which applies a full event
reconstruction.

At the hardware trigger stage, events are required to have a muon with high \pt or a hadron, photon or electron with high transverse energy in the calorimeters. For hadrons, the transverse energy threshold at the hardware trigger stage is 3.5\gev.
The software trigger requires a two-, three- or four-track secondary vertex with a significant displacement from any PV. 
At least one charged particle must have transverse momentum $\pt > 1.6\gevc$ and be inconsistent with originating from a PV.
A multivariate algorithm~\cite{BBDT,LHCb-PROC-2015-018} is used to identify secondary vertices consistent with the decay of a \bquark hadron.
In the offline selection, trigger signals are associated with reconstructed particles.
Selection requirements can therefore be made on the trigger selection itself
and on whether the decision was due to the signal candidate, other particles produced in the $pp$ collision, or to a combination of both.
Triggered data further undergo a centralised, offline processing step
to deliver physics-analysis-ready data across the entire \lhcb physics programme~\cite{Stripping}.

Simulation is required to model the effects of the detector acceptance and the imposed selection requirements.
In the simulation, $pp$ collisions are generated using \pythia~\cite{Sjostrand:2006za} with a specific \lhcb configuration~\cite{LHCb-PROC-2010-056}. Decays of unstable particles are described by \evtgen~\cite{Lange:2001uf}, in which final-state radiation is generated using \photos~\cite{davidson2015photos}.
The interaction of the generated particles with the detector, and its response, are implemented using the \geant toolkit~\cite{Allison:2006ve,Agostinelli:2002hh} as described in Ref.~\cite{LHCb-PROC-2011-006} and further calibrated with various control samples for the differences between data and simulation, as described in Sec.~\ref{sec:efficiency}.

\section{Event selection}
\label{sec:selection}
Offline requirements are used to select the signal and normalisation candidates. 
Four charged hadron tracks displaced from any PV are selected, each required to have momentum in the range $p\in[3,100]\gevc$ and $\pt > 500\mevc$. 
Two of the oppositely charged tracks must be consistent with the kaon mass hypothesis and two with the pion mass hypothesis, which are distinguished by requirements on the difference in the log-likelihood of the two different mass hypotheses, based on information from the RICH detectors.
For the signal modes, all four tracks must be consistent with originating from the same vertex and $K^\pm \pi^\mp$ pairs are required to lie within the elastic scattering region from the $K\pi$ to $K\eta$ thresholds, corresponding to a mass in the range $[633, 1042]\mevcc$, for reasons outlined in Sec.~\ref{sec:propagators}.
The four tracks are combined to produce a $B$ candidate that must originate from the PV and have a mass in the range $[5000, 5800]\mevcc$.
Signal mode candidates with a three-body mass $m_{\Kp\Km\pimp}$ in the region $[1946.4,1990.4]\mevcc$ are removed to reject the background from \decay{\Bs}{\Dsmp\pipm} decays.

The normalisation candidates have identical track selection requirements to the signal modes, except for the requirement that all tracks should originate from the PV. Instead, both kaon tracks and one of the pion tracks are required to be displaced from the remaining pion track to form a \DmorDsm candidate with a mass in the range $[1796.2, 2068.5]\mevcc$.
Candidates are required to have $m_{\Kp\Km}<1800\mevcc$ to reject \decay{\Dz}{\Kp\Km} backgrounds. Stricter particle identification~(PID) requirements are placed on the kaon with the same charge as the pion in the \DmorDsm decay, providing the \DmorDsm mass lies in the range $m_{\Kp\Km\pim} \in [1839, 1899]\mevcc$, in order to reject misidentified backgrounds from \decay{\BdorBs}{\DmorDsm [\to {\Kp\pim\pim}]\pip} decays. 
A mass constraint is also applied to the \DmorDsm candidates through a kinematic fit procedure.

Data candidates are further refined using a boosted decision tree (BDT) classifier implemented in the \textsc{XGBoost} package\cite{XGBoost}, trained with a variety of kinematic, topological and isolation variables. For \decay{\BdorBs}{\Kstarz\Kstarzb} candidates, the BDT is trained with a signal proxy comprised of simulated \decay{\Bs}{\Kstarz\Kstarzb} decays, using an amplitude model based on previous LHCb studies~\cite{LHCb-PAPER-2019-004}, while for the normalisation modes a simulated sample of \decay{\Bs}{\Dsm\pip} decays is used. The background proxy is formed from the upper-mass sideband $[5600,5800]\mevcc$ from each of the signal and normalisation data samples. Separate BDTs are trained for the signal and normalisation modes, as well as in each data-taking period: Run 1 (2011--2012) and Run 2 (2015--2018). The $k$-fold cross-validation method is used with $k=5$~\cite{kFold}.

The \B-candidate kinematics in the simulation samples are corrected using a gradient-boosted algorithm from the \hepml package \cite{Rogozhnikov:2016bdp} to match the background-subtracted distributions of \Pp and \pt of the \B candidate in data, as well as the distribution of the track-fit \chisq of the final-state particles. 
The background-subtracted distributions are determined using the \textit{sPlot} procedure~\cite{Pivk:2004ty,sweights} from a preliminary fit to the signal and normalisation data samples selected with a loose cut on the BDT output to reject combinatorial background.

To reduce contamination from misidentified backgrounds from \decay{\Bd}{\rhoz\Kstarz}, \decay{\Bd}{\phiz\Kstarz} and \decay{\Lb}{\proton\Km\pip\pim} decays in the signal modes, stricter PID requirements are applied to the final-state tracks. 
Where available, simulated samples are generated according to known amplitude models~\cite{LHCb-PAPER-2018-042,LHCb-PAPER-2014-005}, otherwise uniformity in phase space is assumed. 
These requirements are optimised alongside the combinatorial-background BDT output score, using the basin-hopping algorithm \cite{2020SciPy-NMeth,basin-hopping}, for the significance figure of merit, defined as $S/\sqrt{S+B}$, where $S$ $(B)$ is the expected number of signal (background) events in the region of the \Bd mass peak, which is expected to have a lower yield.
Initial signal and background yields and the bounds of the \Bd signal region are determined from the aforementioned preliminary fit to data. The optimisation is constrained to reject at least 90\% of the dominant misidentified \decay{\Bd}{\rhoz\Kstarz} background.
The normalisation modes optimise only the BDT requirement with the same figure of merit, but subsequently the same PID cuts used in the signal modes are applied.

\section{Four-body mass fits}
\label{sec:mass_fits}
Extended unbinned maximum-likelihood fits are performed on the four-body mass distributions of the selected signal and normalisation candidates. The parameters of each fit component are determined from suitable simulation samples and fixed in the fit to data, except for the positions and widths of the peaks which are allowed to shift and scale, respectively.
Selected signal candidates in the reduced mass range $[5050, 5750]\mevcc$ are used in the fit to data which is further subdivided such that separate models for Run~1 and Run~2 can be determined simultaneously.
For the signal modes, a Hypatia distribution \cite{Santos:2013gra}, with parameters extracted from a simulated sample of \decay{\Bs}{\Kstarz\Kstarzb} decays, is used to model the contribution for both \BsKstKst and \BdKstKst decays, where the distribution for the latter is shifted by the known mass difference between the \Bs and \Bz mesons~\cite{PDG2024}. 
The misidentified backgrounds from \decay{\Bd}{\rhoz\Kstarz} and \decay{\Bd}{\phiz\Kstarz} decays are modelled using Johnson-$S_U$ functions\cite{JohnsonSU}, while partially reconstructed decays are modelled using a kernel density estimate (KDE) from a sample of \decay{\Bs}{\Kp\pim\Km\pip\piz} decays generated with the fast simulation package~\textsc{RapidSim}~\cite{ Cowan:2016tnm}.
The combinatorial background is modelled using a uniform distribution in Run~1 data, while an exponential distribution with a slope parameter determined in the fit to the data is adopted for Run~2. 

The normalisation samples are split into \BdDpi and \BsDspi candidates by requiring the \DmorDsm candidate mass to be in the range $[1849, 1893]\mevcc$ for the former, and $[1941, 1993]\mevcc$ for the latter. 
The two samples are then modelled separately. 
The fit to the \BdDpi mode uses the mass range $[5200, 5500]\mevcc$ and consists of Hypatia distributions to model the \BdDpi component and contributions from misreconstructed \BdDK decays. 
The fit to the \BsDspi mode uses the mass range $[5290, 5600]\mevcc$ and consists of Hypatia distributions to describe the \BsDspi component and the background from \BsDsK, \BdDspi and \BsDsstpi decays where the photon is not reconstructed.
Due to the large overlap and small expected yields of the \decay{\Bd}{\Dsm\pip} and \decay{\Bs}{\Dssm\pip} backgrounds, these components are combined into a single ``low-mass'' shape whose yield is determined in the fit to data.
In both normalisation mode fits an exponential distribution with a slope parameter determined from data is used to model the combinatorial background.

\begin{figure}[tb]
    \centering
    \includegraphics[width=0.49\linewidth,clip,trim=0.7cm 0.0cm 0.7cm 0.7cm]{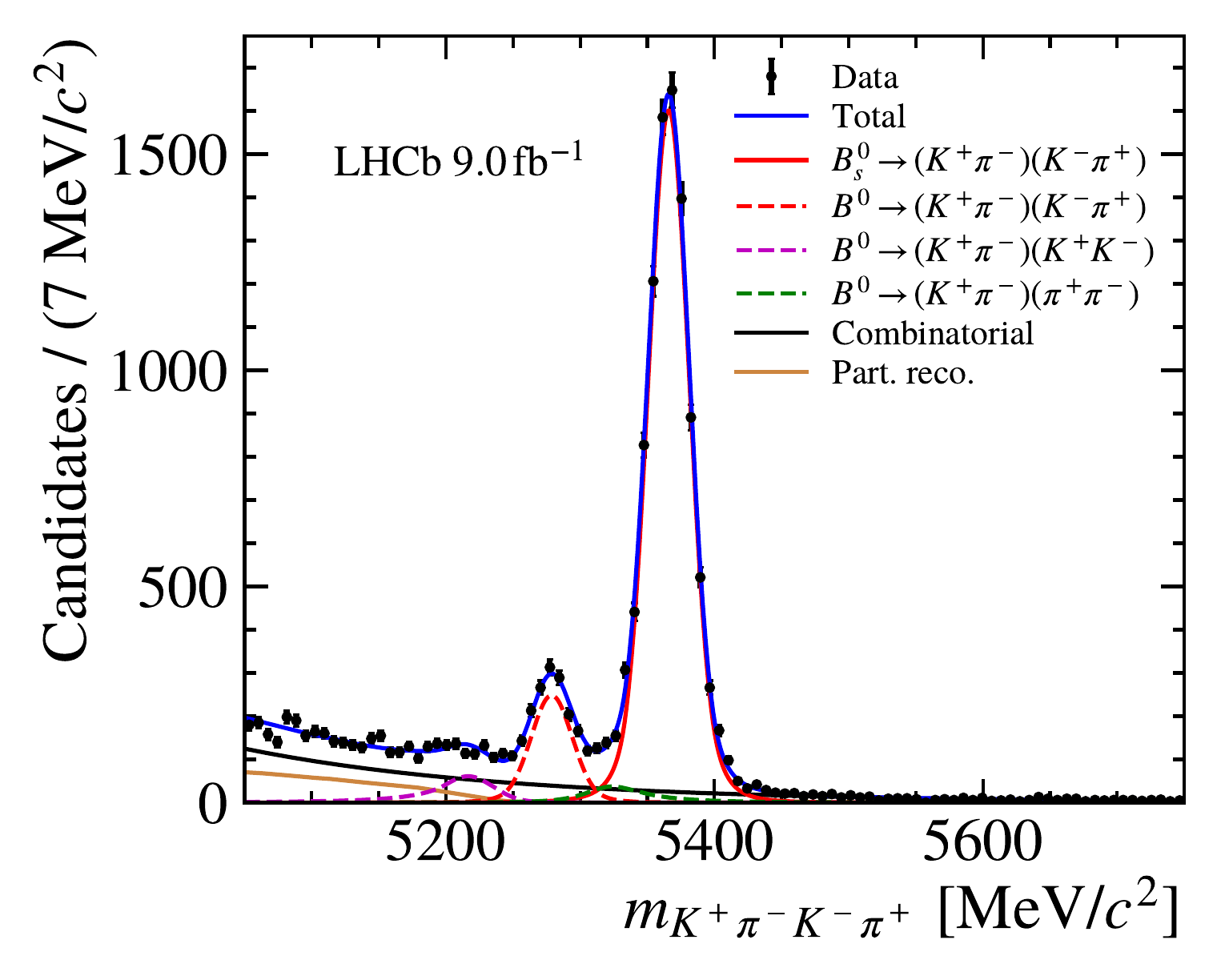} \\
    \includegraphics[width=0.49\linewidth,clip,trim=0.7cm 0.7cm 0.7cm 0.4cm]{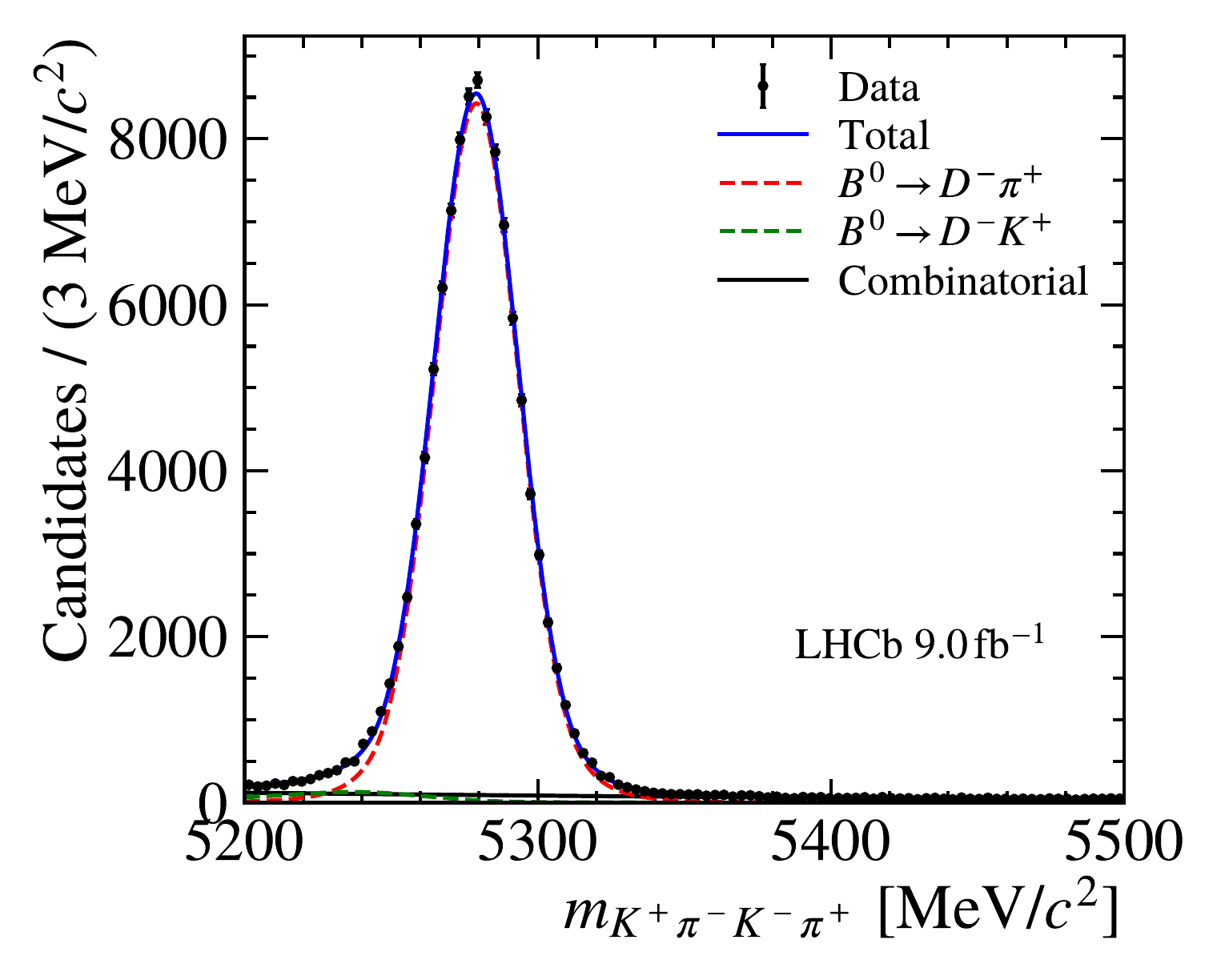}
    \includegraphics[width=0.49\linewidth,clip,trim=0.7cm 0.7cm 0.7cm 0.4cm]{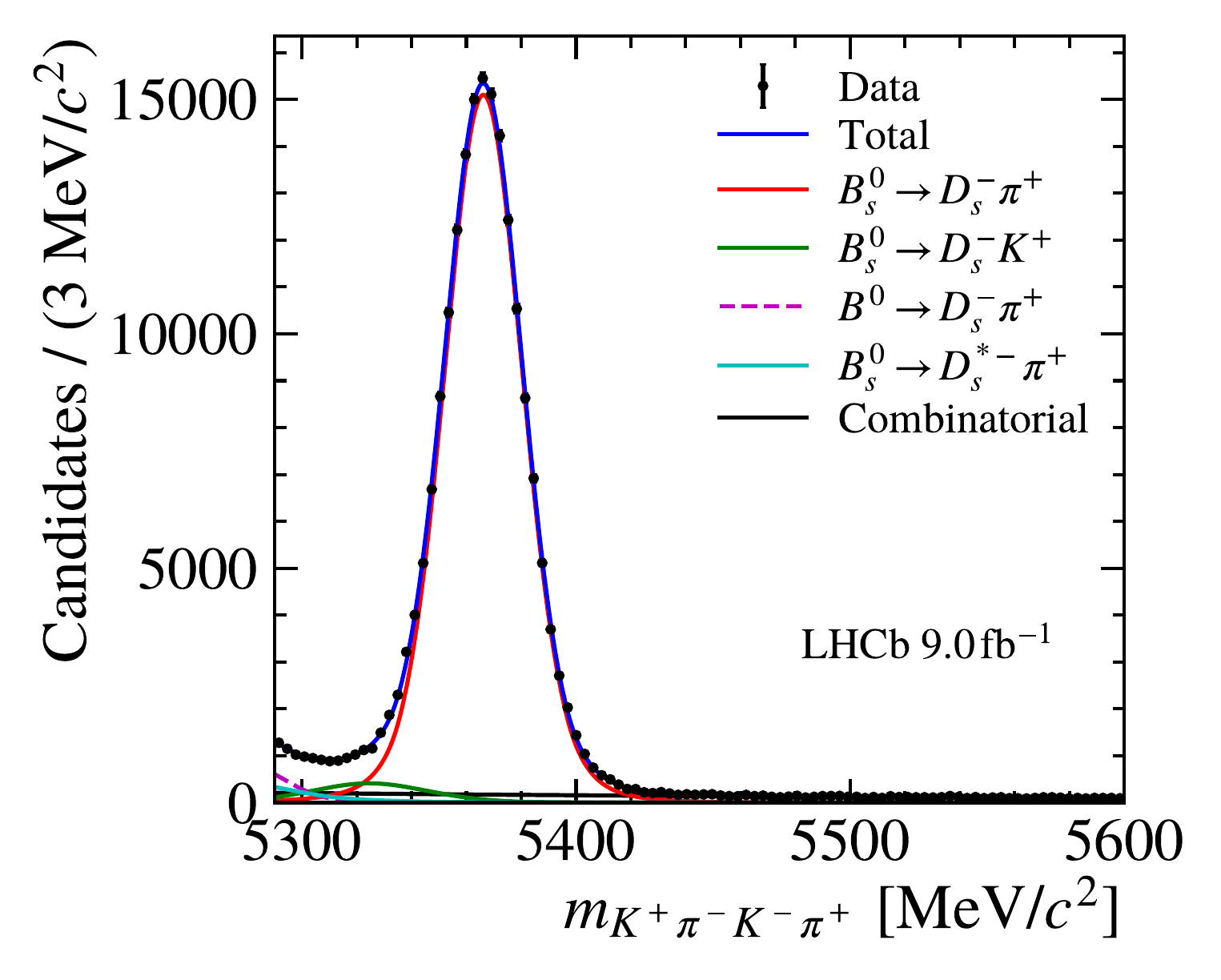}
    \caption{Distributions and fit results of the~$K^+\pi^-K^-\pi^+$~mass for the (top)~signal modes, (bottom left)~$B^0$~normalisation mode and (bottom right)~$B_s^0$~normalisation mode. Equivalent plots on a logarithmic scale are shown in Fig.~\ref{fig:mass_fits_log} in Appendix~\ref{sec:app_log_scale_mass_plots}.}
    \label{fig:mass_fits}
\end{figure}

The results of the fits to the selected signal and normalisation candidates are shown in Fig.~\ref{fig:mass_fits}.
The extracted yields of the signal modes are $9190 \pm 114$ and $1424 \pm 59$ for the \BsKstKst and \BdKstKst candidates, respectively, while the extracted yields of the normalisation modes are $180\,800 \pm 500$ and $105\,700 \pm 300$ for the \BsDspi and \BdDpi candidates, respectively, where all uncertainties are statistical only. Systematic uncertainties are considered in Sec.~\ref{sec:systematics}.

\section{Amplitude analysis}
\label{sec:amplitude_model}
Following the convention in Ref.~\cite{PDG2024}, the differential decay width of the \BdorBs meson  can be expressed as
\begin{equation}
\begin{aligned}
	\frac{\textrm{d}^2\Gamma (\Phi_4, t, q_{\rm tag})}{\textrm{d}\Phi_4\,\textrm{d}t} \propto e^{-t/\tau} \Big[ &\big(|A(\Phi_4)|^2+|\Ab(\Phi_4)|^2\big) \coshdgt \\
    -&2 \, \Real \big( A^* (\Phi_4) \, \Ab (\Phi_4) \big) \sinhdgt \\
	+&q_{\rm tag} \big( |A (\Phi_4)|^2 - |\Ab (\Phi_4)|^2 \big) \cosdmt\\
    +&2 \, q_{\rm tag} \, \Imag \big( A^* (\Phi_4) \, \Ab (\Phi_4) \big) \sindmt \Big] \; ,
\end{aligned}
\label{eq:timeevolution}
\end{equation}
where $\Phi_4$ is the set of four-momenta of the final-state particles characterising the phase-space position of the four-body $B_{(s)}^0\to\Kp\pim\Km\pip$ decay,  $t$ is the decay time, and $q_{\rm tag}$ denotes the initial-state flavour: $+1$ for \Bds, and  $-1$ for \Bdsb states. 
The amplitudes $A(\Phi_4)$ and $\Ab(\Phi_4)$, detailed in Sec.~\ref{sec:static_amplitude}, are the static decay amplitudes of the \Bds and \Bdsb meson, respectively. 
There are three decay-time-relevant parameters: $\tau$ denotes the average \BdorBs lifetime, while $\Delta\Gamma$ and $\Delta m$ are the width and mass differences between the high- and low-mass eigenstates associated with the \BdorBs meson, respectively. These parameters are fixed to their known values~\cite{PDG2024}. 
In principle, $\Ab(\Phi_4)$ also absorbs the contribution from the \BdorBs--\BdorBsbar mixing amplitude $q/p$. This is assumed to be unity as $|q/p|\sim 1$ in the \B sector, $\Delta\Gamma \sim 0$ in \Bd decays, and the \Bs--\Bsb mixing phase cancels the weak phase of \BsKstKst decays in the SM~\cite{Descotes-Genon:2011rgs}.

Although an untagged time-integrated study is performed, the impact of certain parameters related to decay time remains in an incoherent \bquark\bquarkbar-production environment. After time and \B-flavour integration, the final signal probability density function~(PDF) with detector acceptance considered, is given by
\begin{align}
\begin{split} 
	\mathcal{P}_{\rm sig} (\Phi_4) & \propto \sum_{q_{\rm tag}=\pm1} \int_{t_1}^{t_2} \textrm{d}t \, 
    \frac{\textrm{d}^2\Gamma(\Phi_4, t,q_{\rm tag})}{\textrm{d}\Phi_4\,\textrm{d}t}
    \, \epsilon(\Phi_4)\epsilon( t)  \\
    & \propto 
    \bigg[ \big(|A (\Phi_4)|^2+|\Ab (\Phi_4)|^2\big)c_t - 2 \, \Real \big( A^* (\Phi_4) \, \Ab (\Phi_4) \big) \, s_t \bigg]\, \epsilon(\Phi_4) \; ,
\end{split}
\label{eq:time-integrated-pdf}
\end{align}
with
\begin{align}
\begin{split}
    c_t \equiv& \int_{t_1}^{t_2} \epsilon(t)\, e^{-t/\tau} \cosh ({\Delta\Gamma t}/{2}) \textrm{d}t,  \\
    s_t \equiv & \int_{t_1}^{t_2} \epsilon(t)\, e^{-t/\tau} \sinh ({\Delta\Gamma t}/{2}) \textrm{d}t \,.
\end{split}
    \label{eq:aman:rt}
\end{align}
The decay-time integral range is chosen to be $[0,\, 15]\ps$, corresponding to approximately ten times the \Bds lifetime, $\epsilon(\Phi_4)$ is the efficiency as a function of phase space, and $\epsilon(t)$ is the decay-time efficiency.
Details about the efficiency functions are given in Sec.~\ref{sec:efficiency}.

\subsection{Static decay amplitudes}
\label{sec:static_amplitude}

The static~\mbox{\decay{\Bds}{(\Kp\pim)(\Km\pip)}} decay amplitude is constructed within the isobar approximation, where decay processes are assumed to factorise out from multibody unitarity into a topology of subsequent two-body decay processes~\cite{isobar1,isobar2,isobar3}. The total amplitude is taken as the coherent sum over components, each described by a function~$A_i$ that parametrises quasi-two-body intermediate processes,
\begin{equation}
    \label{eq:aman:isobar}
    A(\Phi_4) \equiv \sum_{i=1}^N a_i A_i(\Phi_4)\,,
\end{equation}
where $a_i$ are the complex free parameters of the model, representing the relative contributions of each component $i$. 
The function~$A_i$ contains only strong dynamics, and is generically parametrised, removing the component index $i$ for brevity, as
\begin{equation}
    A(\Phi_4) \equiv F_{(\Kp\pim)(\Km\pip)}(\Phi_4) [F_{\Kp\pim}(\Phi_4) T_{\Kp\pim}(\Phi_4)] [F_{\Km\pip}(\Phi_4) T_{\Km\pip}(\Phi_4)] Z(\Phi_4) \,,
\end{equation}
where~$F$ is a barrier factor associated with each subsequent two-body decay,  $T$ is a mass propagator, and~$Z$ is the overall angular amplitude. Their explicit formulas are defined in Sec.~\ref{sec:barrier_factors}, Sec.~\ref{sec:propagators}~and Sec.~\ref{sec:angular_amplitude}, respectively. The particles in the subscript denote the final state of the two-body decay involved.
For a~\CP-conjugate final state, the total amplitude corresponding to the~\CP-conjugate decay process is given relative to the particle ordering that determines $A$,
\begin{equation}
    \bar A(\Phi_4) = A(\bar \Phi_4) \,.
\end{equation}
In practice,~$\bar \Phi_4$ indicates the phase-space position with oppositely charged final-state particles of the same species interchanged~($C$) and all three-momenta inverted~($P$), \ie
\begin{equation}
    \Kp(\vec{p}_1) \pim(\vec{p}_2) \Km(\vec{p}_3) \pip(\vec{p}_4) \overset{\CP}{\longrightarrow}   \Kp(-\vec{p}_3) \pim(-\vec{p}_4) \Km(-\vec{p}_1) \pip(-\vec{p}_2)\, ,
    \label{eq:cpordering}
\end{equation}
where $\vec{p}_i$ is the three-momentum associated with each final-state particle.

\subsubsection{Barrier factors}
\label{sec:barrier_factors}

To account for deviations from pointlike interactions, finite sizes are attributed to decaying states via Blatt--Weisskopf penetration factors~\cite{Blatt:1952ije}, 
assuming a square-well interaction potential with effective radius~$r = 4.0 \; \hbar c \; \gev^{-1}$, equivalent to a distance scale of $0.8 \fm$.
The barrier factors also depend on the orbital angular momentum
between the decay products, $L$, and the (breakup) three-momentum of either decay product~$q$ defined in the rest frame of the decaying state. 
The explicit expressions~\cite{VonHippel:1972fg}, up to $L=2$, are
\begin{equation}
F(q, L) \equiv
\left\{
\begin{array}{c l}
1 & \text{if } L=0 \,,\\[4pt]
\dfrac{1}{\sqrt{1+(qr)^2}} & \text{if } L=1 \,,\\[6pt]
\dfrac{1}{\sqrt{9+3(qr)^2+(qr)^4}} & \text{if } L=2 \,.
\end{array}
\right.
\end{equation} 

In general, an additional momentum factor $q^L$ is required to ensure the correct threshold behaviour of the amplitude when the angular amplitudes are constructed using the helicity formalism~\cite{PDG2024,Chung:186421}. In the covariant tensor formalism used in this analysis, described in Sec.~\ref{sec:angular_amplitude}, this factor is already included in the angular amplitude, thus no further implementation is necessary.

\subsubsection{Mass propagators}
\label{sec:propagators}

Intermediate resonant contributions in the \Kpm\pimp system  are described as functions of its rest-frame energy squared,~$s$, by the relativistic Breit--Wigner~(BW) propagator
\begin{equation}
    T(s) \equiv \frac{1}{m_0^2 - s - i
    \sqrt{s} \Gamma(s)} \,,
\end{equation}
where the imaginary part is given by
\begin{equation}
    \sqrt{s} \Gamma(s) = m_0 \Gamma_0 \frac{m_0}{\sqrt{s}} \left( \frac{q}{q_0} \right)^{2L + 1} \frac{F(q, L)^2}{F(q_0, L)^2} \,,
\end{equation}
for decays into two stable particles~\cite{BW}. 
The total energy-dependent width,~$\Gamma(s)$, is normalised to give the pole width,~$\Gamma_0$, when evaluated at the pole mass,~$m_0$, while~$q_0$ is the value of the breakup momentum at the pole.

For the intermediate $S$-wave contribution in the $\Kpm\pimp$ system, precise amplitude parametrisations are available for isospin $I=1/2$ based on dispersively constrained studies of $\pi K$-$\pi K$ and $\pi\pi$-$KK$ scattering data that satisfy the principles of unitarity, analyticity and crossing symmetry. In general, the scattering amplitude is given by
\begin{equation}
    S_{\pi K\textrm{-}\pi K}(s) = 1 + iT_{\pi K\textrm{-}\pi K}(s) \,,
\end{equation}
where $T_{\pi K\textrm{-}\pi K}(s)$ denotes the dynamic interaction part. As this amplitude analysis is conducted in the fully elastic region of $K\pi$, unitarity essentially reduces the scattering amplitude to an energy-dependent phase, $\delta_0^{1/2}$, through
\begin{equation}
    T_{\pi K\textrm{-}\pi K}(s) = \frac{1}{\cot \delta_0^{1/2}(s) - i} \,.
    \label{eq:swavetheory}
\end{equation}
The parametrisation of this elastic phase is taken from the constrained fit to data~(CFD) model determined in Ref.~\cite{Pelaez:2020gnd}. 

Watson's theorem~\cite{Watson:1952ji}, a unitarity argument in the elastic region that would ordinarily equate this scattering phase to that of $K\pi$ production in multibody particle decay, does not necessarily apply to~\mbox{\decay{\Bds}{(\Kp\pim)(\Km\pip)}} transitions, as the additional vertex contained within the associated penguin topologies may induce further phase motion. 
Therefore, the scattering amplitude is modulated by a production amplitude as
\begin{equation}
    T(s) \equiv P_{\pi K\textrm{-}\pi K}(s) S_{\pi K\textrm{-}\pi K}(s) \,.
    \label{eq:swave}
\end{equation}
The form of the production amplitude $P_{\pi K\textrm{-}\pi K}(s)$ is determined directly from the data. For this purpose, a complex-valued polynomial of the form
\begin{equation}
    P_{\pi K\textrm{-}\pi K}(s) \equiv |1 + \sum_{i=1} c_i^{\rm abs} X^i(s)| \exp(i\sum_{j=1} c_j^{\rm arg} X^i(s)) \,,
\end{equation}
is chosen. Details of the real coefficients $c_i^{\rm abs}$ and $c_i^{\rm arg}$ and the orders of the polynomial functions are described in Sec.~\ref{sec:fit_strategy}.  
For reasons of stability,~$X(s)$ transforms the value of~$s$ onto the region spanned by~$X(s) = [-1, +1]$, defined as
\begin{equation}
    X(s) = 2 \frac{\sqrt{s}-\sqrt{s_{\rm min}}}{\sqrt{s_{\rm max}} - \sqrt{s_{\rm min}}} - 1\, ,
\end{equation}
where $\sqrt{s_{\rm min}}$ and $\sqrt{s_{\rm max}}$ denote the lower and upper boundaries of the $\Kpm\pimp$ mass range, respectively. 
These are chosen to coincide with the $\Kpm\pimp$ threshold, where \mbox{$\sqrt{s_{\rm min}} = m_{\Kp} + m_{\pim} = 633\mevcc$}, and the $\Kz \eta$ threshold, where \mbox{$\sqrt{s_{\rm max}} = m_{\Kz}+m_\eta = 1042\mevcc$}.

\subsubsection{Angular amplitudes}
\label{sec:angular_amplitude}
Previous measurements from the \lhcb collaboration utilised the helicity formalism to construct the angular amplitudes~\cite{LHCb-PAPER-2019-004,LHCb-PAPER-2017-048}, where decays of a pseudoscalar $B$ meson to two vector states can be described by three independent helicity amplitudes, $A_L$, $A_+$, $A_-$.
The explicit angular amplitudes were expressed within the \emph{transversity} basis in which the states become \CP-eigenstates via the transformations 
$A_\parallel = (A_+ + A_-)/\sqrt{2}$, $A_\perp = (A_+ - A_-)/\sqrt{2}$,
while $A_L$ remains unchanged.
A problem with describing the amplitude in the transversity basis is the ambiguity of the barrier factor at the production vertex of the vector-vector state.
While the~\CP-odd~$A_\perp$ amplitude can be definitively identified as the~$L=1$ configuration, the~\CP-even~$A_L$ and~$A_\parallel$ amplitudes are linear superpositions of~$L=0$ and~$L=2$.
Typically,~$L=0$ is chosen as the baseline model, while~$L=2$ is considered as an alternative for evaluating systematic uncertainties. 
As an otherwise irreducible source of uncertainty, such an approach is ultimately unsustainable, as already seen in the time-dependent amplitude analysis of~\mbox{\decay{\Bds}{(\Kp\pim)(\Km\pip)}} decays conducted with the LHCb Run\,1 data, where this issue constitutes a leading systematic uncertainty~\cite{LHCb-PAPER-2017-048}.

To mitigate this effect, the covariant Zemach (Rarita--Schwinger) tensor formalism~\cite{Rarita:1941mf,Zemach:1965ycj} is employed in the amplitude model of this analysis, in which the spin amplitude~$Z(\Phi_4)$ is defined to be associated with a specific orbital angular momentum. An initial- or final-state particle, with spin $J$, four-momentum $p$ and polarisation index $\lambda$, is represented by a  rank-$J$ polarisation tensor~$\epsilon_{\mu_1...\mu_J}(p,\lambda)$, with Lorentzian indices $\mu_1...\mu_J$.
The Rarita--Schwinger conditions assert this tensor to be symmetric, traceless, and
orthogonal to $p$. 
This reduces the $4^J$ elements of the polarisation tensor to $2J+1$ independent elements in accordance with the number of degrees of freedom available to a spin-$J$ state. 
The projection operator,
\begin{equation}
    P_{\mu_1...\mu_J, \nu_1...\nu_J}(p) = \sum_\lambda \epsilon_{\mu_1...\mu_J}(p,\lambda) \epsilon^*_{\nu_1...\nu_J}(p,\lambda) \,,
\end{equation}
is the fundamental object with which all spin amplitudes are built. 

Contracted with an arbitrary tensor, the projection operator retains the tensor component that satisfies the Rarita--Schwinger conditions.
Therefore, in a two-body process with relative orbital angular momentum~$L$ and relative momentum~$q \equiv p_a - p_b$ between decay products, the rank-$L$ tensor that describes orbital angular momentum is identified as
\begin{equation}
    L_{\mu_1...\mu_L}(p,q) = P_{\mu_1...\mu_L, \nu_1...\nu_L}(p) q^{\nu_1}...q^{\nu_L} \,.
\end{equation}

For a given vertex in the decay chain, a Lorentz scalar describing the spin amplitude is produced by contracting all polarisation tensors at that vertex with any orbital tensor required for conservation of total angular momentum. 
In cases where the transition is~$P$-odd, it is necessary to include the Levi-Civita totally antisymmetric tensor via~$\varepsilon_{abcd} p^{d}$ to ensure the correct behaviour of the spin amplitude under parity transformation. 
This procedure avoids the need for manual insertion of a~\CP eigenvalue that accompanies \Bdsb amplitudes expressed in the transversity basis~\cite{LHCb-PAPER-2019-004,LHCb-PAPER-2017-048}. 
The overall angular amplitude is then the product of all two-body spin amplitudes in the decay chain, summing over all unobservable polarisation indices. 

\begin{table}[tb]
    \centering
    \caption{\label{tab:aman:topo}
        Angular amplitudes expressed with the covariant tensor formalism. 
        Square brackets associated with a topology indicate the relative orbital angular momentum between the intermediate states.
    }

    \renewcommand{\arraystretch}{1.2}
    \begin{tabular}
    {@{\hspace{0.50cm}}l@{\hspace{0.125cm}}
     @{\hspace{0.125cm}}l@{\hspace{0.125cm}}
     @{\hspace{0.125cm}}l@{\hspace{0.25cm}}
     @{\hspace{0.25cm}}r@{\hspace{0.50cm}}} \hline
    \multicolumn{3}{l}{\Bds decay topology} & Angular amplitude $Z(\Phi_4)$\\ \hline

    $\decay{B}{V \Vb[S]}$, & $\decay{V}{K^+_{1} \pi^-_{2}}$, & $\decay{\Vb}{K^-_{3} \pi^+_{4}}$ & $L_a(p_{V}, q_{V}) L^a(p_{\Vb}, q_{\Vb})$\\
    $\decay{B}{V \Vb[P]}$, & $\decay{V}{K^+_{1} \pi^-_{2}}$, & $\decay{\Vb}{K^-_{3} \pi^+_{4}}$ & $\varepsilon_{abcd} p_{B}^d L^c(p_{B}, q_{B}) L^b(p_{V}, q_{V}) L^a(p_{\Vb}, q_{\Vb})$\\
    $\decay{B}{V \Vb[D]}$, & $\decay{V}{K^+_{1} \pi^-_{2}}$, & $\decay{\Vb}{K^-_{3} \pi^+_{4}}$ & $L_{ab}(p_{B}, q_{B}) L^b(p_{V}, q_{V}) L^a(p_{\Vb}, q_{\Vb})$\\
    $\decay{B}{\VSb}$, & $\decay{V}{K^+_{1} \pi^-_{2}}$, & $\decay{\Sb}{K^-_{3} \pi^+_{4}}$ & $L_a(p_{B}, q_{B}) L^a(p_{V}, q_{V})$\\
    $\decay{B}{\SVb}$, & $\decay{S}{K^+_{1} \pi^-_{2}}$, & $\decay{\Vb}{K^-_{3} \pi^+_{4}}$ & $L_a(p_{B}, q_{B}) L^a(p_{\Vb}, q_{\Vb})$\\
    $\decay{B}{S\Sb}$, & $\decay{S}{K^+_{1} \pi^-_{2}}$, & $\decay{\Sb}{K^-_{3} \pi^+_{4}}$ & $1$\\

    \hline
    \end{tabular}
\end{table}

Considering only vector~($V$) and scalar~($S$) intermediate states in the baseline model, six decay topologies are possible, listed in Table~\ref{tab:aman:topo}. 
The final-state particle with the lowest index necessary to calculate $q$ for any two-body decay sequence constitutes $p_a$. 
The corresponding angular amplitudes for $\Ab$ are obtained by applying the transformation described in Eq.~\ref{eq:cpordering}. These spin amplitudes are calculated with the \textsc{qft++} software package~\cite{Williams:2008wu}. 
In a flavour-tagged analysis, it would be possible to disentangle scalar amplitudes produced in the spectator interaction, thereby deepening knowledge of the underlying processes governing $S$-wave production. 
However, since this analysis is flavour integrated, the \VSb and \SVb topologies are replaced by \CP eigenstates to allow comparison with previous results. This is achieved through the linear combinations,
\begin{equation}
    A^\pm_{\VS} \equiv \frac{1}{\sqrt{2}} \left( A_{\VSb} \pm A_{\SVb} \right) \,,
\end{equation}
where $A^+_\VS$ is \CP-odd and $A^-_\VS$ \CP-even. These definitions are consistent with those in previous analyses~\cite{LHCb-PAPER-2019-004,LHCb-PAPER-2017-048}.

\subsection{Efficiency model}
\label{sec:efficiency}

While the amplitude directly takes the four-momenta of the final-state particles as arguments, the efficiency across the phase space, $\epsilon(\Phi_4)$, is modelled with five variables, $\cos\theta_{\Kp\pim}$,  $\cos\theta_{\Km\pip}$, $\varphi$, $m_{\Kp\pim}$ and $m_{\Km\pip}$, which constitutes the minimum set of variables to define a point in the phase space. The definitions of these variables are the mass of the $\Kpm\pimp$ system,~$m_{\Kpm\pimp}$, the cosine of the angle between the \Kpm momentum and the opposite of the momentum of the \Kmp\pipm pair, both defined in the \Kpm\pimp pair rest frame,~$\cos\theta_{\Kpm\pimp}$ and the angle between the plane defined by the $\Kp$ and $\pim$ momenta and that defined by the $\Km$ and $\pip$ momenta, both in the \BdorBs rest frame,~$\varphi$, as shown in Fig.~\ref{fig:transversity}.
\begin{figure}[b]
    \centering
    \includegraphics[width=0.6\linewidth]{figs/Fig3.pdf}
    
    \caption{Definitions of the three angular variables $(\cos\theta_{K^+\pi^-}, \cos\theta_{K^-\pi^+}, \varphi)$ in the transversity basis.}
    \label{fig:transversity}
\end{figure}

The efficiency model is determined separately for~\Bz and~\Bs decays, using simulated samples generated uniformly within the phase space of the considered $m_{\Kpm\pimp}$ region. 
These simulated samples are further subdivided by data-taking year and hardware-trigger category. 
Nonuniformities in these distributions arise from the detector geometry, trigger and tracking algorithms, PID selections, and other background-rejection requirements such as those imposed on the~BDT classifier to discriminate against combinatorial background. 
The efficiency map is subsequently calibrated with control samples to account for differences between data and simulation arising in the aforementioned effects. 

The hardware-trigger efficiency correction is calculated using pions from~\mbox{\decay{\Dz}{\Km\pip}} decays, arising from promptly produced~\mbox{\decay{\Dstarp}{\Dz [\to\Km\pip]\pip}} decays, and affects two disjoint subsets of the selected candidates: those where the trigger requirements were satisfied by hadronic calorimeter deposits as a result of the signal decay and those where the requirements were satisfied only by deposits from the rest of the event. 
In the first case, the probability to satisfy the trigger requirements is calculated using calibration data as a function of the transverse energy of each final-state particle of a given species, the dipole-magnet polarity, and the hadronic calorimeter region. 
In the second subset, a smaller correction is applied following the same procedure, to account for the requirement that these tracks did not activate the hardware trigger. 
The effect of overlapping energy clusters deposited in the hadronic calorimeter causing an erroneous trigger decision is also corrected using the procedure developed for Refs.~\cite{LHCb-PAPER-2014-035,LHCb-PAPER-2014-036}.

Prompt~\mbox{\decay{\jpsi}{\mup \mun}} decays provide a calibration of the tracking efficiency~\cite{LHCb-DP-2013-002}, by comparing the performance of the track-finding algorithm including all subdetectors to that obtained when excluding information from at least one subdetector. 
This calibration is determined as a function of the track $p$ and $\eta$, as well as the multiplicity of the event, and is assumed to factorise with respect to the tracks of the final-state particles so that the efficiency for each track is multiplied to form the overall efficiency. 
The PID efficiency is calculated from calibration data corresponding to the~\mbox{\decay{\Dstarp}{\Dz [\to\Km\pip]\pip}} decay, where pions and kaons can be identified without the use of PID information~\cite{LHCb-PUB-2016-021}. 
The PID efficiencies for the background-subtracted pions and kaons are parametrised in bins of their $p$ and $\pt$, as well as the number of tracks in the event. 
Corrections to the signal efficiency of the BDT that suppresses combinatorial background are determined with the \sPlot technique~\cite{Pivk:2004ty,sweights}. 
Input variables to the classifier from simulation are calibrated to match those of background-subtracted data, whose weights derive from fits to the~\Bds-candidate mass. 

\begin{figure}[tb]
    \centering
    \includegraphics[width=0.5\linewidth]{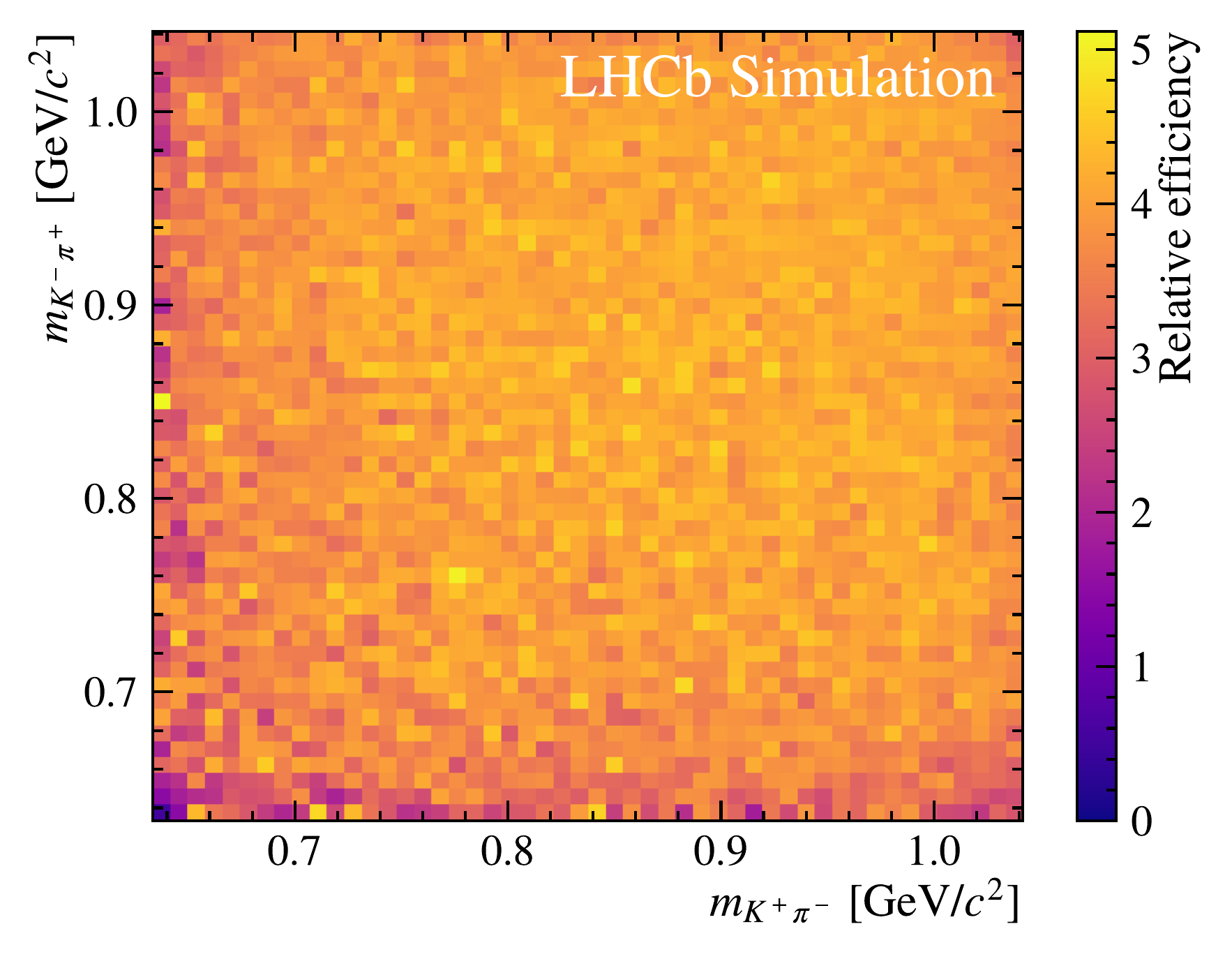}%
    \includegraphics[width=0.5\linewidth]{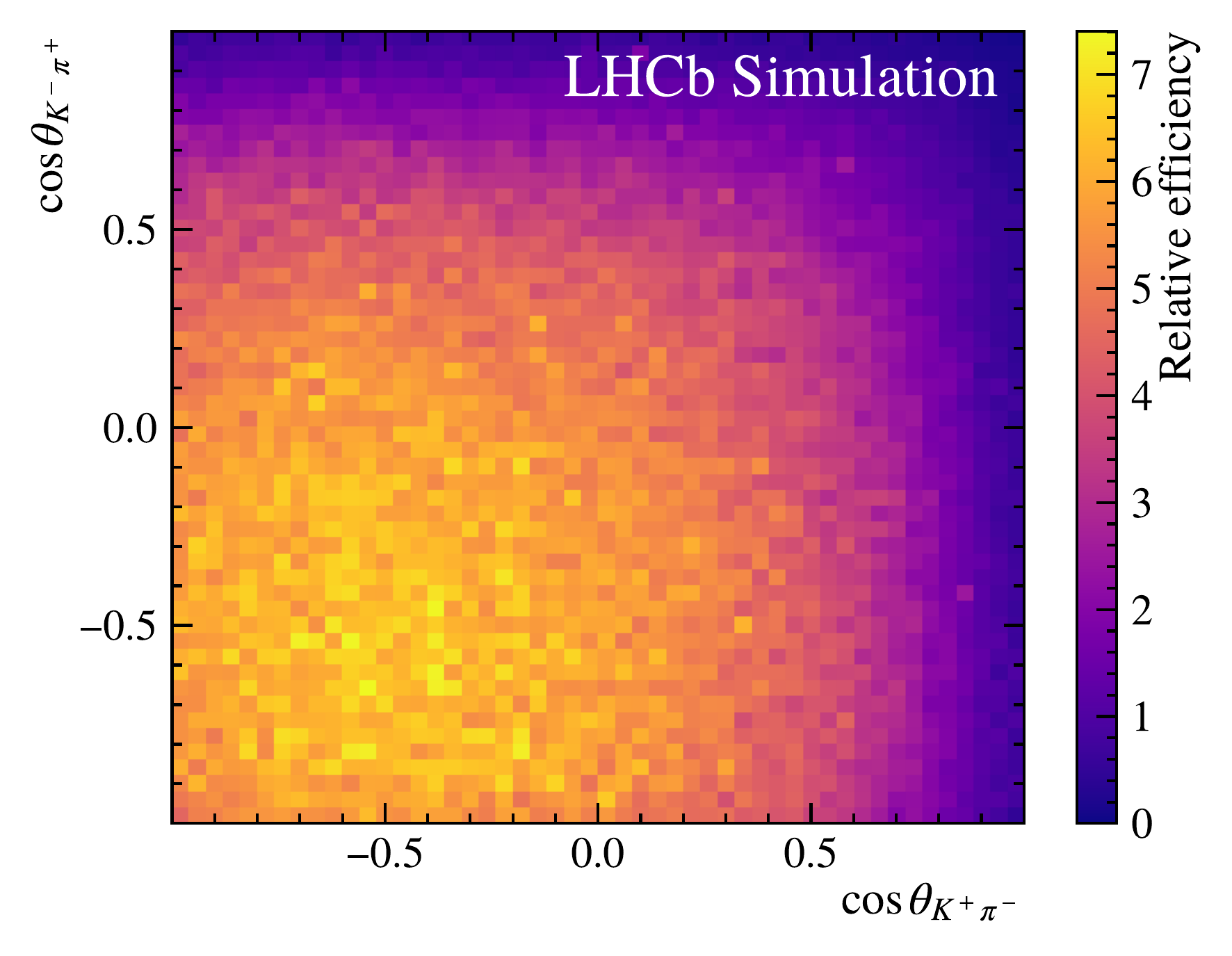}

    \caption{Examples of two-dimensional efficiency-model projections before KDE smoothing for~$B_s^0$~decays combined over data-taking year and hardware-trigger category.}
    \label{fig:efficiency}
\end{figure}

A continuous model for the efficiency is obtained via a KDE technique~\cite{Kelley2021} to reduce the impact of statistical fluctuations in regions of low efficiency arising from limited simulation sample size. While the efficiency of $\varphi$ is found to be uniform, the remaining transversity variables are visibly impacted by detector effects, examples of which are shown in Fig.~\ref{fig:efficiency}. Inefficiencies in the $K\pi$ mass arise at the kinematic threshold where the kaon and pion are produced at rest in the $K\pi$ rest frame, while the helicity angles are most impacted by the \pt trigger threshold when the kaon and pion are more aligned with the \Bds flight direction.

A notable improvement over the previous time-integrated amplitude analysis~\cite{LHCb-PAPER-2019-004} is the inclusion of experimental effects in the calculation of $c_t$ and $s_t$ from the time-integrated PDF defined in Eq.~\ref{eq:aman:rt}, whereas the previous analysis did not include the efficiency term in the integral.
Due to the finite range of~$t$ imposed in the selection criteria and heavy suppression of the detection efficiency at low~$t$ values, \CP-even amplitudes are disproportionally affected due to their shorter effective lifetimes in the presence of nonzero~$\Delta\Gamma$. 
As the fit model is only sensitive to the ratio of integrated quantities,~$r_t \equiv s_t/c_t$, the time model is largely insensitive to both the detector resolution in decay time and any differences between data and simulation. 
Therefore,~$c_t$ and $s_t$ for~\Bs decays are determined from the generated~$t$ distribution of simulated events passing the full selection procedure, again separated by data-taking year and hardware-trigger category, with the known values of $\tau_s$ and $\Delta\Gamma_s$~\cite{PDG2024,HFLAV23}. 
Experimental effects increase $r_t$ by about 32–45\% relative to the value without detection effects.
For \Bd decays, $r_t$ is set to unity on account of the negligible value of $\Delta\Gamma_d$~\cite{HFLAV23}.

\subsection{Fit strategy}
\label{sec:fit_strategy}

The raw baseline, or covariant, amplitude fit is performed utilising the \sFit method~\cite{Xie:2009rka}. 
The \Kp\pim\Km\pip mass regions in which the \Bd and \Bs amplitude analyses are performed, are chosen to be $m_{\Kp\pim\Km\pip} \in [5237, 5307]\mevcc$ and $m_{\Kp\pim\Km\pip} \in [5307, 5419]\mevcc$, respectively.
Signal projection weights~(\emph{sWeights}) from the \sPlot technique~\cite{Pivk:2004ty} are calculated with the \textsc{sweights} package~\cite{sweights} in each of these mass regions in order to remove the low background levels remaining near the \Bd and \Bs signal peaks, as seen in Fig.~\ref{fig:mass_fits}.
In these disjoint samples, a mass constraint is applied so that the sum of the four-vectors of the final-state particles produces the known mass~\cite{PDG2024} of the appropriate \Bds initial state. 
The log-likelihood function is constructed as
 \begin{equation}
     \label{eq:likelihood}
     -2\log\mathcal{L}(\vec{\theta}) = -2\frac{\sum_{i=1}^N w_i}{\sum_{i=1}^N w_i^2}\sum_{i=1}^{N} w_i \log \mathcal{P}_{\rm sig}(\Phi^i_4|\vec{\theta})\,,
 \end{equation}
where $w_i$ are the \emph{sWeights},~$N$ is the number of events entering the fit, 
and $\vec{\theta}$ are the parameters of the signal PDF (defined in Eq.~\ref{eq:time-integrated-pdf}) determined from the fit. 

This analysis improves upon previous measurements by refining the method used to evaluate the normalisation integral~\cite{LHCb-PAPER-2017-048,LHCb-PAPER-2019-004}. Previous studies used simulated samples of \mbox{\decay{\Bds}{\Kstarz \Kstarzb}} decays for integration. However, limited coverage in phase-space regions dominated by the $K\pi$ $S$-wave led to the restricted simulation sample size becoming the dominant source of systematic uncertainty. 
In this analysis, this limitation is mitigated by using a large simulation sample generated uniformly in the phase space without experimental effects. 

The minimisation is based on the~\textsc{Mint2} interface~\cite{Mint2} to the \textsc{Minuit2} minimisation
algorithm~\cite{James:1975dr,Brun:1997pa}. 
The presence of the \emph{sWeights} in the log-likelihood function is known to produce incorrect estimates of the statistical uncertainties computed by \textsc{Minuit2}~\cite{Langenbruch:2019nwe,sweights}.
Consequently, the uncertainties for the parameters from the covariant amplitude fit are instead estimated by a bootstrapping procedure with data~\cite{Efron:1979bxm}.

The~$\Kstarz$ resonance is the only intermediate vector contribution which needs to be included and its BW parameters are determined separately from the fit to the \Bd and~\Bs samples. 
For the production amplitude of the $K\pi$ $S$-wave model, an abductive-heuristic approach is adopted to set the order of the polynomials representing the magnitude and phase energy dependence. 
Through repeated fits to the data, the polynomial order is iteratively increased, terminating before reaching the point at which the change in $-2\log\mathcal{L}$ no longer corresponds to an improvement of at least three standard deviations~($\sigma$). 
In the~\Bd sample, two polynomial orders are necessary for both the magnitude and phase, while for the~\Bs sample, two orders are still sufficient for the phase, while four orders are needed in the magnitude. 
Based on previous studies, the next-leading charmless contribution to the final state is expected to arise from the \theKstarzTwo resonance~\cite{LHCb-PAPER-2019-004,LHCb-PAPER-2017-048}. However, since its residual contribution in the elastic region is estimated to be negligible, the inclusion of this state is considered as a systematic uncertainty.

The~\mbox{\decay{\Bds}{\Kstarz \Kstarzb}} contribution in a relative $S$-wave configuration serves as the reference amplitude, with its complex coupling fixed to unity along the real axis. Without flavour tagging, it is not possible in this final state to measure the phase difference between the overall~\CP-even and \CP-odd amplitudes, thus the phase of the~\mbox{\decay{\Bds}{\Kstarz \Kstarzb}} contribution in a relative $P$-wave configuration is also set to zero. 
Since amplitude analyses involve multidimensional parameter spaces, the initial parameter values may result in local, rather than global, minima of the $-2\log \mathcal{L}$ function.
To find the global minimum, a large number of fits are performed where the initial values of the complex couplings are randomised.
The fit result with the smallest $-2\log \mathcal{L}$ value out of the ensemble is then taken as the baseline result. 
No secondary solutions are found within 25 (13) units of $-2\log \mathcal{L}$ of this baseline for the \Bs (\Bd) amplitude fit. 

\subsection{Measurement quantities}

As the measured complex coefficients of the amplitude model~$a_i$ defined in Eq.~\ref{eq:aman:isobar} are convention dependent, they have limited physical meaning. 
Instead, it is useful to compare \CP-averaged fit fractions for each intermediate component~$i$, defined from the static decay amplitudes as
\begin{equation}
\label{equation:fitFracs}
    \mathcal{F}_i \equiv \frac{\int_{m} [|a_i A_i(\Phi_{4})|^2 + |a_i \Ab_i(\Phi_{4})|^2]~{\rm d}\Phi_4}{\int_{m} [|\sum_i a_i A_i(\Phi_4)|^2 + |\sum_i a_i \Ab_i(\Phi_4)|^2]~{\rm d}\Phi_4} \,,
\end{equation}
where the notation $\int_m$ indicates the integral is restricted to the analysis region, \mbox{$m_{\Kpm\pimp} \in [633, 1042]\mevcc$}. 
These fit fractions alone will not sum to unity if there is net constructive or destructive interference. 
Such effects are described by the interference fit fractions, given by
\begin{equation}
    \mathcal{I}_{i\,<\,j} \equiv \frac{\int_m [2\Real (a_i a_j^{*} A_i(\Phi_4) A_j^*(\Phi_4)) + 2\Real (a_i a_j^{*} \Ab_i(\Phi_4) \Ab_j^*(\Phi_4))]~{\rm d}\Phi_4}{\int_m [|\sum_i a_i A_i(\Phi_4)|^2 + |\sum_i a_i \Ab_i(\Phi_4)|^2]~{\rm d}\Phi_4} \,,
    \label{equation:intFracs}
\end{equation}
where the sum over all fit and interference fractions is unity by definition. Fit fractions of compound intermediate states such as that of the total \CP-even $V\Vb$ contribution,~$\mathcal{F}_{V\Vb}^{S+D}$, are obtained by summing over the fit fractions associated with their individual~$S$- and~$D$-wave relative orbital-angular-momentum configurations and the interference fraction between them. The strong-phase differences, $\delta_i \equiv \arg(a_i)$, are also determined.

To facilitate comparisons with other measurements and theoretical predictions, the results from the fit based on the covariant tensor formalism for the $\BdorBs\to\Kst\Kstb$ system are translated into parameters of the transversity basis. 
An ensemble of pseudoexperiments containing only the $V\overline{V}$ static amplitudes is generated from the covariant fit result. 
The input yield is varied according to its fit fraction as measured in the baseline result to propagate the overall statistical uncertainty, while the model parameters are distributed according to their systematic covariance matrix (see Sec.~\ref{sec:systematics}). 
A three-dimensional fit in the transversity angular variables is performed in each pseudosample, using the amplitude
\begin{align}
    A_{V\overline{V}} (\theta_{\Kp\pim}, \theta_{\Km\pip}, \varphi) & \propto 
A_L \cos\theta_{\Kp\pim} \cos\theta_{\Km\pip} \nonumber\\ 
    & + \frac{A_\parallel}{\sqrt{2}} \sin\theta_{\Kp\pim} \sin\theta_{\Km\pip} \cos\varphi + \frac{A_\perp}{\sqrt{2}} \sin\theta_{\Kp\pim} \sin\theta_{\Km\pip} \sin\varphi \,.
\label{eq:transversity_amplitude}
\end{align}
Fit fractions of each amplitude in this basis are extracted as
\begin{equation}
        f_{L,\parallel,\perp} = \frac{|A_{L,\parallel,\perp}|^2}{|A_L|^2+|A_\parallel|^2 + |A_\perp|^2} \,,
\end{equation}
while the only observable phase is $\delta_{\parallel} \equiv \arg(A_{\parallel})$.

This procedure is cross-checked using the fact that the perpendicular amplitude in the transversity basis is the $P$-wave amplitude from the covariant tensor formalism, and their fit fractions with respect to the total $V\overline{V}$ contribution should be consistent. 
The two results are in good agreement.
Furthermore, the uncertainty on the sum of the longitudinal and parallel fit fractions in the transversity basis is in close agreement with the equivalent sum of the $S$- and $D$-wave configurations in the covariant tensor formalism. 
This gives high confidence in the robustness of the statistical procedure for determining the values of the quantities in the transversity basis.

\section{Determination of branching fractions}
\label{sec:branching_fractions}

The branching fractions of \BdsKstKst decays are measured relative to \BdorBsDpi decays via a ratio of efficiency-corrected yields, according to

\begin{align}
\begin{split}
    \frac{ \br{\BdsKstKst} }{ \br{\BdorBsDpi} \br{\DmorDsmKKpi} } = &\frac{N_{\BdsKstKst}}{N_{\BdorBsDpi}}\cdot \frac{\epsilon_{\BdorBsDpi}}{\epsilon_{\BdsKstKst}}\\
    &\cdot \frac{1}{ \br{\Kstarz\to\Kp\pim}^2} \,,
    \end{split}
    \label{eq:bfs_dpi}
\end{align}
where $N$ and $\epsilon$ are signal yields and total efficiencies, respectively, while $\br{\Kstarz\to \Kp\pim}$ is set to 2/3 based on isospin symmetry and the assumption that \Kstarz mesons decay only in the $K\pi$ channel.  
The total efficiencies are obtained from the corresponding simulation samples, which are calibrated in order to correct for known data-simulation differences from sources already described in Sec.~\ref{sec:efficiency}. As these signal simulation samples are produced with the amplitude model from Run~1~\cite{LHCb-PAPER-2019-004}, they are also reweighted by the amplitude results of this analysis. 
Simulation for the normalisation samples is based on known \decay{\DmorDsm}{\Km\Kp\pim} amplitude models~\cite{CLEO:2008msk,BaBar:2010wqe}.

Uncertainties on the efficiencies are included as part of the statistical uncertainties and other effects on the assumptions made are addressed as additional sources of systematic uncertainty, detailed in Sec.~\ref{sec:systematics}.
As a cross-check, the ratio of the efficiency-corrected yields of the two normalisation decays, $N_{\BdDpi}/N_{\BsDspi}$, is computed for each data-taking period and found to be in excellent agreement with previous \lhcb measurements~\cite{LHCb-PAPER-2011-022,LHCb-PAPER-2020-046}.

The \BsKpiKpi yield, $N^{\rm Mix}_{\BsKpiKpi}$,  measured from the fits to the \Bs-candidate mass in Sec.~\ref{sec:mass_fits}, includes contributions from \Bs--\Bsb mixing according to Eq.~\ref{eq:time-integrated-pdf}, and signal components other than $\Kstar\Kstarb$ as discussed in Sec.~\ref{sec:amplitude_model}. Furthermore, the \BsKstKst simulation is also affected by \Bs--\Bsb mixing, producing the total efficiency $\epsilon^{\rm Mix}_{\BsKstKst}$. 
As decay rates are described purely with static amplitudes as defined in Sec.~\ref{sec:static_amplitude}, both $N^{\rm Mix}_{\Bs\to(\Kp\pim)(\Km\pip)}$ and $\epsilon^{\rm Mix}_{\BsKstKst}$ first need to be corrected with the results of the amplitude analysis, following a similar procedure to that discussed in Ref.~\cite{DeBruyn:2012wj}. In principle, this issue would also apply to the \BsDspi normalisation channel. However, no correction is applied to the associated branching fraction, since the other measurements entering the world average~\cite{PDG2024} do not include this effect. For consistency with those results, and given that the expected impact is at the subpercent level, the uncorrected value is used.

The corrected \BsKstKst yield, still subject to efficiency effects, is therefore obtained through reweighting $N^{\rm Mix}_{\Bs\to(\Kp\pim)(\Km\pip)}$ by the fractional contribution of the static \BsKstKst component 
propagated through time with the relation,
\begin{align}
    N_{\BsKstKst} = &  
    N^{\rm Mix}_{\Bs\to(\Kp\pim)(\Km\pip)} \times \nonumber \\
    & \frac{\int_{t_1}^{t_2} \epsilon(t)\, e^{-t/\tau} \textrm{d}t \int_m \epsilon(\Phi_4)\big[|A_{V\overline{V}}(\Phi_4)|^2 + |\Ab_{V\overline{V}}(\Phi_4)|^2\big]~\textrm{d}\Phi_4}{
    \int_m 
    \epsilon(\Phi_4)\big[(|A(\Phi_4)|^2 + |\Ab(\Phi_4)|^2) c_t - 2 \Real \big(A(\Phi_4)^* \Ab(\Phi_4) \big) s_t\big]~\textrm{d}\Phi_4} \,.
    \label{eq:bf:Ncorr}
\end{align}
The total efficiency from \BsKstKst simulation $\epsilon^{\rm Mix}_{\BsKstKst}$ 
undergoes a similar reweighting procedure to obtain the efficiency without mixing-induced effects, $\epsilon_{\BsKstKst}$. 
For this purpose, the relative efficiency model $\epsilon(\Phi_4)\epsilon(t)$, weighted by the average of the physical \BsKstKst time evolution from Eq.~\ref{eq:timeevolution}, serves as a reweighting proxy that removes the effects of the underlying generated model for $\epsilon^{\rm Mix}_{\BsKstKst}$, as
\begin{align}
\eta^{\rm Mix}_{\BsKstKst}& \nonumber\\
\propto&\frac{\int_m \big[ (|A_{VV}(\Phi_4)|^2 + |\bar A_{VV}(\Phi_4)|^2)c_t - 2 \Real(A_{VV}(\Phi_4)^* \bar A_{VV}(\Phi_4)) s_t \big] \epsilon(\Phi_4)~\textrm{d}\Phi_4}{\int \big[(|A_{VV}(\Phi_4)|^2 + |\bar A_{VV}(\Phi_4)|^2)c_t^{\rm Phys} - 2 \Real(A_{VV}(\Phi_4)^* \bar A_{VV}(\Phi_4)) s_t^{\rm Phys}\big]~\textrm{d}\Phi_4 } \,,
\label{eq:bf:effunweight}
\end{align}
where $c_t^{\rm Phys}$ and $s_t^{\rm Phys}$ are the analogue of Eq.~\ref{eq:aman:rt} with the time efficiency $\epsilon(t)$ removed and integrated over all time. Similarly, the reweighting proxy for $\epsilon_{\BsKstKst}$ is given by
\begin{equation}
    \eta_{\BsKstKst} \propto
    \frac{\int_{t_1}^{t_2} \epsilon(t)\, e^{-t/\tau} \textrm{d}t\int_m \big[ (|A_{VV}(\Phi_4)|^2 + |\bar A_{VV}(\Phi_4)|^2) \big] \epsilon(\Phi_4)~\textrm{d}\Phi_4}{\int_{0}^{\infty} e^{-t/\tau} \textrm{d}t\int \big[(|A_{VV}(\Phi_4)|^2 + |\bar A_{VV}(\Phi_4)|^2)\big]~\textrm{d}\Phi_4 } \,,
\label{eq:bf:effreweight}
\end{equation}
where the total \BsKstKst efficiency without mixing-induced effects is obtained through $\epsilon_{\BsKstKst} = (\eta_{\BsKstKst} / \eta^{\rm Mix}_{\BsKstKst})\,\epsilon^{\rm Mix}_{\BsKstKst}$.

The branching-fraction ratio of the two signal modes is determined with the single ratio
\begin{equation}
    \frac{ \br{\BdKstKst}}{\br{\BsKstKst}} = \frac{N_{\BdKstKst}}{N_{\BsKstKst}} \frac{\epsilon_{\BsKstKst}}{\epsilon_{\BdKstKst}} \frac{f_s}{f_d} \,,
    \label{eq:bfs_kst}
\end{equation}
where $f_s/f_d$ is the ratio of \Bs to \Bd fragmentation fractions~\cite{LHCb-PAPER-2020-046}. The \BdKstKst signal yield, $N_{\BdKstKst}$, can be computed with corrections analogous to Eq.~\ref{eq:bf:Ncorr}, 
with simplifications based on a vanishing $\Delta\Gamma$ in the \Bz system. For this reason, the sequential application of Eqs.~\ref{eq:bf:effunweight} and~\ref{eq:bf:effreweight} also have no effect on the \BdKstKst efficiency as expected. 
For comparative purposes, the branching-fraction-like quantity based on,
\begin{equation}
    \frac{ \mathcal{B}^{\rm Mix}(\BsKstKst) }{ \br{\BsDspi} } = \frac{N^{\rm Mix}_{\BsKstKst}}{N_{\BsDspi}} \frac{\epsilon_{\BsDspi}}{\epsilon^{\rm Mix}_{\BsKstKst}} \frac{1}{ \br{\Kstarz\to\Kp\pim}^2} \,,
    \label{eq:bfs_dpi_mix}
\end{equation}
that includes the contribution mediated by \Bs--\Bsb mixing is also reported. In this case, Eq.~\ref{eq:bf:Ncorr} is amended to provide the efficiency-corrected $N^{\rm Mix}_{\BsKstKst}$ by including the mixing term in the $V \overline V$ contribution. 
For measurements that include the \Bs--\Bsb mixing-induced contribution in incoherent \bquark\bquarkbar production, the {\it a posteriori} correction appearing in the literature,
\begin{equation}
    \eta^{\rm Mix} = \frac{1 - y^2}{1 + y(1-2f_{\perp})} \,,
\end{equation}
where $y = \DGs /(2\Gs)$, has an equivalent effect of removing the mixing term~\cite{Descotes-Genon:2011rgs}. 
However, the uncertainty on the cumulative effect of the integral ratios of Eqs.~\ref{eq:bf:Ncorr}--\ref{eq:bf:effreweight} is preferred, as it may be determined by propagation of the total covariance matrix through the raw amplitude model alone. Instead, with the use of $\eta^{\rm Mix}$, the systematic correlations between $y$, the covariant integrals of Eq.~\ref{eq:bf:Ncorr} and $f_{\perp}$ from the transversity basis, would also need to be accounted for.

\section{Systematic uncertainties}
\label{sec:systematics}
Three main classes of systematic uncertainties are considered: those that affect the mass-fit yields, the efficiency calculation and amplitude-model sources that only affect the amplitude fits. 
For any systematic uncertainty impacting the amplitude model, a covariance matrix is constructed including the amplitude fit parameters in the covariant basis.
These matrices are then summed for each systematic category in question to produce a total systematic covariance matrix. 

In order to propagate the systematic uncertainties to the transversity basis parameters, pseudoexperiments are generated from the total statistical and systematic covariance matrix with the same number of events as in the baseline fit, to preserve the statistical uncertainty. 
The spread of values across the ensemble is then distributed with variance $\sigma_\mathrm{tot}^2 = \sigma_\mathrm{stat}^2 + \sigma_\mathrm{syst}^2$.
The value of $\sigma_\mathrm{stat}^2$ is already determined from bootstrapped samples and thus a subtraction in quadrature allows for the systematic uncertainty to be estimated. 

\subsection{Four-body mass fits}
\label{sec:systematics-mass_fit}

There are three sources of systematic uncertainty considered for the four-body mass fits described in Sec.~\ref{sec:mass_fits}: imperfect modelling of the signal mass shapes, imperfect modelling of the background mass shapes and the \textit{sWeights} used to project out the signal component for the amplitude analysis.
For the signal mass shape, the four parameters describing the tails of the Hypatia distribution are varied within their uncertainties from the values determined from the fits to the simulation samples. The mass fits are repeated and the spread of results determines the systematic uncertainty value.
In addition, an alternative parametrisation of the signal shape, a Johnson-$S_U$ distribution is used instead, and the difference with the baseline fit result to data is taken as the systematic uncertainty.
For the combinatorial background shape, alternative parametrisations of an exponential distribution in Run~1 data or a second-order Bernstein polynomial in Run~2 data are used to derive a systematic uncertainty. 
Misidentified backgrounds modelled using Johnson-$S_U$ functions are instead modelled with double-sided Crystal Ball functions~\cite{Skwarnicki:1986xj}. For the partially reconstructed background shape, the bandwidth of the KDE is halved. 
The construction of the \emph{sWeights} assumes that the four-body mass is uncorrelated with the fit basis used in the amplitude fit. This assumption is tested by repeating the mass fits and recalculating the \emph{sWeights} in independent bins of the angular $\cos\theta_{\Km\pip}$ and $\cos\theta_{\Kp\pim}$ variables. The amplitude fit is repeated with the new set of \emph{sWeights} and the difference with respect to the baseline fit is taken as the systematic uncertainty.

\subsection{Efficiency}
\label{sec:systematics-efficiency}

Several sources of systematic uncertainty are considered in relation to the efficiency. Previous \lhcb analyses of the same decay modes were subject to a notably large uncertainty due to the limited size of the simulation samples \cite{LHCb-PAPER-2017-048}. The limited size of the signal simulation samples is included in the statistical uncertainty of the efficiency used in the branching-fraction calculations, while the large size of the normalisation samples results in a negligible systematic uncertainty for the amplitude analysis.
The KDE method used to model the efficiency is tested by changing to a parabolic kernel as opposed to the Gaussian default, and by halving the bandwidth.

To account for other potential biases in the method used to correct the hardware-trigger efficiency, an alternative method using \decay{\Bz}{\jpsi [\to\mumu]\Kp\pim} decays, requiring a positive trigger decision on the muons from the \jpsi decay, is used to apply corrections to the simulation~\cite{LHCb-PAPER-2017-010, LHCb-PAPER-2018-045}. 
The systematic uncertainties related to the trigger and tracking corrections are then computed by redetermining the efficiency maps when the associated corrections are varied within their uncertainties. 
A systematic uncertainty due to the PID corrections is estimated by using two alternative binning schemes and taking the largest resulting difference as the uncertainty.
To account for potential mismodelling of the combinatorial BDT response in the simulation samples, the \emph{sWeighted} BDT distribution in data is compared to that of the simulation. Weights are derived to account for the differences between data and simulation, which are then used to calculate the efficiency of the BDT requirement.
The corrective weights for the $p$ and \pt distributions of the \B candidate are further tested by deriving an alternative set of kinematic weights by changing the hyperparameters of the gradient boosted reweighter used to derive the weights.

\subsection{Amplitude analysis}
\label{sec:systematics-aman}

Potential intrinsic biases introduced by the amplitude fitter are explored by generating and fitting an ensemble of 400 pseudoexperiments and comparing the distributions of the resulting fit parameters with the baseline fit results. The uncertainty coverage is deemed to be reasonable and any significant biases are accounted for as systematic uncertainties.
Fixed parameters used in the $S$-wave parametrisation, given by Eq.~\ref{eq:swavetheory}, and the decay-time-relevant parameters from Eq.~\ref{eq:aman:rt} are varied within their uncertainties with an ensemble of 100 pseudoexperiments. The root-mean-square deviations of the parameters with respect to the baseline values are taken as the size of the systematic uncertainty.

\begin{table}[tb]
    \centering
    \caption{Absolute size of the systematic uncertainties relating to the amplitude fits to (top) \Bs and (bottom) \Bd data. The uncertainties due to the barrier factors are not included as they change the scale of the amplitude parameters and so are handled separately. Fit fraction values, $\mathcal{F}$, are quoted in percent and strong phases, $\delta$, in radians.}
    \label{tab:systematics:aman}
    \renewcommand{\arraystretch}{1.2}
    \resizebox{\linewidth}{!}{
    \begin{tabular}{l l | rrrrrrrrr | rr}
        \hline
        \hline
        & Parameter & \emph{sWeights} & KDE & KDE & Tracking & Fit bias & $S$-wave & Decay-time & $S$-wave & Tensor & Total & Total \\[-0.4em]
        &  &  & bandwidth & kernel &  &  & lineshape &  & shape & resonance & syst & stat \\
        \hline
    \multirow{9}{*}{\Bs} 
      & $\mathcal{F}_{VV}^{S+D}$ & $0.05$ & $0.07$ & $0.04$ & $<0.01$ & $0.05$ & $<0.01$ & $0.03$ & $0.14$ & $0.18$ & 0.26 & $0.63$  \\
      & $\mathcal{F}_{VV}^{P}$   & $0.19$ & $0.06$ & $0.08$ & $<0.01$ & $0.22$ & $<0.01$ & $0.13$ & $0.15$ & $0.08$ & 0.38 & $0.56$ \\
      & $\mathcal{F}_{VS}^{+}$   & $0.17$ & $0.05$ & $0.14$ & $<0.01$ & $0.19$ & $<0.01$ & $0.03$ & $0.61$ & $2.19$ & 2.29 & $0.59$ \\
      & $\mathcal{F}_{VS}^{-}$   & $0.48$ & $0.11$ & $0.04$ & $<0.01$ & $0.50$ & $<0.01$ & $0.11$ & $0.81$ & $0.26$ & 1.11 & $0.89$  \\
      & $\mathcal{F}_{SS}$       & $0.06$ & $0.02$ & $0.01$ & $<0.01$ & $0.05$ & $<0.01$ & $0.02$ & $0.09$ & $0.19$ & 0.23 & $0.56$ \\
      & $\delta_{VV}^{D}$        & $<0.01$ & $<0.01$ & $<0.01$ & $<0.01$ & $<0.01$ & $<0.01$ & $<0.01$ & $0.01$ & $<0.01$ & 0.01 & $<0.01$  \\
      & $\delta_{VS}^{+}$        & $0.02$ & $0.01$ & $0.03$ & $<0.01$ & $0.01$ & $0.01$ & $<0.01$ & $0.09$ & $0.84$ & 0.84 & $0.11$   \\
      & $\delta_{VS}^{-}$        & $0.01$ & $<0.01$ & $<0.01$ & $<0.01$ & $0.01$ & $0.01$ & $<0.01$ & $0.12$ & $0.01$ & 0.12 & $0.04$   \\
      & $\delta_{SS}$            & $0.02$ & $0.01$ & $0.02$ & $<0.01$ & $0.02$ & $0.01$ & $<0.01$ & $0.25$ & $0.01$ & 0.25 & $0.06$   \\
    \hline 
    \multirow{9}{*}{\Bd} 
      & $\mathcal{F}_{VV}^{S+D}$ & $0.05$ & $0.79$ & $2.06$ & $<0.01$ & $0.26$ & $0.03$ & $0.08$ & $0.14$ & $0.72$ & 2.34 & $2.04$ \\
      & $\mathcal{F}_{VV}^{P}$   & $0.19$ & $0.38$ & $0.45$ & $<0.01$ & $0.24$ & $<0.01$ & $0.12$ & $0.15$ & $0.28$ & 0.74 & $1.35$ \\
      & $\mathcal{F}_{VS}^{+}$   & $0.17$ & $0.39$ & $0.76$ & $<0.01$ & $0.27$ & $0.02$ & $0.05$ & $0.61$ & $0.11$ & 1.10 & $2.17$  \\
      & $\mathcal{F}_{VS}^{-}$   & $0.48$ & $0.93$ & $2.47$ & $<0.01$ & $0.11$ & $0.01$ & $0.08$ & $0.81$ & $0.72$ & 2.89 & $2.27$  \\
      & $\mathcal{F}_{SS}$       & $0.06$ & $0.12$ & $0.10$ & $<0.01$ & $0.12$ & $0.02$ & $0.01$ & $0.09$ & $0.10$ & 0.24 & $1.00$ \\
      & $\delta_{VV}^{D}$        & $<0.01$ & $0.01$ & $<0.01$ & $<0.01$ & $0.01$ & $<0.01$ & $<0.01$ & $0.01$ & $<0.01$ & 0.02 & $0.04$ \\
      & $\delta_{VS}^{+}$        & $0.02$ & $0.11$ & $0.21$ & $<0.01$ & $0.02$ & $0.01$ & $<0.01$ & $0.09$ & $<0.01$ & 0.25 & $0.22$  \\
      & $\delta_{VS}^{-}$        & $0.01$ & $0.03$ & $0.03$ & $<0.01$ & $0.01$ & $0.01$ & $<0.01$ & $0.12$ & $<0.01$ & 0.13 & $0.11$  \\
      & $\delta_{SS}$            & $0.02$ & $0.07$ & $0.10$ & $<0.01$ & $0.01$ & $0.01$ & $<0.01$ & $0.25$ & $<0.01$ & 0.28 & $0.15$  \\
    \hline
    \hline
    \end{tabular}}
\end{table}

The amplitude fits are also repeated using the LASS parametrisation \cite{Aston:1987ir} for the $S$-wave lineshape as an alternative to the scattering amplitude approach.
Contributions from tensor resonances, which are ignored in the amplitude fit model, are tested by including the lowest-lying tensor resonance in \Kp\pim, the \theKstarzTwo, in the dominant tensor topologies for these decays, the tensor-vector and vector-tensor amplitudes in a $P$-wave configuration. 
Such contributions may be more susceptible to statistical fluctuations in data, since the \theKstarzTwo pole mass resides well outside the analysis region. 
Alternative values of the effective meson radius, $r$, used in the Blatt--Weisskopf barrier factors are trialled by repeating the amplitude fit 100 times with uniformly distributed values of $r$ for the scattering and production barrier factors in the range $3$--$5\gev^{-1}\hbar c$. 
As these factors change the scale of each intermediate contribution, a correction procedure based on the convention-independent fit fractions and strong-phase differences is applied to determine the systematic uncertainties of the baseline complex couplings.

\subsection{Summary of systematic uncertainties}
The systematic uncertainties affecting only the amplitude analysis results are summarised in Table~\ref{tab:systematics:aman} and the size of the systematic uncertainties on the branching-fraction-related observables are presented in Table~\ref{tab:systematics:both}. 
For the branching-fraction-related quantities, several sources also impact the amplitude analysis thereby inducing systematic correlations between measurements. Through joint consideration, these are accounted for in the total systematic uncertainty. As such, a broad indication of the systematic uncertainty deriving from the amplitude analysis in Table~\ref{tab:systematics:both} is obtained through its evaluation in isolation. 
For \Bs decays, the systematic uncertainty is smaller than the statistical uncertainty, with the dominant systematic component arising from the signal shape modelling in the four-body mass fits.
Conversely, the \Bd measurement quantities are systematically limited. The dominant sources of systematic uncertainty arise from the KDE efficiency map and the \textit{sWeights} method. For the former, the relatively larger uncertainty arises from the limited size of the phase-space simulation sample used to train the KDE. The uncertainty associated with the \textit{sWeights} procedure is larger for the \Bd than for the \Bs sample due to the worse signal-to-background ratio of the less abundant \BdKpiKpi decay.

\begin{table}
\centering
\caption{
Sources of systematic uncertainty affecting branching-fraction-related quantities.
}
\label{tab:systematics:both}
\renewcommand{\arraystretch}{1.2}
\resizebox{\textwidth}{!}{
\begin{tabular}{l cccc}
\hline
\hline
\noalign{\vskip 3pt}
    Source & $\br{\BdKstKst}$ & $\br{\BsKstKst}$ & $\frac{\br{\BdKstKst}}{\br{\BsKstKst}}$ & $L_{\Kst\Kstb}$ \\[-0.3em]
& $\times 10^{-8}$ & $\times 10^{-7}$ & $\times 10^{-3}$ &\\
    \hline
    Tail parameters & 0.46 & 0.68 & 0.69 & 0.16\\
    Signal model & 0.76 & 0.83 & 1.98 & 0.21\\
    Background models & 3.40 & 0.28 & 4.89 & 0.37\\
    Barrier factors & 0.44 & 0.57 & 1.18 & 0.10\\
    BDT & 0.35 & 0.56 & 0.16 & 0.01\\
    PID weights & 0.53 & 0.02 & 0.64 & 0.17\\
    Kinematic weights & 1.13 & 0.61 & 2.26 & 0.27\\
    Trigger correction & 0.09 & 0.44 & 0.38 & 0.02\\

    \hline
    Total combined syst & 3.77 & 1.56 & 5.95 & 0.56\\
    Amplitude syst & 2.07 & 1.07 & 4.84 & 0.21\\
    \hline
    Total syst & 4.30 & 1.89 & 7.03 & 0.48\\
    Total stat & 2.93 & 2.47 & 3.66 & 0.55\\
    \hline
    \hline
\end{tabular}}
\end{table}

\section{Results}
\label{sec:results}

\begin{figure}
    \centering
    \includegraphics[width=0.32\linewidth]{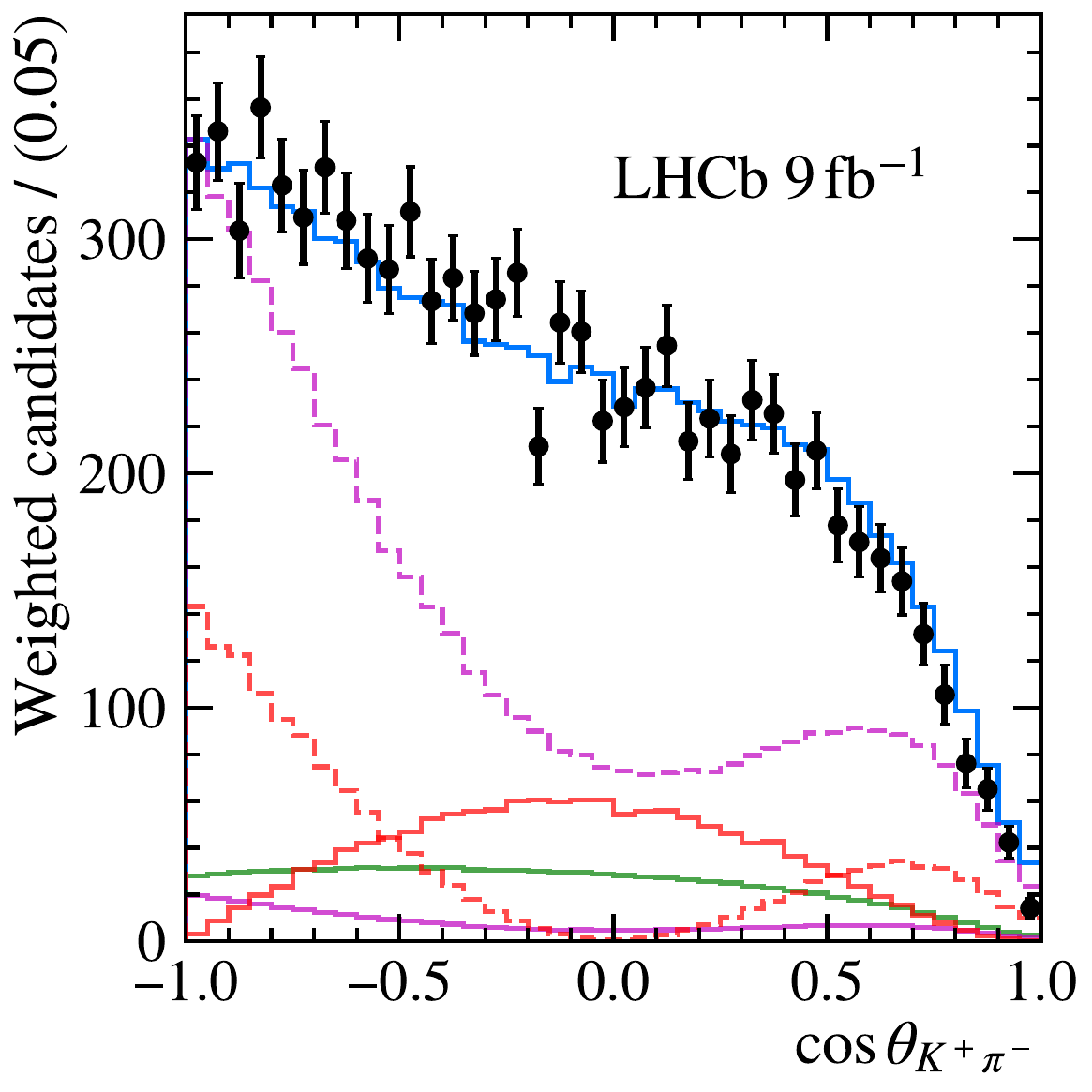}
    \includegraphics[width=0.32\linewidth]{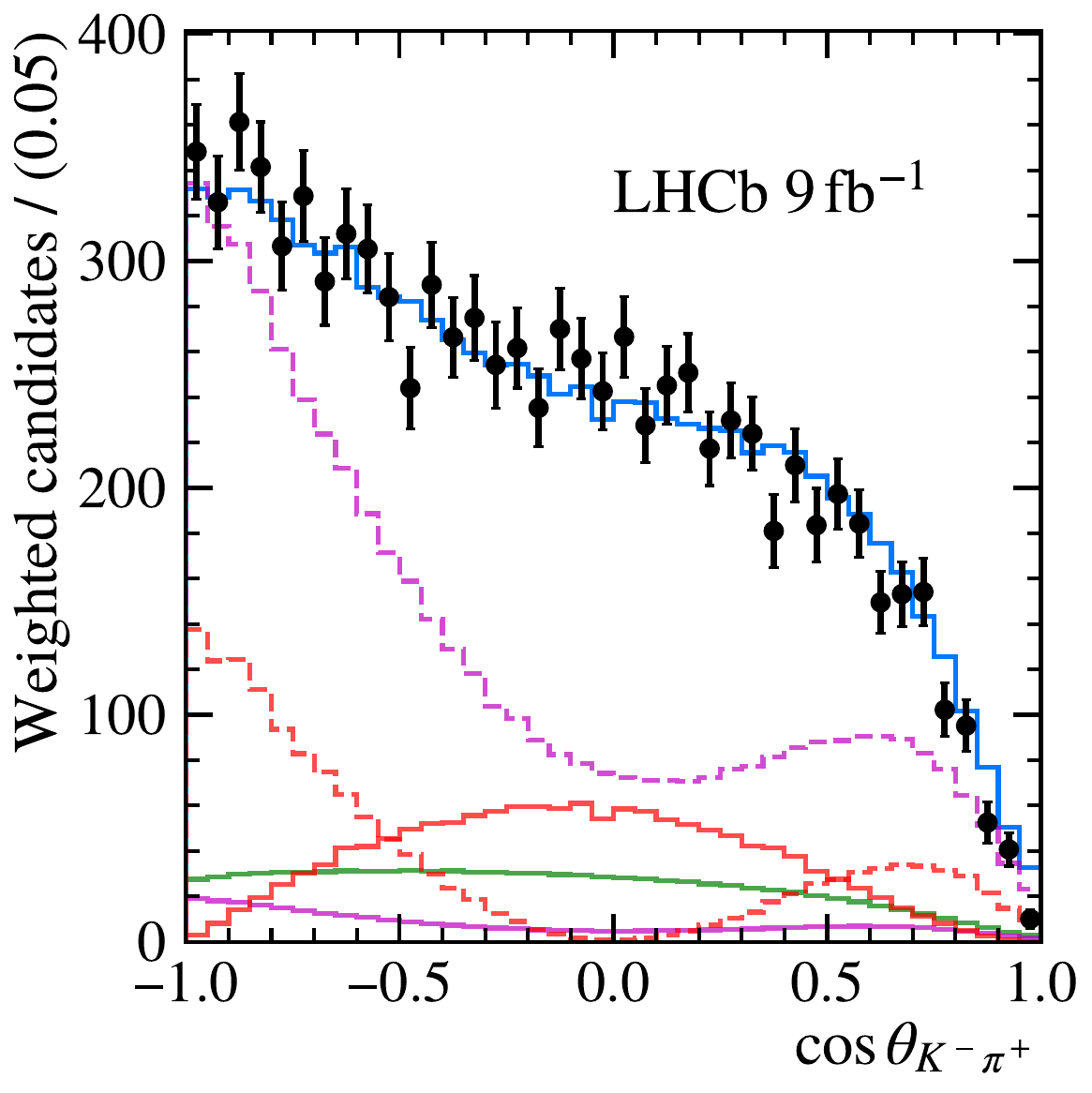}
    \includegraphics[width=0.32\linewidth]{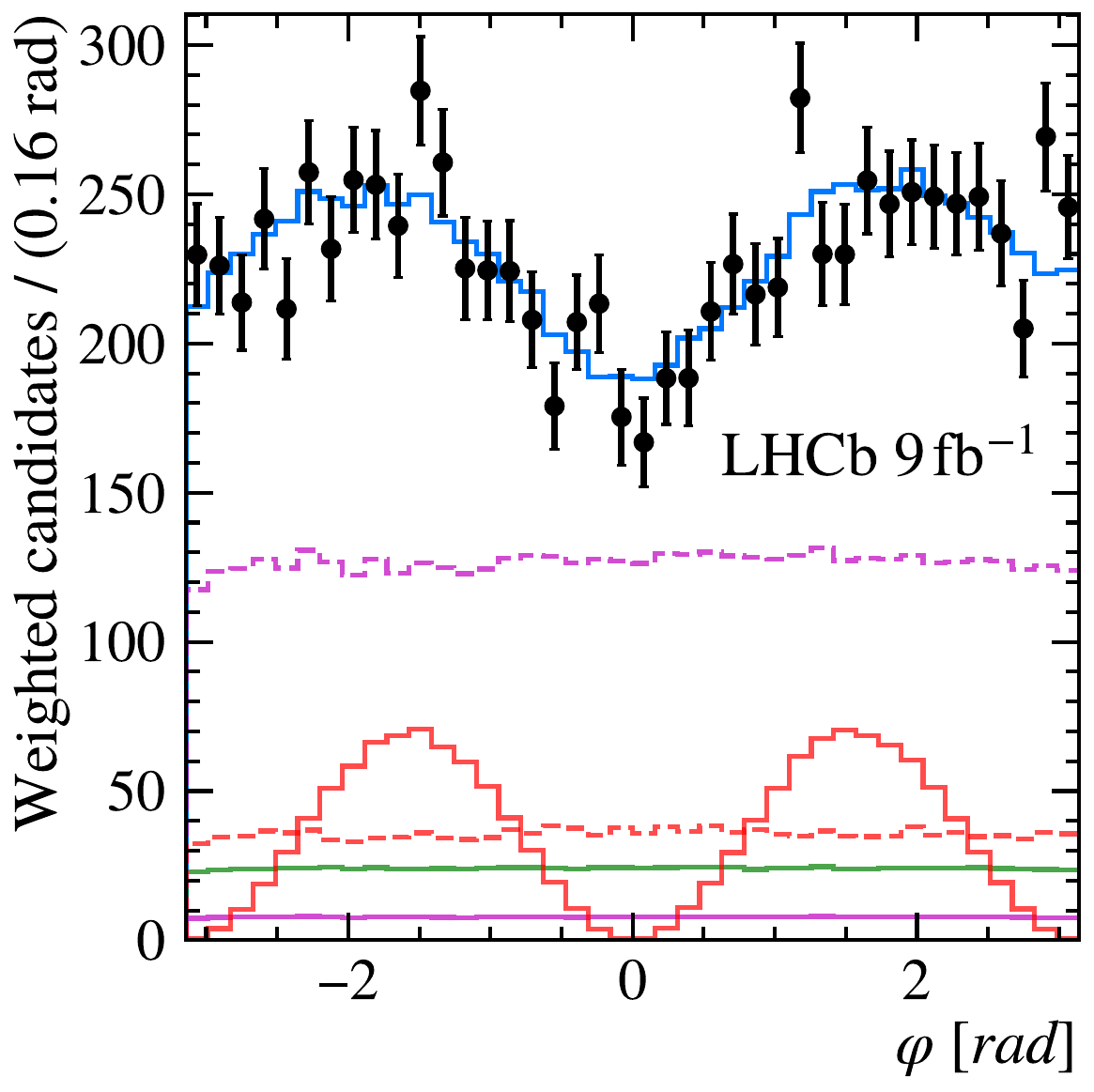} \\
    \includegraphics[width=0.32\linewidth]{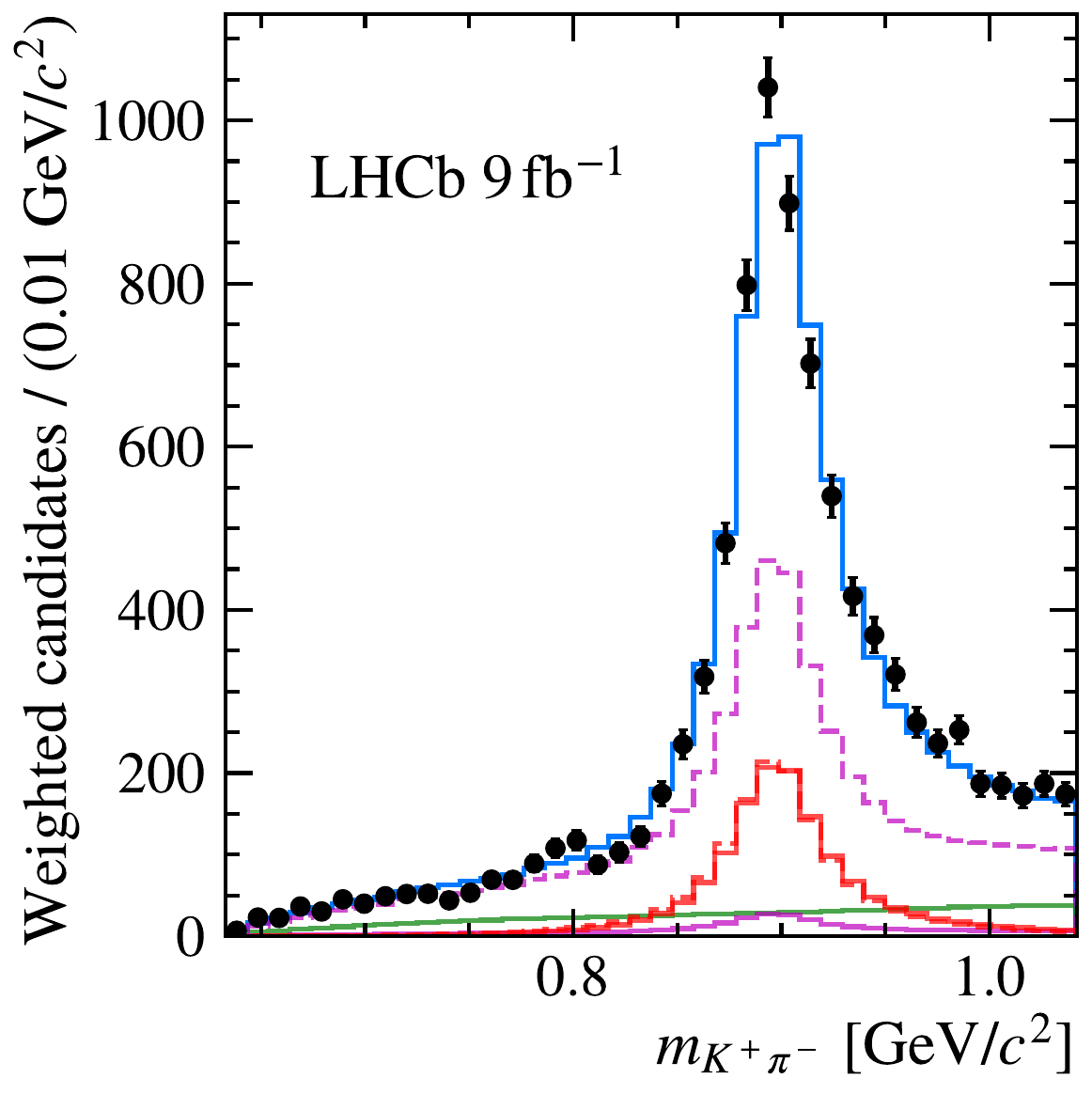}
    \includegraphics[width=0.32\linewidth]{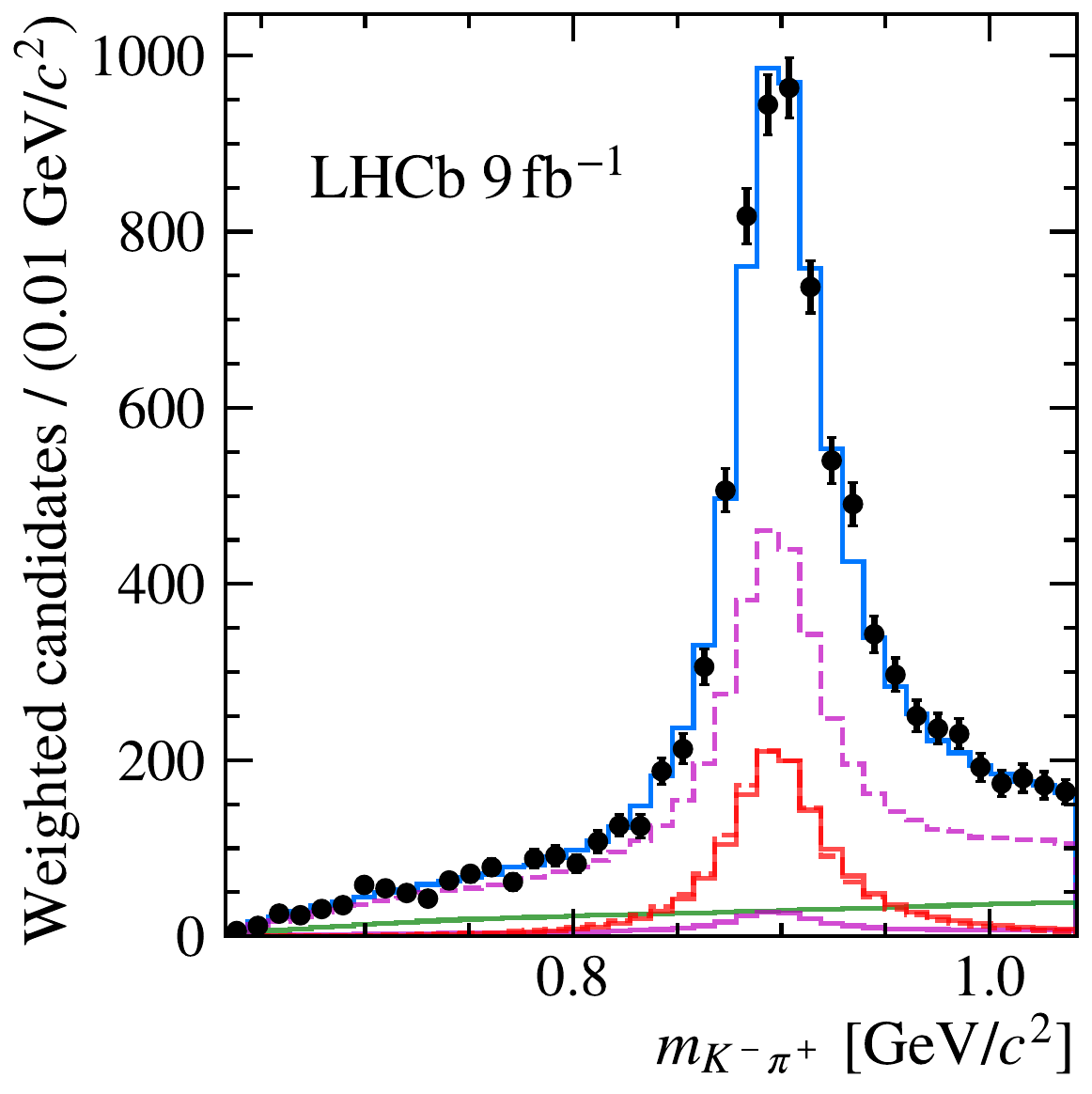} 
    \includegraphics[width=0.32\linewidth,clip,trim=4.5cm 4.2cm 4.5cm 5cm]{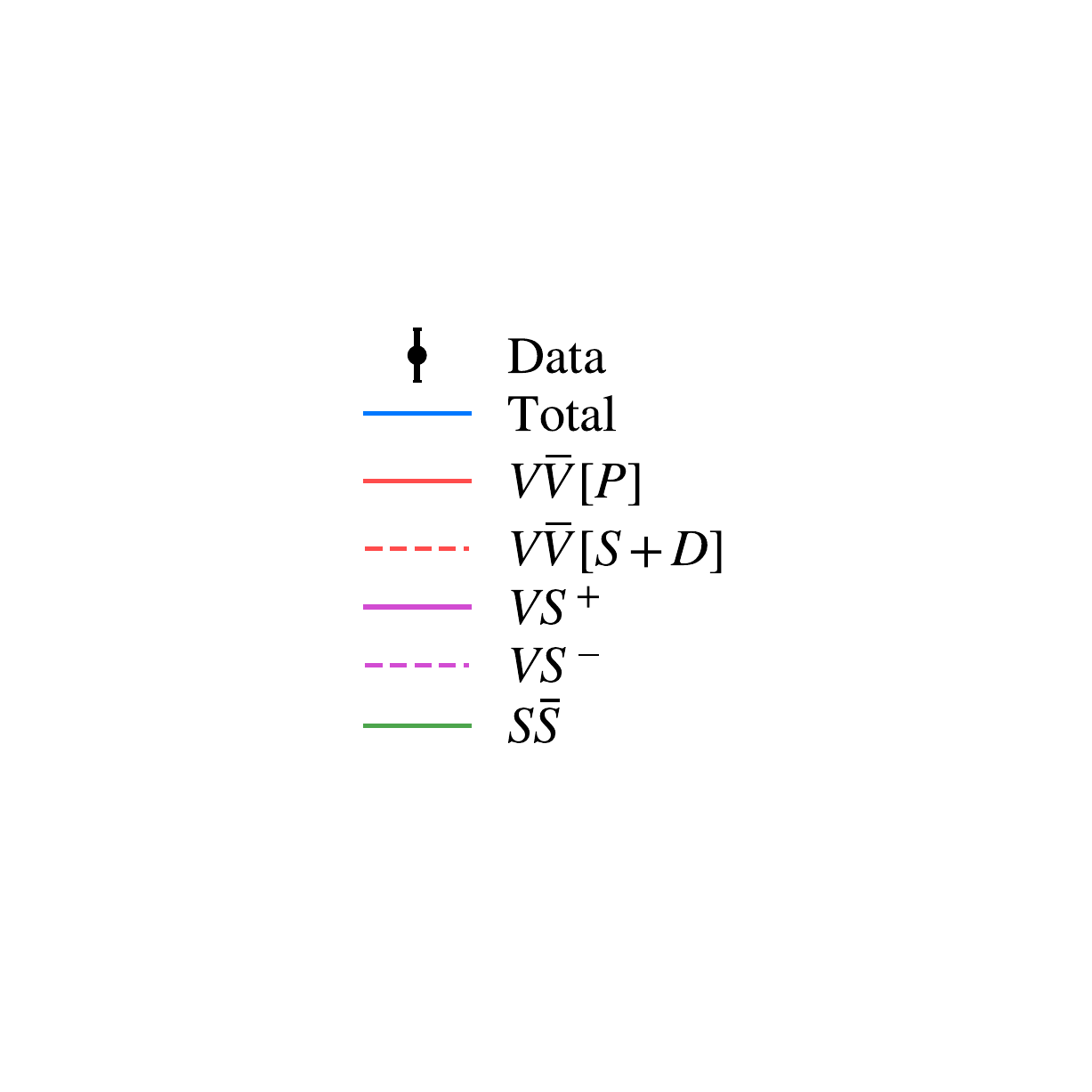} 
    \caption{Background-subtracted distributions and fit results for the~$B_s^0 \to (K^+\pi^-)(K^-\pi^+)$~data candidates in the transversity basis of (top)~the angular variables and (bottom)~the~$K^\pm\pi^\mp$~masses.}
    \label{fig:aman_res_Bs}
\end{figure}

\begin{figure}
    \centering
    \includegraphics[width=0.32\linewidth]{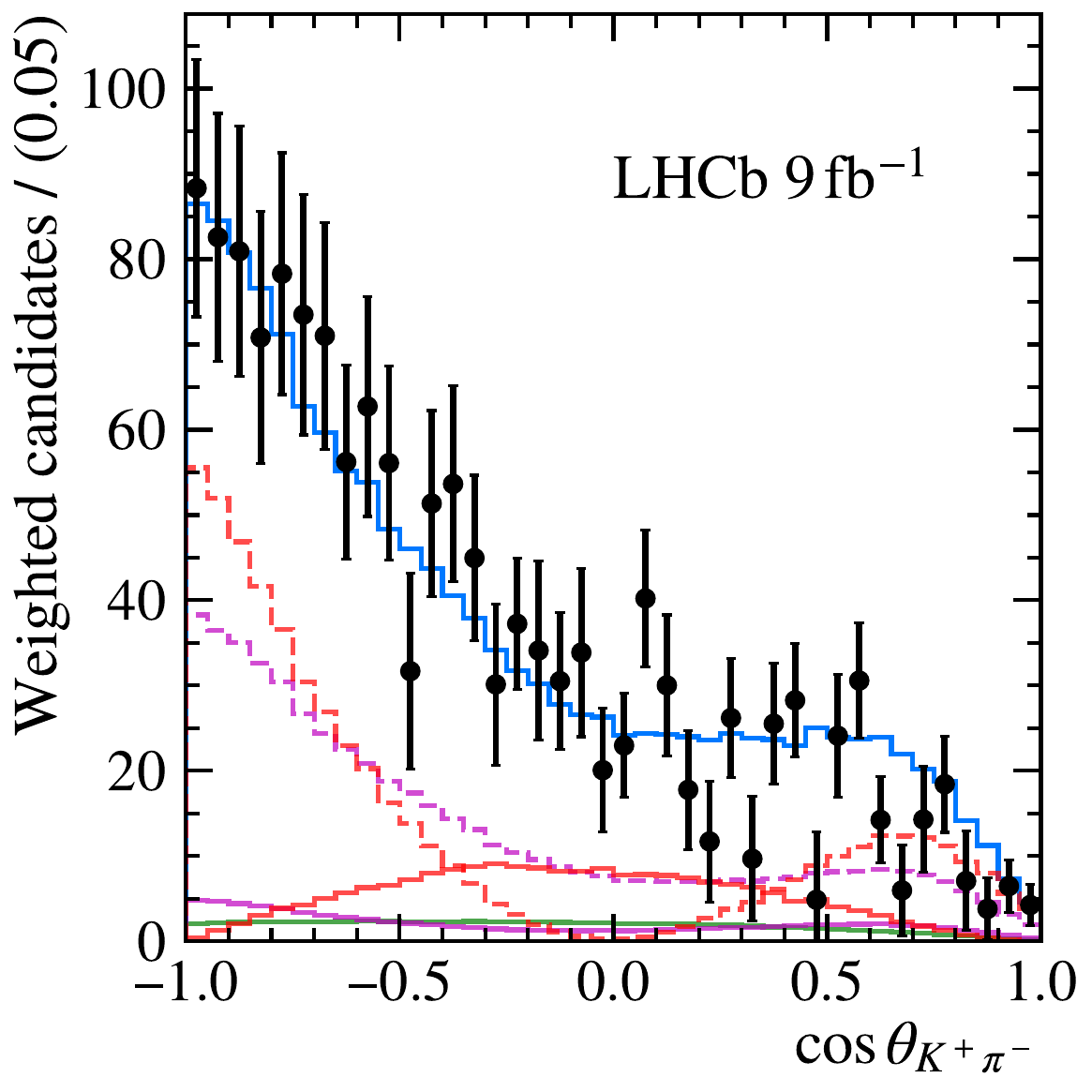}
    \includegraphics[width=0.32\linewidth]{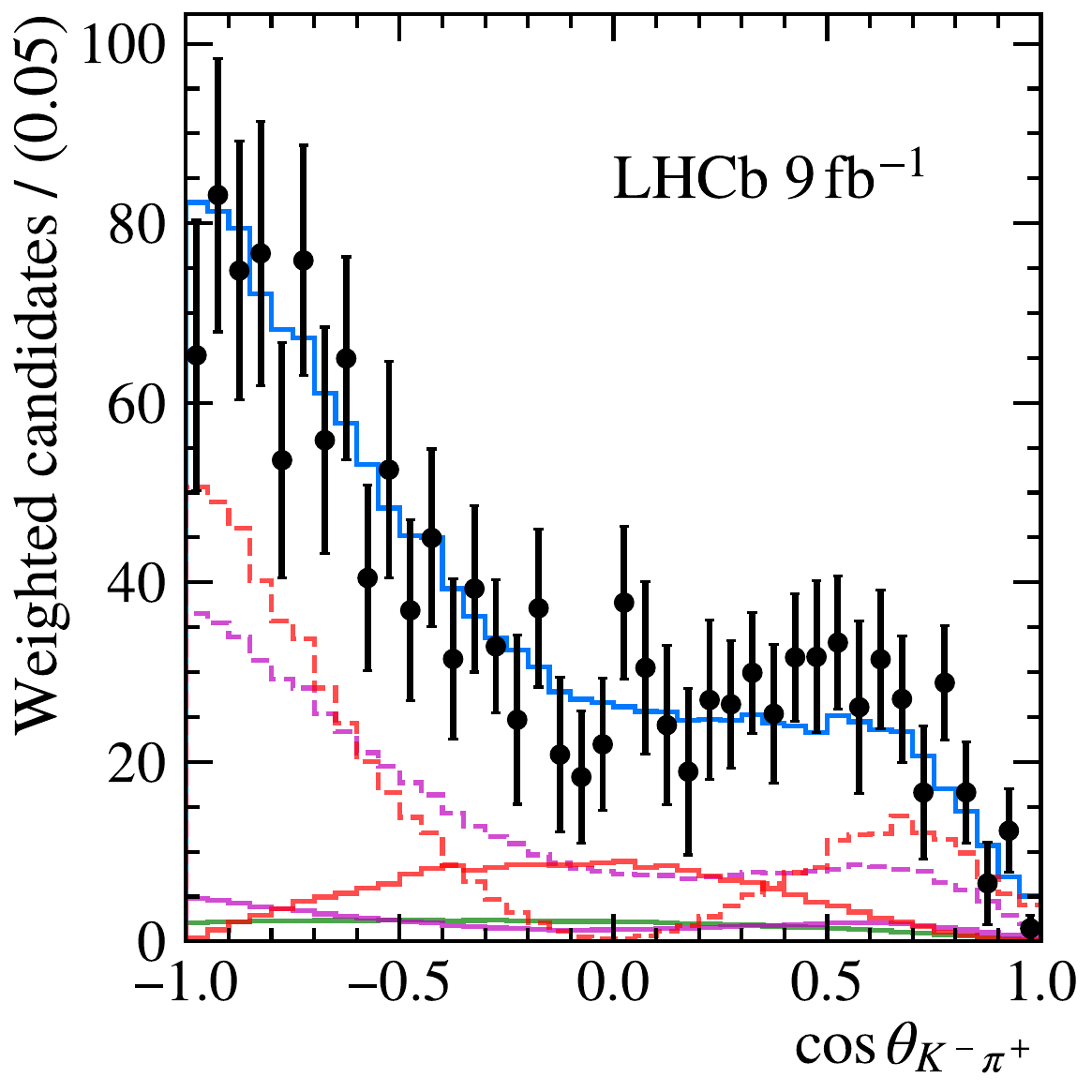}
    \includegraphics[width=0.32\linewidth]{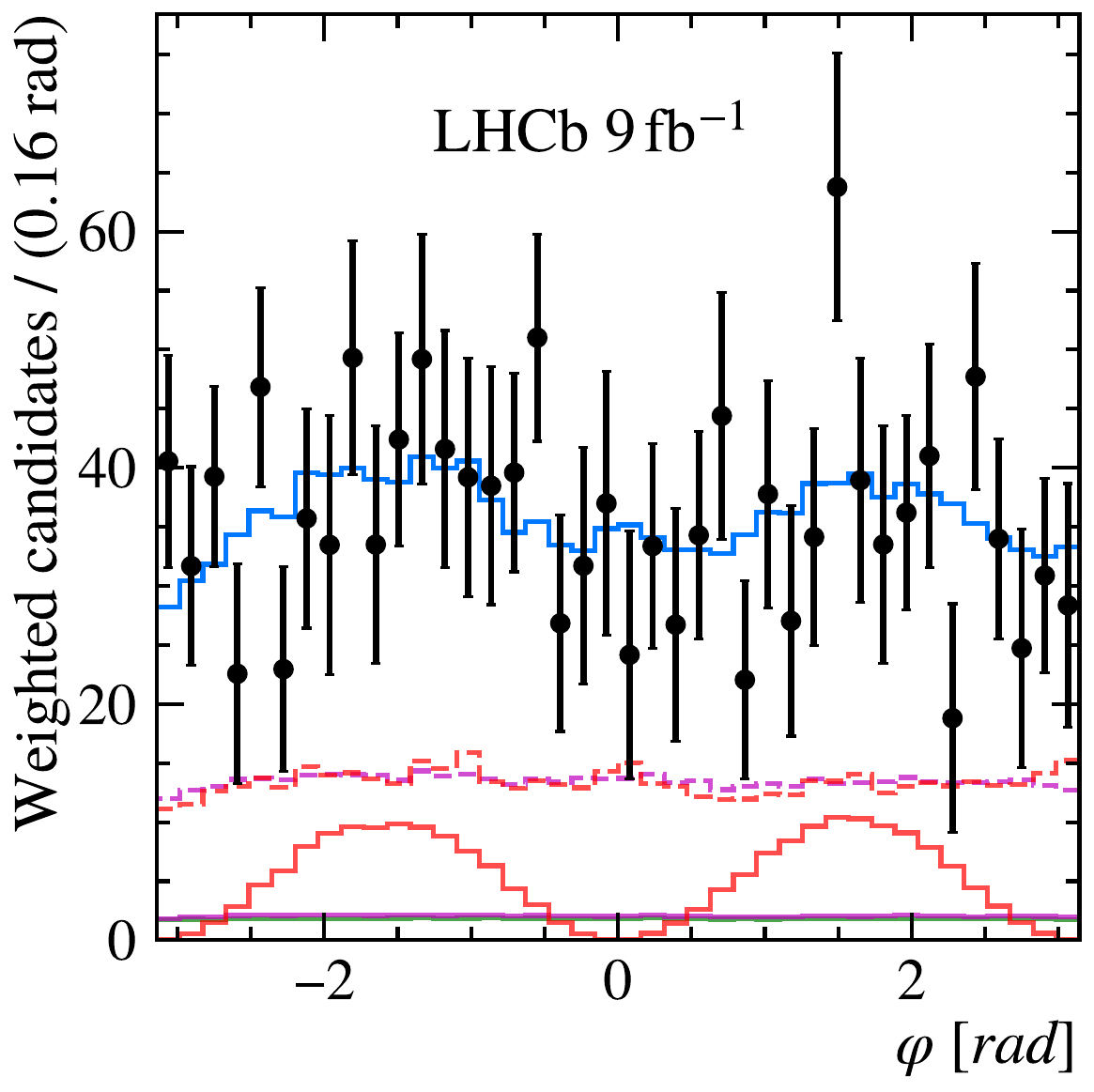} \\
    \includegraphics[width=0.32\linewidth]{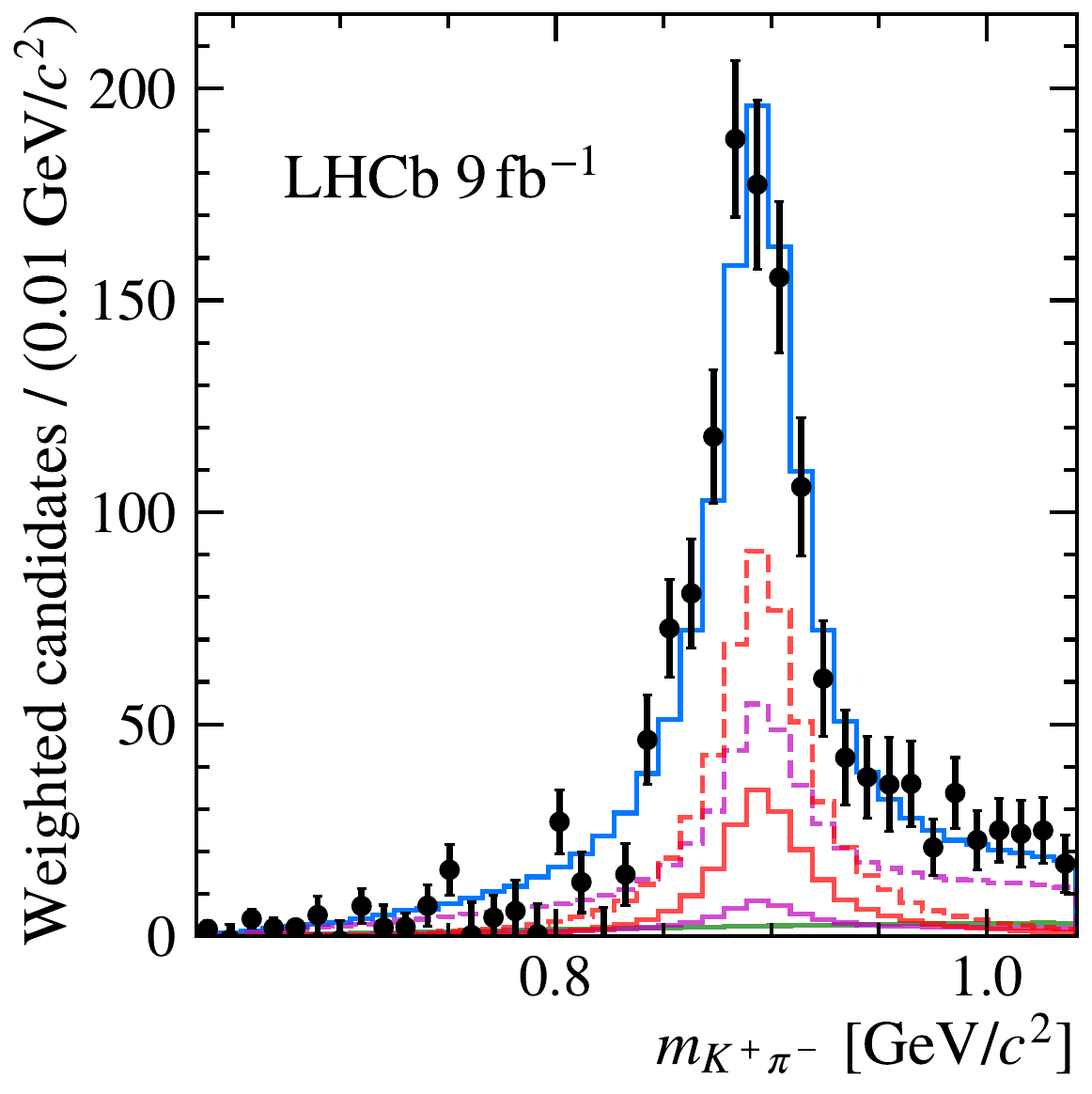}
    \includegraphics[width=0.32\linewidth]{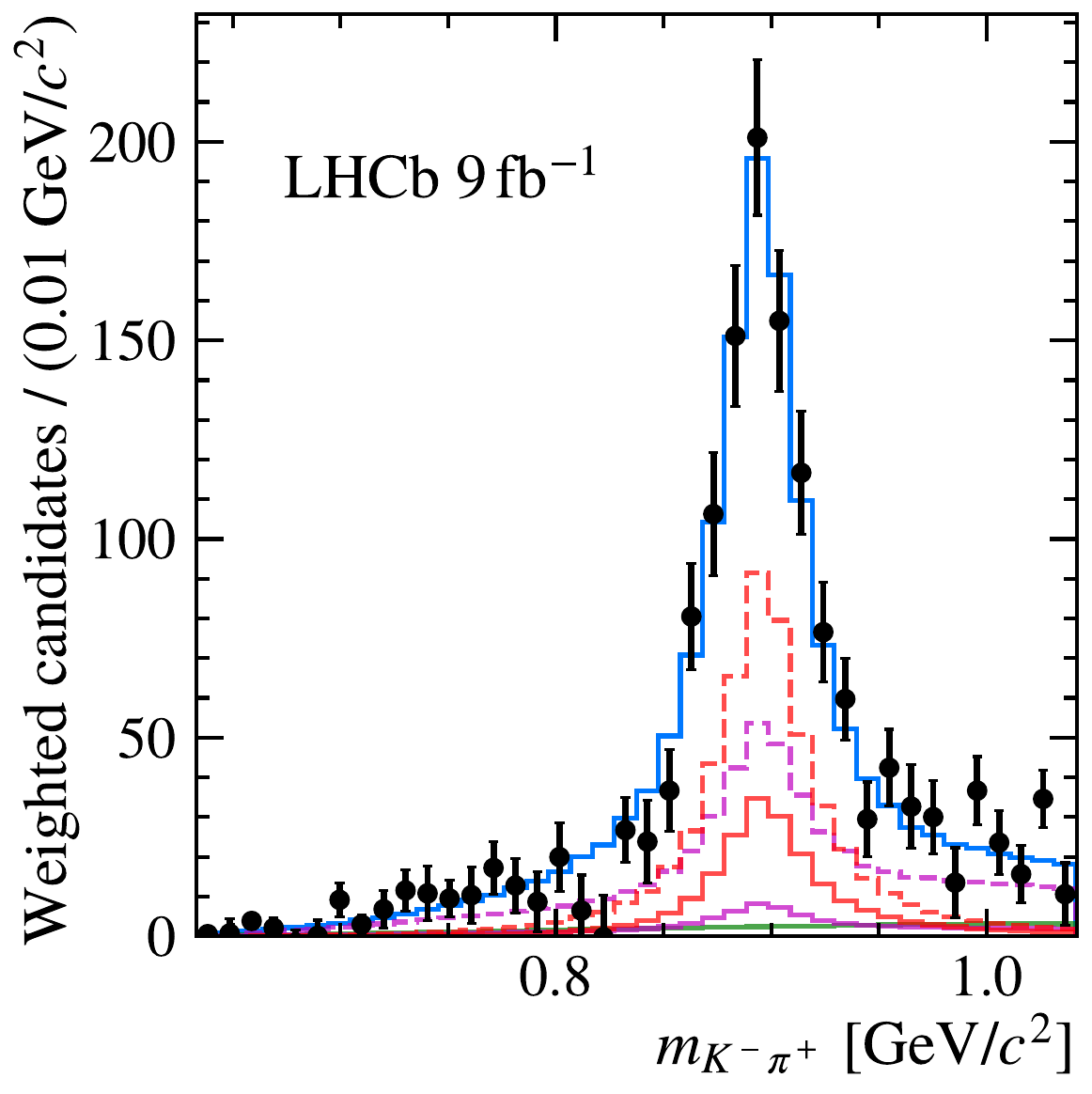} 
    \includegraphics[width=0.32\linewidth,clip,trim=4.5cm 4.2cm 4.5cm 5cm]{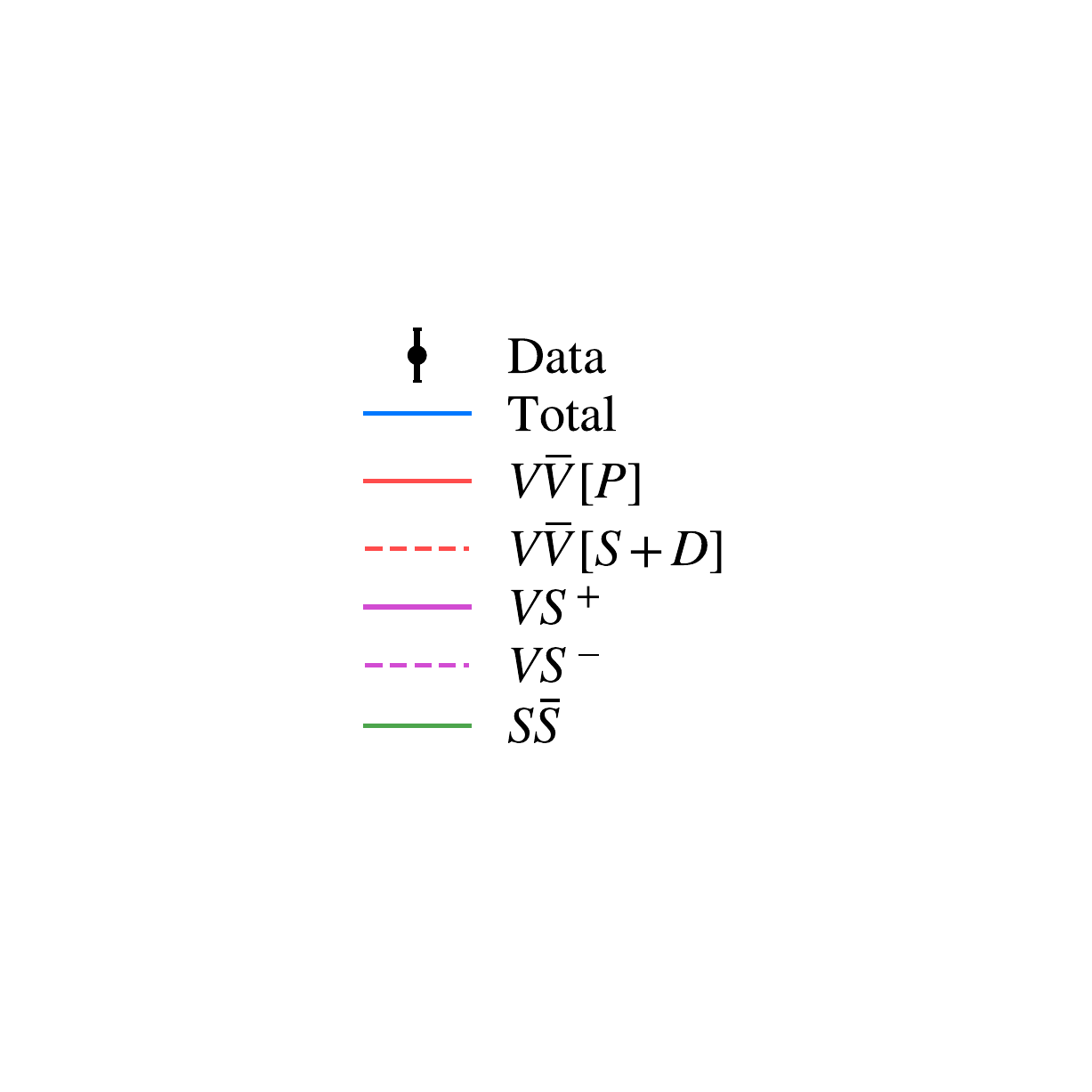} 
    \caption{Background-subtracted distributions and fit results for the~$B^0 \to (K^+\pi^-)(K^-\pi^+)$~data candidates in the transversity basis of (top)~the angular variables and (bottom)~the~$K^\pm\pi^\mp$~masses.}
    \label{fig:aman_res_Bd}
\end{figure}

The results of the amplitude fit projected onto the five variables in the transversity basis are shown in Figs.~\ref{fig:aman_res_Bs} and \ref{fig:aman_res_Bd}, with the numerical values reported in Table~\ref{tab:aman_res}. 
Numerous projections into various slices of each transversity variable reveal no obvious source of mismodelling in the phase space. 
The corresponding correlation matrices can be found in Appendix~\ref{sec:app_correlations}, while other notable hadronic results are presented in Appendix~\ref{sec:app_hadronic}.
The longitudinal polarisation fractions are measured to be
\begin{equation*}
    \begin{alignedat}{7}
    \fLBs &= 0.159 \pm 0.010 \pm 0.007, \\
    \fLBd &= 0.600 \pm 0.022 \pm 0.017 ,
    \end{alignedat}
\end{equation*}
where the first uncertainty is statistical and the second is systematic.
Similar to previous analyses~\cite{LHCb-PAPER-2019-004,LHCb-PAPER-2017-048}, the longitudinal polarisation fraction $f_L$ for \decay{\Bs}{\Kstarz\Kstarzb}  is found to be significantly lower than the corresponding value for \decay{\Bd}{\Kstarz\Kstarzb} decays.

Using Eqs.~\ref{eq:bfs_dpi} and \ref{eq:bfs_kst}, the vector-vector branching-fraction ratios 
are measured to be
\begin{equation*}
    \begin{alignedat}{7}
        \frac{\mathcal{B}^{\rm Mix}(\decay{\Bs}{\Kstarz\Kstarzb})}{\mathcal{B}(\BsDspi)\br{\DsKKpi}} &= 0.0589~&&\pm~0.0016\stat&&\pm~0.0012\syst&&,\\
        \frac{\mathcal{B}(\decay{\Bs}{\Kstarz\Kstarzb})}{\mathcal{B}(\BsDspi)\br{\DsKKpi}} &= 0.0586~&&\pm~0.0016\stat&&\pm~0.0012\syst&&,\\
        \frac{\mathcal{B}(\decay{\Bd}{\Kstarz\Kstarzb})}{\mathcal{B}(\BdDpi)\br{\DKKpi}} &= 0.0195~&&\pm~0.0012\stat&&\pm~0.0019\syst&&,\\
        \frac{\mathcal{B}(\decay{\Bd}{\Kstarz\Kstarzb})}{\mathcal{B}(\decay{\Bs}{\Kstarz\Kstarzb})} &= \mathrm{0.054}\phantom{0}~&&\pm~0.004\phantom{0}\stat&&\pm~0.007\phantom{0}\syst\\
        &\multicolumn{5}{l}{~$\pm~0.002~(f_s/f_d)$,}
    \end{alignedat}
\end{equation*}
where the uncertainties are statistical, systematic, and due to knowledge of the fragmentation-fraction ratio, respectively.
Using the known values for the total branching fractions of the normalisation decays, $\mathcal{B}(\decay{\Bs}{\Dsm\pip})\times\mathcal{B}(\decay{\Dsm}{\Kp\Km\pim}) = (1.60\pm0.08)\times10^{-4}$ and $\mathcal{B}(\decay{\Bd}{\Dm\pip})\times\mathcal{B}(\decay{\Dm}{\Kp\Km\pim}) = (2.43\pm0.09)\times10^{-5}$\cite{PDG2024}, the vector-vector branching fractions are determined to be
\begin{equation*}
    \begin{alignedat}{7}
        \mathcal{B}^{\rm Mix}(\decay{\Bs}{\Kstarz\Kstarzb}) &= (0.942~&&\pm~0.025\stat&&\pm~0.019\syst&&\pm~0.036~(\mathrm{ext}))\times10^{-5},\\
        \mathcal{B}(\decay{\Bs}{\Kstarz\Kstarzb}) &= (0.938~&&\pm~0.025\stat&&\pm~0.019\syst&&\pm~0.036~(\mathrm{ext}))\times10^{-5},\\
        \mathcal{B}(\decay{\Bd}{\Kstarz\Kstarzb}) &= (4.73~&&\pm~0.30\phantom{0}\stat&&\pm~0.43\phantom{0}\syst&&\pm~0.16\phantom{0}~(\mathrm{ext}))\times10^{-7},\\
    \end{alignedat}
\end{equation*}
where the uncertainty due to knowledge of external branching fractions is propagated separately. These values are in good agreement with the world average branching fractions \BfBdorBsKst, but with an increase in precision by a factor 4.4 and 5.7 for \BfBdKst and \BfBsKst, respectively. 
They are also in good agreement with recent phenomenological predictions~\cite{Yu:2024vlz}.

The theory-motivated $L_{\Kstarz\Kstarzb}$ observable is determined, using Eq.~\ref{eq:LKK}, to be
\begin{equation*}
L_{\Kstarz\Kstarzb} = 4.92 \pm 0.55\stat \pm 0.48 \syst \pm 0.02~(\mathrm{ext})\pm0.10~(f_s/f_d).
\end{equation*}
Here the external uncertainty refers to the uncertainties of the masses and lifetimes of the \Bd and \Bs mesons which are used to compute $\mathcal{G}$. The value of \LKK is in good agreement with the value calculated with the Run 1 \lhcb amplitude analysis results~\cite{LHCb-PAPER-2019-004,Alguero:2020xca,Biswas:2023pyw,Biswas:2025drz} and considerably more precise.
This measurement confirms the tension between the experimental results and the most recent theory predictions~\cite{Biswas:2025drz}, computed within the framework of QCD factorisation, at the level of $4.4\sigma$.

\begin{table}[tb]
    \centering
    \caption{Fit fractions and phases from the \Bs and \Bd amplitude fits, including the corresponding values for the vector-vector transversity amplitudes. The first uncertainty is statistical and the second is systematic.}
    \label{tab:aman_res}
    \renewcommand{\arraystretch}{1.2}
    \resizebox{\linewidth}{!}{
    \begin{tabular}{llc@{\:$\pm$\:}c@{\:$\pm$\:}l lc@{\:$\pm$\:}c@{\:$\pm$\:}l lc@{\:$\pm$\:}c@{\:$\pm$\:}l}
    \hline
    \hline
        & Parameter & \mc{3}{c}{Value} & Parameter & \mc{3}{c}{Value} & Parameter & \mc{3}{c}{Value} \\
    \hline
        \multirow{5}{*}{\Bs} & $\mathcal{F}^{S+D}_{VV}$\,(\%) & $15.6$ & $0.63$ & $0.26$ & $\mathcal{F}^P_{VV}$\,(\%) & $15.55$ & $0.56$ & $0.38$ & $\mathcal{F}^+_{VS}$\,(\%) & $3.57$ & $0.59$ & $2.3$ \\
        & $\mathcal{F}^{-}_{VS}$\,(\%) & $54.75$ & $0.89$ & $1.1$ & $\mathcal{F}_{SS}$\,(\%) & $10.55$ & $0.56$ & $0.23$ & $\delta^D_{VV}$\,(rad) & $-0.074$ & $0.005$ & $0.008$\\
        & $\delta^+_{VS}$\,(rad) & $-1.92$ & $0.11$ & $0.84$ & $\delta^-_{VS}$\,(rad) & $-2.98$ & $0.04$ & $0.12$ & $\delta_{SS}$\,(rad) & $0.12$ & $0.06$ & $0.25$\\
    \cline{2-13}
        & $f_L$\,(\%) & $15.9$ & $1.0$ & $0.7$ & $f_\parallel$\,(\%) & $34.2$ & $1.3$ & $1.0$ & $f_{\perp}$\,(\%) & $50.0$ & $1.4$ & $0.3$ \\
        & $\delta_\parallel$\,(rad) & $0.73$ & $0.08$ & $0.12$ &  \mc{3}{l}{} & & \mc{3}{l}{}\\
    \hline
    \hline
        \multirow{5}{*}{\Bd} & $\mathcal{F}^{S+D}_{VV}$\,(\%) & $37$ & $2$ & $2$ & $\mathcal{F}^P_{VV}$\,(\%) & $14$ & $1$ & $0.7$ & $\mathcal{F}^+_{VS}$\,(\%) & $6$ & $2$ & $1$\\
        & $\mathcal{F}^{-}_{VS}$\,(\%) & $38$ & $2$ & $3$ & $\mathcal{F}_{SS}$\,(\%) & $5$ & $1$ & $0.2$ & $\delta^D_{VV}$\,(rad) & $-0.10$ & $0.04$ & $0.02$ \\
        & $\delta^+_{VS}$\,(rad) & $2.1$ & $0.2$ & $0.3$ & $\delta^-_{VS}$\,(rad) & $-0.4$ & $0.1$ & $0.1$ & $\delta_{SS}$\,(rad) & $-1.7$ & $0.2$ & $0.3$ \\
    \cline{2-13}
        & $f_L$\,(\%) & $60$ & $2$ & $2$ & $f_\parallel$\,(\%) & $17$ & $2$ & $2$ & $f_{\perp}$\,(\%) & $24$ & $2$ & $2$ \\
        & $\delta_\parallel$\,(rad) & $0.3$ & $0.2$ & $0.1$ & \mc{3}{l}{} & & \mc{3}{l}{}\\
    \hline
    \hline
    \end{tabular}}
\end{table}

\section{Summary}
\label{sec:conclusions}
An untagged, decay-time-integrated analysis of the \BdsKpiKpi decays in the 
$\kaon^*(892)^0 \Kb{}^*(892)^0$ region is presented, using $pp$ collision data collected by the \lhcb experiment, corresponding to an integrated luminosity of 9\invfb. The longitudinal polarisation fraction for the \BsKstKst decay is measured to be significantly lower than that of the \BdKstKst decay, contrary to the expectation from U-spin symmetry and QCD factorisation. Updated measurements of the branching fractions \BfBsKst and \BfBdKst are provided, and the U-spin observable \LKK is calculated to be $4.4\sigma$ away from the SM prediction. 
These measurements provide valuable input for refining theoretical form-factor calculations, helping to constrain hadronic effects that impact precision tests of the SM. The observed deviations from SM expectations may possibly indicate contributions from physics beyond the SM, but may also indicate poorly constrained SM effects, providing motivation for continued theoretical and experimental scrutiny, not only in the \BdsKstKst decay modes studied here, but also in related \BVV modes. With the increased statistical precision anticipated with the LHCb Run~3 dataset, substantially improved precision will be possible. 
In conjunction with future reductions in the theoretical uncertainty, this will enable more stringent tests of flavour symmetries and the opportunity to confirm or resolve the current tensions.

\emph{Data availability:} The data that support the findings of this article are openly available~\cite{LHCb-PAPER-2025-046-cds}.

\section*{Acknowledgements}
%
%
\noindent We express our gratitude to our colleagues in the CERN
accelerator departments for the excellent performance of the LHC. We
thank the technical and administrative staff at the LHCb
institutes.
We acknowledge support from CERN and from the national agencies:
ARC (Australia);
CAPES, CNPq, FAPERJ and FINEP (Brazil); 
MOST and NSFC (China); 
CNRS/IN2P3 (France); 
BMFTR, DFG and MPG (Germany);
INFN (Italy); 
NWO (Netherlands); 
MNiSW and NCN (Poland); 
MCID/IFA (Romania); 
MICIU and AEI (Spain);
SNSF and SER (Switzerland); 
NASU (Ukraine); 
STFC (United Kingdom); 
DOE NP and NSF (USA).
We acknowledge the computing resources that are provided by ARDC (Australia), 
CBPF (Brazil),
CERN, 
IHEP and LZU (China),
IN2P3 (France), 
KIT and DESY (Germany), 
INFN (Italy), 
SURF (Netherlands),
Polish WLCG (Poland),
IFIN-HH (Romania), 
PIC (Spain), CSCS (Switzerland), 
and GridPP (United Kingdom).
We are indebted to the communities behind the multiple open-source
software packages on which we depend.
Individual groups or members have received support from
Key Research Program of Frontier Sciences of CAS, CAS PIFI, CAS CCEPP, 
Minciencias (Colombia);
EPLANET, Marie Sk\l{}odowska-Curie Actions, ERC and NextGenerationEU (European Union);
A*MIDEX, ANR, IPhU and Labex P2IO, and R\'{e}gion Auvergne-Rh\^{o}ne-Alpes (France);
Alexander-von-Humboldt Foundation (Germany);
ICSC (Italy); 
Severo Ochoa and Mar\'ia de Maeztu Units of Excellence, GVA, XuntaGal, GENCAT, InTalent-Inditex and Prog.~Atracci\'on Talento CM (Spain);
SRC (Sweden);
the Leverhulme Trust, the Royal Society and UKRI (United Kingdom).

\clearpage
\section*{Appendices}

\appendix

\section{Logarithmic scale invariant-mass fit plots}
\label{sec:app_log_scale_mass_plots}

\begin{figure}[h]
    \centering
    \includegraphics[width=0.49\linewidth]{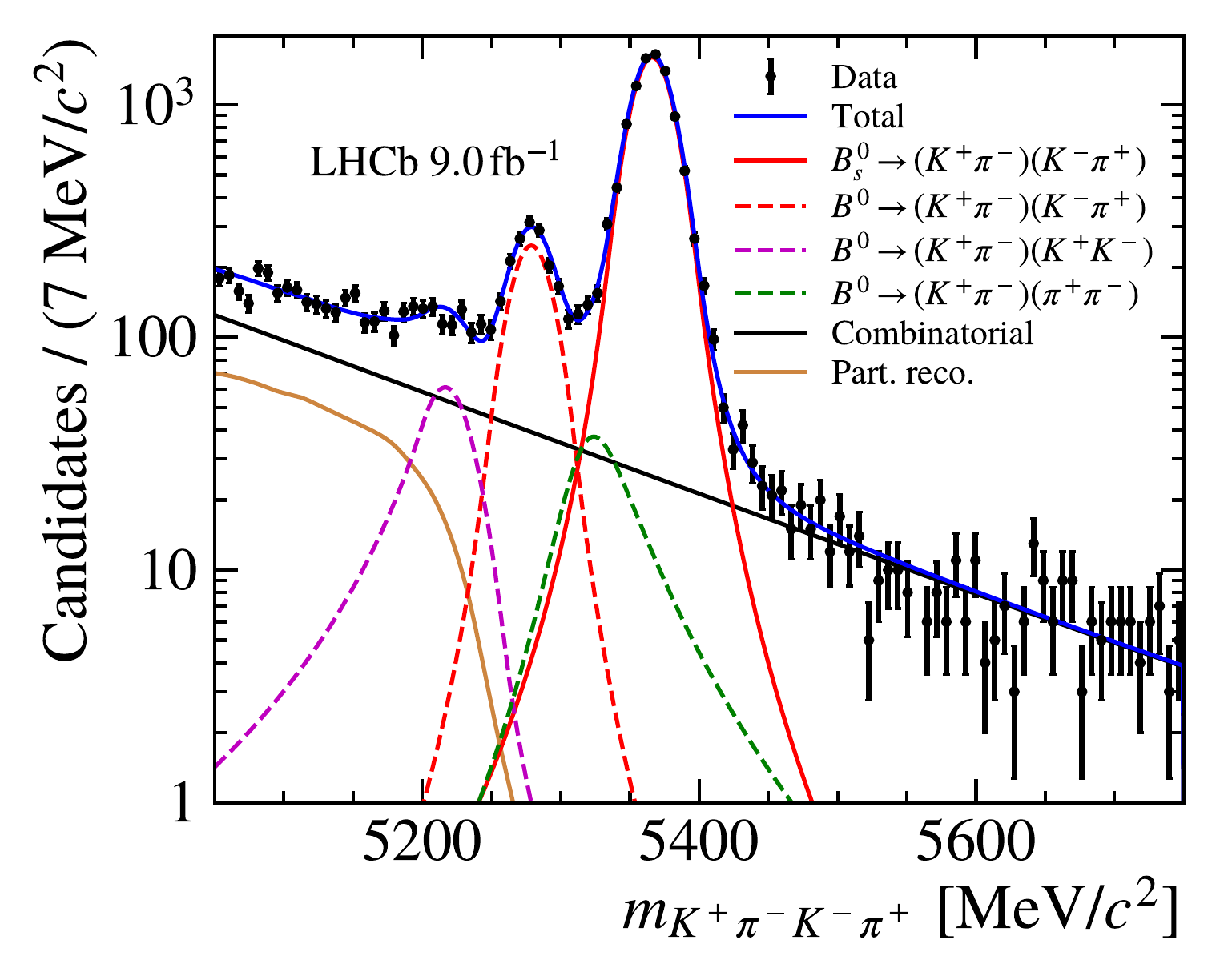} \\
    \includegraphics[width=0.49\linewidth]{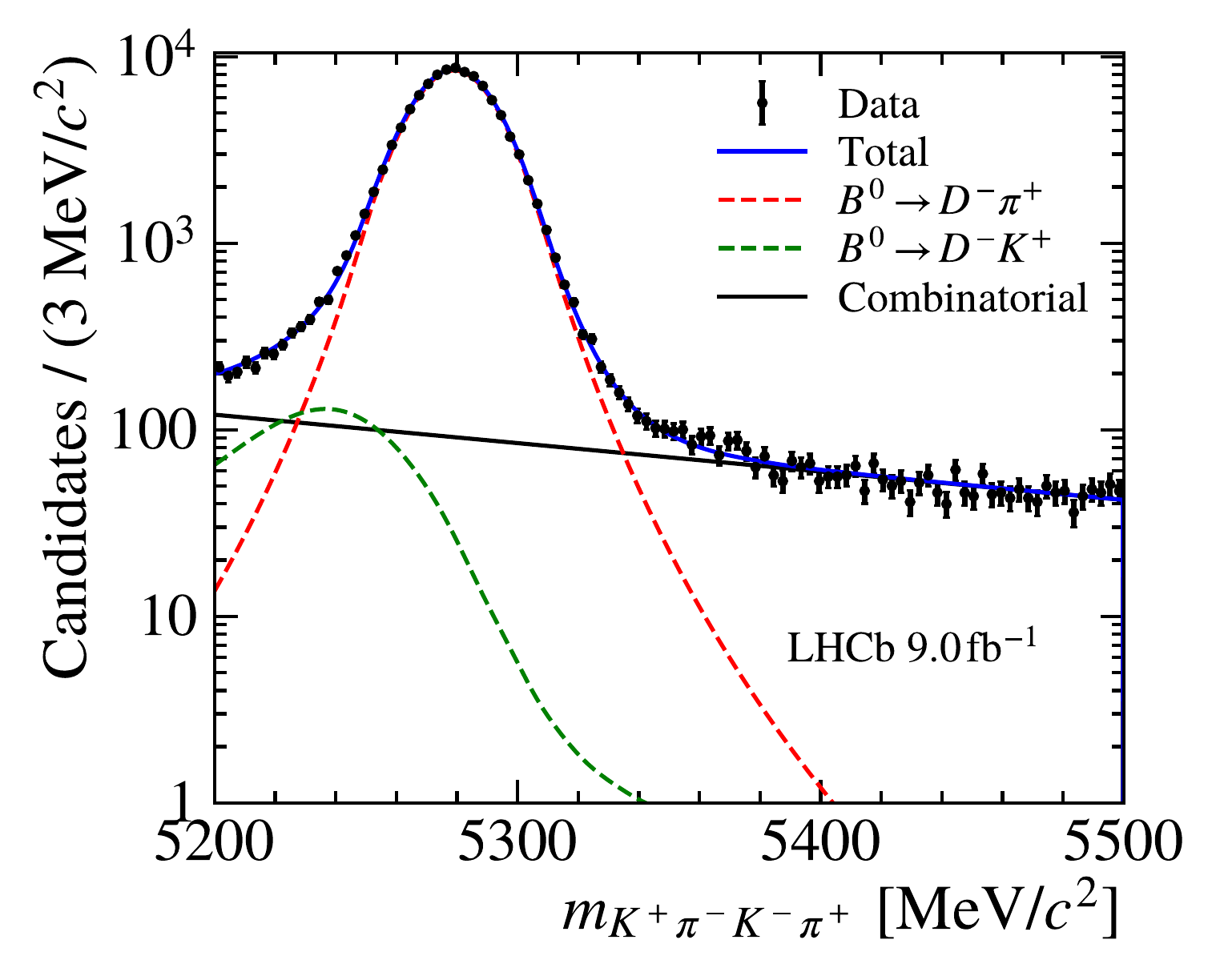}
    \includegraphics[width=0.49\linewidth]{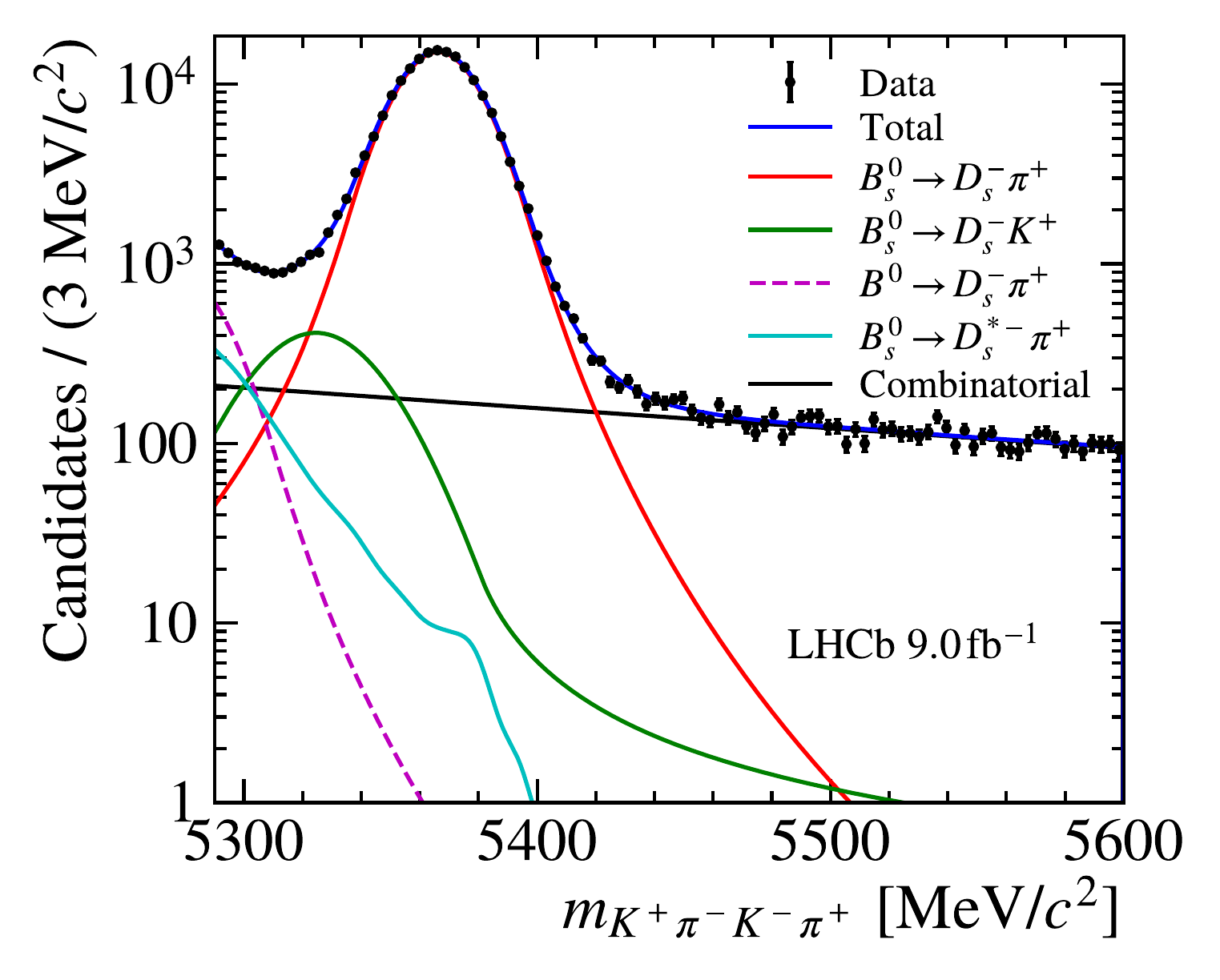}
    \caption{Distributions and fit results of the $K^+\pi^-K^-\pi^+$ mass for the (top)~signal modes, (bottom left)~$B^0$~normalisation mode and (bottom right)~$B_s^0$~normalisation mode. Equivalent plots on a linear scale are shown in Fig.~\ref{fig:mass_fits}.}
    \label{fig:mass_fits_log}
\end{figure}

The results of the fits to the selected signal and normalisation candidates, as described in Sec.~\ref{sec:mass_fits}, are shown with a logarithmic $y$-axis in Fig.~\ref{fig:mass_fits_log}.

\FloatBarrier

\section{Correlation matrices}
\label{sec:app_correlations}

The statistical and systematic correlation matrices for the \Bd and \Bs amplitude fits, in both the covariant and transversity basis, are shown in Tables~\ref{tab:corr_covariant_Bs_stat}--\ref{tab:corr_transversity_Bd_syst}.

\begin{table}
    \renewcommand{\arraystretch}{1.2}
    \centering
    \caption{Statistical uncertainty correlation matrix for the \BsKstKst amplitude fit in the covariant basis.}
    \label{tab:corr_covariant_Bs_stat}
    \begin{tabular}{l | r r r r r r r r r}
        & $\mathcal{F}^{S+D}_{VV}$ & $\mathcal{F}^{P}_{VV}$ & $\mathcal{F}^+_{VS}$ & $\mathcal{F}^-_{VS}$ & $\mathcal{F}_{SS}$ & $\delta^D_{VV}$ & $\delta^+_{VS}$ & $\delta^-_{VS}$ & $\delta_{SS}$ \\
    \hline
    $\mathcal{F}^{S+D}_{VV}$ & $1.00$ & $-0.04$ & $-0.00$ & $-0.43$ & $-0.36$ & $-0.23$ & $-0.02$ & $0.13$ & $0.10$ \\
    $\mathcal{F}^{P}_{VV}$ &  & $1.00$ & $-0.02$ & $-0.36$ & $-0.15$ & $-0.09$ & $-0.02$ & $-0.02$ & $-0.07$ \\
    $\mathcal{F}^+_{VS}$ &  &  & $1.00$ & $-0.57$ & $-0.02$ & $-0.12$ & $-0.26$ & $-0.01$ & $-0.00$ \\
    $\mathcal{F}^-_{VS}$ &  &  &  & $1.00$ & $-0.30$ & $0.29$ & $0.17$ & $0.01$ & $0.03$ \\
    $\mathcal{F}_{SS}$ &  &  &  &  & $1.00$ & $-0.03$ & $0.02$ & $-0.13$ & $-0.09$ \\
    $\delta^D_{VV}$ &  &  &  &  &  & $1.00$ & $0.06$ & $-0.29$ & $-0.25$ \\
    $\delta^+_{VS}$ &  &  &  &  &  &  & $1.00$ & $0.10$ & $0.09$ \\
    $\delta^-_{VS}$ &  &  &  &  &  &  &  & $1.00$ & $0.80$ \\
    $\delta_{SS}$ &  &  &  &  &  &  &  &  & $1.00$ \\
    \end{tabular}
\end{table}

\begin{table}
    \renewcommand{\arraystretch}{1.2}
    \centering
    \caption{Systematic uncertainty correlation matrix for the \BsKstKst amplitude fit in the covariant basis.}
    \label{tab:corr_covariant_Bs_syst}
    \begin{tabular}{l | r r r r r r r r r}
        & $\mathcal{F}^{S+D}_{VV}$ & $\mathcal{F}^{P}_{VV}$ & $\mathcal{F}^+_{VS}$ & $\mathcal{F}^-_{VS}$ & $\mathcal{F}_{SS}$ & $\delta^D_{VV}$ & $\delta^+_{VS}$ & $\delta^-_{VS}$ & $\delta_{SS}$ \\
    \hline
    $\mathcal{F}^{S+D}_{VV}$ & $1.00$ & $1.00$ & $-0.16$ & $-0.74$ & $-0.56$ & $0.84$ & $0.16$ & $0.72$ & $-0.86$ \\
    $\mathcal{F}^{P}_{VV}$ &  & $1.00$ & $-0.15$ & $-0.74$ & $-0.56$ & $0.83$ & $0.16$ & $0.73$ & $-0.86$ \\
    $\mathcal{F}^+_{VS}$ &  &  & $1.00$ & $-0.07$ & $-0.05$ & $0.07$ & $-0.37$ & $-0.10$ & $0.10$ \\
    $\mathcal{F}^-_{VS}$ &  &  &  & $1.00$ & $-0.08$ & $-0.80$ & $-0.24$ & $-0.83$ & $0.96$ \\
    $\mathcal{F}_{SS}$ &  &  &  &  & $1.00$ & $-0.32$ & $0.19$ & $-0.04$ & $0.09$ \\
    $\delta^D_{VV}$ &  &  &  &  &  & $1.00$ & $0.14$ & $0.64$ & $-0.84$ \\
    $\delta^+_{VS}$ &  &  &  &  &  &  & $1.00$ & $0.22$ & $-0.28$ \\
    $\delta^-_{VS}$ &  &  &  &  &  &  &  & $1.00$ & $-0.88$ \\
    $\delta_{SS}$ &  &  &  &  &  &  &  &  & $1.00$ \\
    \end{tabular}
\end{table}

\begin{table}
    \renewcommand{\arraystretch}{1.2}
    \centering
    \caption{Statistical uncertainty correlation matrix for the \BdKstKst amplitude fit in the covariant basis.}
    \label{tab:corr_covariant_Bd_stat}
    \begin{tabular}{l | r r r r r r r r r}
        & $\mathcal{F}^{S+D}_{VV}$ & $\mathcal{F}^{P}_{VV}$ & $\mathcal{F}^+_{VS}$ & $\mathcal{F}^-_{VS}$ & $\mathcal{F}_{SS}$ & $\delta^D_{VV}$ & $\delta^+_{VS}$ & $\delta^-_{VS}$ & $\delta_{SS}$ \\
    \hline
    $\mathcal{F}^{S+D}_{VV}$ & $1.00$ & $-0.20$ & $-0.28$ & $-0.43$ & $-0.18$ & $-0.27$ & $-0.16$ & $0.04$ & $0.01$ \\
    $\mathcal{F}^{P}_{VV}$ &  & $1.00$ & $-0.19$ & $-0.13$ & $-0.09$ & $-0.12$ & $0.07$ & $0.18$ & $0.17$ \\
    $\mathcal{F}^+_{VS}$ &  &  & $1.00$ & $-0.55$ & $-0.34$ & $0.30$ & $0.09$ & $-0.08$ & $-0.22$ \\
    $\mathcal{F}^-_{VS}$ &  &  &  & $1.00$ & $0.13$ & $-0.03$ & $0.06$ & $0.05$ & $0.18$ \\
    $\mathcal{F}_{SS}$ &  &  &  &  & $1.00$ & $0.07$ & $-0.09$ & $-0.24$ & $-0.12$ \\
    $\delta^D_{VV}$ &  &  &  &  &  & $1.00$ & $0.13$ & $-0.72$ & $-0.52$ \\
    $\delta^+_{VS}$ &  &  &  &  &  &  & $1.00$ & $0.08$ & $-0.13$ \\
    $\delta^-_{VS}$ &  &  &  &  &  &  &  & $1.00$ & $0.62$ \\
    $\delta_{SS}$ &  &  &  &  &  &  &  &  & $1.00$ \\
    \end{tabular}
\end{table}

\begin{table}
    \renewcommand{\arraystretch}{1.2}
    \centering
    \caption{Systematic uncertainty correlation matrix for the \BdKstKst amplitude fit in the covariant basis.}
    \label{tab:corr_covariant_Bd_syst}
    \begin{tabular}{l | r r r r r r r r r}
        & $\mathcal{F}^{S+D}_{VV}$ & $\mathcal{F}^{P}_{VV}$ & $\mathcal{F}^+_{VS}$ & $\mathcal{F}^-_{VS}$ & $\mathcal{F}_{SS}$ & $\delta^D_{VV}$ & $\delta^+_{VS}$ & $\delta^-_{VS}$ & $\delta_{SS}$ \\
    \hline
    $\mathcal{F}^{S+D}_{VV}$ & $1.00$ & $0.94$ & $-0.24$ & $-0.72$ & $-0.79$ & $-0.30$ & $0.37$ & $-0.57$ & $0.64$ \\
    $\mathcal{F}^{P}_{VV}$ &  & $1.00$ & $-0.43$ & $-0.63$ & $-0.78$ & $-0.44$ & $0.28$ & $-0.56$ & $0.67$ \\
    $\mathcal{F}^+_{VS}$ &  &  & $1.00$ & $0.54$ & $-0.17$ & $0.46$ & $0.08$ & $0.70$ & $-0.77$ \\
    $\mathcal{F}^-_{VS}$ &  &  &  & $1.00$ & $0.14$ & $0.32$ & $-0.22$ & $0.91$ & $-0.91$ \\
    $\mathcal{F}_{SS}$ &  &  &  &  & $1.00$ & $0.15$ & $-0.32$ & $0.00$ & $-0.10$ \\
    $\delta^D_{VV}$ &  &  &  &  &  & $1.00$ & $0.11$ & $0.40$ & $-0.43$ \\
    $\delta^+_{VS}$ &  &  &  &  &  &  & $1.00$ & $-0.00$ & $0.09$ \\
    $\delta^-_{VS}$ &  &  &  &  &  &  &  & $1.00$ & $-0.95$ \\
    $\delta_{SS}$ &  &  &  &  &  &  &  &  & $1.00$ \\
    \end{tabular}
\end{table}

\begin{table}
    \renewcommand{\arraystretch}{1.2}
    \centering
    \caption{Statistical uncertainty correlation matrix for the \BsKstKst amplitude fit in the transversity basis.}
    \label{tab:corr_transversity_Bs_stat}
    \begin{tabular}{l | r r r r }
        & $f_L$ & $f_\parallel$ & $f_\perp$ & $\delta_\parallel$ \\
    \hline
        $f_L$ & $1.00$ & $-0.27$ & $-0.45$ & $0.14$ \\
        $f_\parallel$ &  & $1.00$ & $-0.73$ & $0.03$ \\
        $f_\perp$ &  &  & $1.00$ & $-0.12$ \\
        $\delta_\parallel$ &  &  &  & $1.00$ \\
    \end{tabular}
\end{table}

\begin{table}
    \renewcommand{\arraystretch}{1.2}
    \centering
    \caption{Systematic uncertainty correlation matrix for the \BsKstKst amplitude fit in the transversity basis.}
    \label{tab:corr_transversity_Bs_syst}
    \begin{tabular}{l | r r r r }
        & $f_L$ & $f_\parallel$ & $f_\perp$ & $\delta_\parallel$ \\
    \hline
        $f_L$ & $1.00$ & $-0.96$ & $0.69$ & $-0.89$ \\
        $f_\parallel$ &  & $1.00$ & $-0.86$ & $0.73$ \\
        $f_\perp$ &  &  & $1.00$ & $-0.29$ \\
        $\delta_\parallel$ &  &  &  & $1.00$ \\
    \end{tabular}
\end{table}

\begin{table}
    \renewcommand{\arraystretch}{1.2}
    \centering
    \caption{Statistical uncertainty correlation matrix for the \BdKstKst amplitude fit in the transversity basis.}
    \label{tab:corr_transversity_Bd_stat}
    \begin{tabular}{l | r r r r }
        & $f_L$ & $f_\parallel$ & $f_\perp$ & $\delta_\parallel$ \\
    \hline
        $f_L$ & $1.00$ & $-0.46$ & $-0.62$ & $0.00$ \\
        $f_\parallel$ &  & $1.00$ & $-0.41$ & $0.21$ \\
        $f_\perp$ &  &  & $1.00$ & $-0.19$ \\
        $\delta_\parallel$ &  &  &  & $1.00$ \\
    \end{tabular}
\end{table}

\begin{table}
    \renewcommand{\arraystretch}{1.2}
    \centering
    \caption{Systematic uncertainty correlation matrix for the \BdKstKst amplitude fit in the transversity basis.}
    \label{tab:corr_transversity_Bd_syst}
    \begin{tabular}{l | r r r r }
        & $f_L$ & $f_\parallel$ & $f_\perp$ & $\delta_\parallel$ \\
    \hline
        $f_L$ & $1.00$ & $-0.49$ & $-0.53$ & $-0.51$ \\
        $f_\parallel$ &  & $1.00$ & $-0.47$ & $-0.19$ \\
        $f_\perp$ &  &  & $1.00$ & $0.70$ \\
        $\delta_\parallel$ &  &  &  & $1.00$ \\
    \end{tabular}
\end{table}

\FloatBarrier

\section{Hadronic results}
\label{sec:app_hadronic}

The measured pole parameters of the \Kst state from the amplitude analysis are given in Table~\ref{tab:kst_pole}. Considering only the statistical error, agreement between the \Bd and \Bs samples is marginal. Log-likelihood scans indicate that the production barrier-factor radius is correlated with the pole mass and width. While set to be identical between \Bd and \Bs decays in the baseline model, this radius may differ in principle. As the pole parameters are measured relatively cleanly atop of what amounts to a model-independent $S$-wave, future studies may permit a simultaneous measurement of the production radius and the \Kst pole, which is reminiscent of Ref.~\cite{BESIII:2015hty}. In turn, this will facilitate measurement of a common set of \Kst pole parameters within a joint analysis of \Bd and \Bs decays.
\begin{table}
    \renewcommand{\arraystretch}{1.2}
    \centering
    \caption{Measured pole parameters of the \Kst state for the \Bd and \Bs decay samples, where the uncertainties are statistical only.}
    \label{tab:kst_pole}
    \begin{tabular}{l | c | c}
        & \Bd & \Bs \\
    \hline
        $m_0\,(\!\gevcc)$ & $0.8963 \pm 0.0006$ & $0.8977 \pm 0.0003$\\
        $\Gamma_0\,(\!\gev)$ & $0.0422 \pm 0.0016$ & $0.0469 \pm 0.0008$\\
    \end{tabular}
\end{table}

\begin{figure}[tb]
    \centering
    \includegraphics[width=0.5\linewidth]{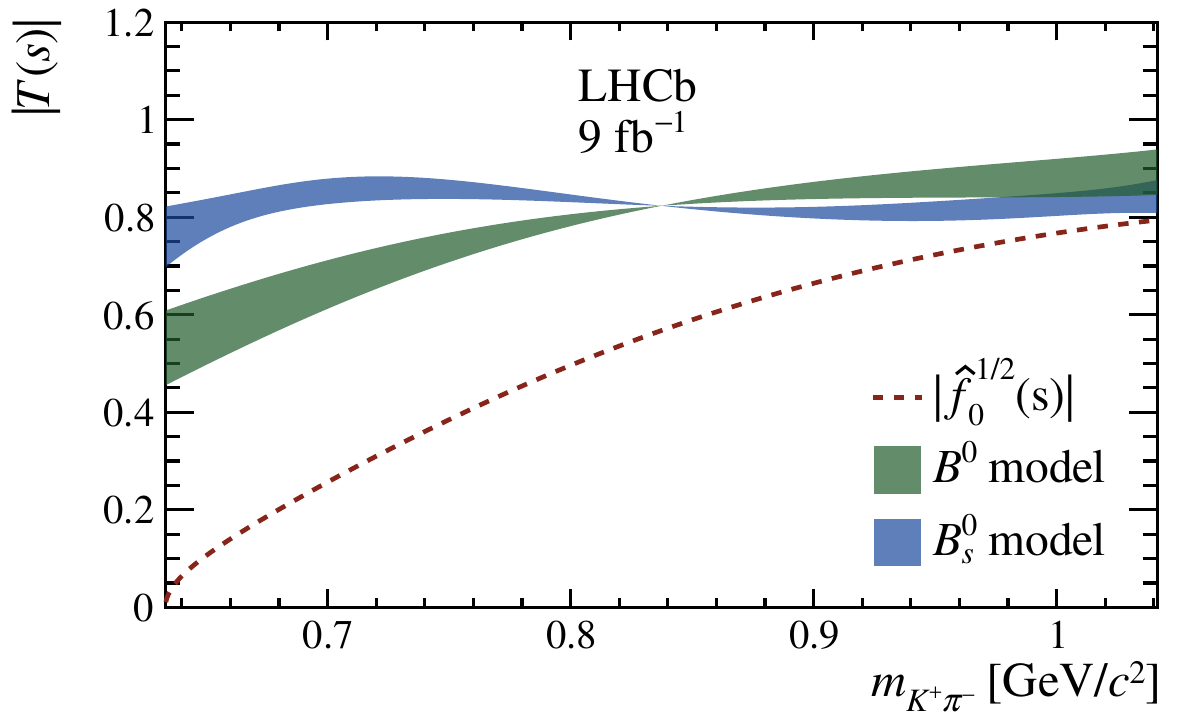}%
    \includegraphics[width=0.5\linewidth]{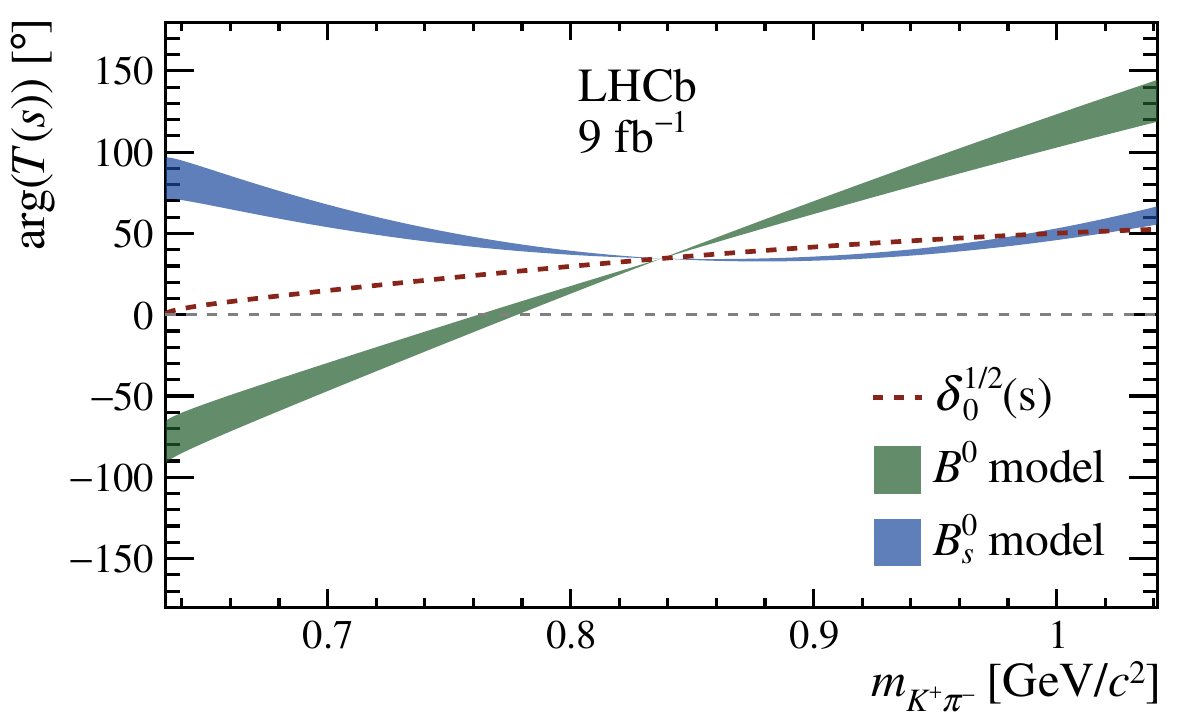}

    \caption{Measured (left)~magnitude and (right)~phase of the $S$-wave in the $K^\pm\pi^\mp$ system produced from $B^0_{(s)}$ decays. The spread indicates the uncertainty at $1\sigma$ and is determined by propagating the total covariance matrix of the raw covariant baseline result. For comparison, the dashed line shows the equivalent partial wave for $I=1/2$ elastic $\pi K$-$\pi K$ scattering~\cite{Pelaez:2020gnd}.}
    \label{fig:Swave}
\end{figure}

Another important outcome of this analysis is the structure of the $S$-wave in the $\Kpm\pimp$ system, which provides a measure of the penguin-induced production amplitude modulating the $\pi K$-$\pi K$ scattering amplitude as shown in Eq.~\ref{eq:swave}. Parameters determined in the raw baseline covariant fit comprising the production amplitude are included as part of Supplemental Material~\cite{supplemental}. 
The impact of the production environment on the $\Kpm\pimp$ mass propagator can be seen in Fig.~\ref{fig:Swave} when compared to the partial wave for $I=1/2$ elastic $\pi K$-$\pi K$ scattering given in Ref.~\cite{Pelaez:2020gnd}. 
A particularly noteworthy difference is the presence of nonzero magnitude and phase contributions at the lower kinematic threshold, showcasing how the penguin topologies involved in the production process lead to lineshapes that would otherwise violate unitarity in scattering.

\clearpage
\addcontentsline{toc}{section}{Supplemental material}
\section*{Supplemental material}
\label{sec:supplemental}

In addition to the results presented in the main body, a supplementary file recording complete expressions of the covariant amplitude model is provided as they are otherwise impractical to publish either in the main text or as Appendixes. This includes the set of fit fractions, interference fractions and strong phase differences. Also included are the set of parameters determined in the transversity basis. This file is given in the JavaScript Object Notation~(JSON) format, which is both machine and human readable, containing parameter names, central values and total uncertainties. An additional JSON file is included containing the correlations between all of the fit parameters, provided as an array of parameter names and a two-dimensional array consisting of the values of the correlations.

\addcontentsline{toc}{section}{References}
\bibliographystyle{LHCb/LHCb}
\bibliography{main,LHCb/standard,LHCb/LHCb-PAPER,LHCb/LHCb-CONF,LHCb/LHCb-DP,LHCb/LHCb-TDR}

\newpage
\centerline
{\large\bf LHCb collaboration}
\begin
{flushleft}
\small
R.~Aaij$^{38}$\lhcborcid{0000-0003-0533-1952},
A.S.W.~Abdelmotteleb$^{58}$\lhcborcid{0000-0001-7905-0542},
C.~Abellan~Beteta$^{52}$\lhcborcid{0009-0009-0869-6798},
F.~Abudin{\'e}n$^{58}$\lhcborcid{0000-0002-6737-3528},
T.~Ackernley$^{62}$\lhcborcid{0000-0002-5951-3498},
A. A. ~Adefisoye$^{70}$\lhcborcid{0000-0003-2448-1550},
B.~Adeva$^{48}$\lhcborcid{0000-0001-9756-3712},
M.~Adinolfi$^{56}$\lhcborcid{0000-0002-1326-1264},
P.~Adlarson$^{86}$\lhcborcid{0000-0001-6280-3851},
C.~Agapopoulou$^{14}$\lhcborcid{0000-0002-2368-0147}
% ...

{\footnotesize \it

% ...
$^{14}$Universit{\'e} Paris-Saclay, CNRS/IN2P3, IJCLab, Orsay, France\\
% ...
$^{38}$Nikhef National Institute for Subatomic Physics, Amsterdam, Netherlands\\
% ...
$^{48}$Instituto Galego de F{\'\i}sica de Altas Enerx{\'\i}as (IGFAE), Universidade de Santiago de Compostela, Santiago de Compostela, Spain\\
% ...
$^{52}$Physik-Institut, Universit{\"a}t Z{\"u}rich, Z{\"u}rich, Switzerland\\
% ...
$^{56}$H.H. Wills Physics Laboratory, University of Bristol, Bristol, United Kingdom\\
% ...
$^{58}$Department of Physics, University of Warwick, Coventry, United Kingdom\\
% ...
$^{61}$School of Physics and Astronomy, University of Glasgow, Glasgow, United Kingdom\\
$^{62}$Oliver Lodge Laboratory, University of Liverpool, Liverpool, United Kingdom\\
% ...
$^{70}$Syracuse University, Syracuse, NY, United States\\
% ...
$^{86}$Department of Physics and Astronomy, Uppsala University, Uppsala, Sweden, associated to $^{61}$\\
% ...
\medskip
}
\end{flushleft}

\end{document}